\documentclass[twocolumn]{aastex61}
\usepackage{lipsum}
\usepackage{amsmath,amstext}
\usepackage[T1]{fontenc}
\usepackage{apjfonts} 
\usepackage[shortcuts]{extdash}
\PassOptionsToPackage{hyphens}{url}\usepackage{hyperref}

\shorttitle{SPIRITS Luminous Infrared Transients}
\shortauthors{Jencson et al.}

\begin{document}


\title{The SPIRITS sample of Luminous Infrared Transients:\\ Uncovering Hidden Supernovae and Dusty Stellar Outbursts in Nearby Galaxies\footnote{This paper includes data gathered with the 6.5~m Magellan Telescopes located at Las Campanas Observatory, Chile.}}

\author{Jacob E.\ Jencson}
\altaffiliation{National Science Foundation Graduate Research Fellow}
\affiliation{Division of Physics, Mathematics and Astronomy, California Institute of Technology, Pasadena, CA 91125, USA}
\author{Mansi M.\ Kasliwal}
\affiliation{Division of Physics, Mathematics and Astronomy, California Institute of Technology, Pasadena, CA 91125, USA}
\author{Scott M.\ Adams}
\affiliation{Division of Physics, Mathematics and Astronomy, California Institute of Technology, Pasadena, CA 91125, USA}
\author{Howard E.\ Bond}
\affiliation{Department of Astronomy \& Astrophysics, Pennsylvania State University, University Park, PA 16802, USA}
\affiliation{Space Telescope Science Institute, 3700 San Martin Dr., Baltimore, MD 21218, USA}
\author{Kishalay De}
\affiliation{Division of Physics, Mathematics and Astronomy, California Institute of Technology, Pasadena, CA 91125, USA}
\author{Joel Johansson}
\affiliation{Department of Physics and Astronomy, Division of Astronomy and Space Physics, Uppsala University, Box 516, SE 751 20 Uppsala, Sweden}
\author{Viraj Karambelkar}
\affiliation{Department of Physics, Indian Institute of Technology Bombay, Mumbai 400076, India}
\author{Ryan M.\ Lau}
\affiliation{Division of Physics, Mathematics and Astronomy, California Institute of Technology, Pasadena, CA 91125, USA}
\affiliation{Institute of Space \& Astronautical Science, Japan Aerospace  Exploration Agency, 3-1-1 Yoshinodai, Chuo-ku, Sagamihara, Kanagawa 252-5210, Japan}
\author{Samaporn Tinyanont}
\affiliation{Division of Physics, Mathematics and Astronomy, California Institute of Technology, Pasadena, CA 91125, USA}
\author{Stuart D.\ Ryder}
\affiliation{Australian Astronomical Observatory, 105 Delhi Road, North Ryde, NSW 2113, Australia}
\affiliation{Department of Physics \& Astronomy, Macquarie University, NSW 2109, Australia}
\author{Ann Marie Cody}
\affiliation{NASA Ames Research Center, Moffet Field, CA 94035, USA}
\author{Frank J.\ Masci}
\affiliation{Caltech/IPAC, Mailcode 100-22, Pasadena, CA 91125, USA}
\author{John Bally}
\affiliation{Astrophysical and Planetary Sciences Department University of Colorado, UCB 389, Boulder, CO 80309, USA}
\author{Nadejda Blagorodnova}
\affiliation{Department of Astrophysics/IMAPP, Radboud University, Nijmegen, The Netherlands}
\author{Sergio Castell\'{o}n}
\affiliation{Las Campanas Observatory, Carnegie Observatories, Casilla 601, La Serena, Chile}
\author{Christoffer Fremling}
\affiliation{Division of Physics, Mathematics and Astronomy, California Institute of Technology, Pasadena, CA 91125, USA}
\author{Robert D. Gehrz}
\affiliation{Minnesota Institute for Astrophysics, School of Physics and Astronomy, University of Minnesota, 116 Church Street SE, Minneapolis, MN 55455, USA}
\author{George Helou}
\affiliation{Caltech/IPAC, Mailcode 100-22, Pasadena, CA 91125, USA}
\author{Charles D. Kilpatrick}
\affiliation{Department of Astronomy and Astrophysics, University of California, Santa Cruz, CA 95064, USA}
\author{Peter A.\ Milne}
\affiliation{University of Arizona, Steward Observatory, 933 N. Cherry Avenue, Tucson, AZ 85721, USA}
\author{Nidia Morrell}
\affiliation{Las Campanas Observatory, Carnegie Observatories, Casilla 601, La Serena, Chile}
\author{Daniel A.\ Perley}
\affiliation{Astrophysics Research Institute, Liverpool John Moores University, IC2, Liverpool Science Park, 146 Brownlow Hill, Liverpool L3 5RF, UK}
\author{M.\ M.\ Phillips}
\affiliation{Las Campanas Observatory, Carnegie Observatories, Casilla 601, La Serena, Chile}
\author{Nathan Smith}
\affiliation{University of Arizona, Steward Observatory, 933 N. Cherry Avenue, Tucson, AZ 85721, USA}
\author{Schuyler D.\ van Dyk}
\affiliation{Caltech/IPAC, Mailcode 100-22, Pasadena, CA 91125, USA}
\author{Robert E.\ Williams}
\affiliation{Department of Astronomy and Astrophysics, University of California, Santa Cruz, CA 95064, USA}
\affiliation{Space Telescope Science Institute, 3700 San Martin Dr., Baltimore, MD 21218, USA}

\correspondingauthor{Jacob E. Jencson}
\email{jj@astro.caltech.edu}

\begin{abstract}
We present a systematic study of the most luminous ($M_{\mathrm{IR}}$ [Vega magnitudes] brighter than $-14$) infrared (IR) transients discovered by the \textit{SPitzer} InfraRed Intensive Transients Survey (SPIRITS) between 2014 and 2018 in nearby galaxies ($D < 35$\,Mpc). The sample consists of nine events that span peak IR luminosities of $M_{[4.5],\mathrm{peak}}$ between $-14$ and $-18.2$, show IR colors between $0.2 < ([3.6]{-}[4.5]) < 3.0$, and fade on timescales between $55$~days $< t_{\mathrm{fade}} < 480$~days. The two reddest events ($A_V > 12$) show multiple, luminous IR outbursts over several years and have directly detected, massive progenitors in archival imaging. With analyses of extensive, multiwavelength follow-up, we suggest the following possible classifications: five obscured core-collapse supernovae (CCSNe), two erupting massive stars, one luminous red nova, and one intermediate-luminosity red transient. We define a control sample of all optically discovered transients recovered in SPIRITS galaxies and satisfying the same selection criteria. The control sample consists of eight CCSNe and one Type~Iax SN. We find that 7 of the 13 CCSNe in the SPIRITS sample have lower bounds on their extinction of $2 < A_V < 8$. We estimate a nominal fraction of CCSNe in nearby galaxies that are missed by optical surveys as high as $38.5^{+26.0}_{-21.9}$\% (90\% confidence). This study suggests that a significant fraction of CCSNe may be heavily obscured by dust and therefore undercounted in the census of nearby CCSNe from optical searches.
\end{abstract}


\section{Introduction}\label{sec:intro}
While there are now several known classes of stellar transient phenomena for which the observable emission is predominantly infrared (IR), exploration of the landscape of IR-dominated transients is just beginning. Often due to the effects of astrophysical dust, a host of eruptive and explosive stellar phenomena may be best observed in the IR. In particular, otherwise optically luminous transients such as supernovae (SNe) may be significantly obscured by dust in their host galaxies and/or local environments. Dust in the immediate circumstellar environment of a luminous transient, which may have condensed in a steady wind of the progenitor or formed during previous eruptive mass-loss events, may also efficiently reprocess shorter-wavelength emission into the IR. Some transients, particularly those associated with cool, low-velocity outflows, are themselves copious dust producers, leading to IR-dominated spectral energy distributions (SEDs).

Since 2014, we have been conducting the \textit{SPitzer} InfraRed Intensive Transients Survey (SPIRITS; PIDs 11063, 13053, 14089; PI M. Kasliwal, \citealp{kasliwal17}) to discover transients in nearby $D \lesssim 35$\,Mpc galaxies using the 3.6 and 4.5~\micron\ imaging bands ([3.6] and [4.5]) of the Infrared Array Camera (IRAC; \citealp{fazio04}) on board the warm \textit{Spitzer Space Telescope} (\textit{Spitzer}; \citealp{werner04,gehrz07}). In this paper, we focus on a thorough investigation of all luminous ($M_{[4.5]}$ [Vega magnitudes] brighter than -14) IR transients discovered by SPIRITS in the last five years (Section~\ref{sec:obsc_sample}). We compare this sample to a well-defined control sample of all optically discovered and spectroscopically classified transients hosted by SPIRITS galaxies and satisfying the same selection criteria (Section~\ref{sec:control_samp}). Our sample of luminous infrared transients may represent diverse origins, including obscured core-collapse supernovae (CCSNe) and other known classes of IR-dominated transients such as stellar mergers or massive star eruptions (MSEs). 

CCSNe, the explosive deaths of stars of initial masses $\gtrsim 8~M_{\odot}$, are now found in numbers exceeding several hundreds of events per year by numerous, primarily optical, searches. Arising from recently formed, massive stars, CCSNe may be subject to significant extinction from the dusty regions of active star formation in their host galaxies. The fraction of CCSNe missed optically owing to the obscuring effects of dust is therefore an important consideration for measurements of the CCSN rate \citep[e.g.,][]{grossan99,maiolino02}. \citet{horiuchi11} claimed that half of all SNe were missing from observed CCSN rate estimates in comparison to the rate of massive-star formation both locally and across cosmic time from redshifts $0 < z < 1$; however, other studies have found better agreement \citep[e.g.,][]{cappellaro15}. Accounting for obscured or otherwise optically dim CCSNe may also resolve this discrepancy \citep[e.g.,][]{mannucci07,mattila12}. Direct searches for CCSNe at wavelengths less sensitive to extinction are thus required to accurately measure the CCSN rate.  Significant work \citep[e.g.,][and references therein; see discussion in Section~\ref{sec:rates}]{varenius17,kool18} has been dedicated to uncovering extinguished CCSNe in the densely obscured and highly star-forming regions of starburst and (ultra)luminous infrared galaxies (U/LIRGS). In contrast, our SPIRITS sample focuses on local galaxies encompassing a wide variety of galaxy morphologies, masses, and star-formation rates with unbiased sampling of all environments within galaxies (Section~\ref{sec:survey}).  

Other categories of luminous IR transients include intermediate-luminosity red transients (ILRTs), luminous red novae (LRNe), and giant eruptions of luminous blue variables (LBVs)\footnote{Distinguishing among classes of hydrogen-rich, intermediate-luminosity transients is difficult, as they may share many observational properties, and the nomenclature referring to these classes varies throughout the literature. We use the terms ILRT \citep[originally suggested by][]{bond09}, LRN \citep[e.g.,][]{kulkarni07}, and ``giant eruption'' or LBV to refer to these classes as described in the main text. In the literature, giant eruptions may also be referred to as ``SN imposters'' or ``$\eta$ Carinae variables'' \citep{humphreys99,vandyk00,pastorello10,smith10,smith14}. ILRTs are sometimes included in this class, though their progenitors are believed to be lower mass than classical LBVs. Our choice of nomenclature is similar to that used by, e.g., \citet{kashi16}, who also adopted the umbrella term ``intermediate-luminosity optical transient'' of \citet{berger09} to refer to all three of these classes together.}. The prototypical objects for ILRTs are the ``impostor'' SN~2008S and the 2008 transient in NGC~300 (NGC~300 OT2008-1; \citealp{bond09}), suggested to be explosions of ${\approx}10$--$15~M_{\odot}$ stars, possibly extreme asymptotic giant branch (AGB) stars, self-obscured by a dusty wind \citep{prieto08,bond09,thompson09}. Extragalactic LRNe are believed to be massive analogs of stellar mergers observed in the Galaxy, including the striking example of the ${\approx}1$--$3~M_{\odot}$ contact binary merger V1309~Sco \citep{tylenda11} and the B-type stellar merger V838~Mon \citep{bond03,sparks08}. While sharing many observational properties with, e.g., ILRTs, a key difference is that LRNe have surviving remnants (e.g., NGC~4490 OT2011-1; \citealp{smith16}). Surviving dust in the circumstellar environments of LBVs may also result in an observed IR excess during some nonterminal giant eruptions (e.g., $\eta$ Carinae; \citealp{humphreys99}, UGC~2773OT; \citealp{smith10}).

It is particularly telling that known examples and even class prototypes of known IR-dominated events have almost exclusively been identified via their optical emission. It has been proposed, for example, that early dust formation in some LRNe may even entirely obscure or dramatically shorten the optical luminosity peak, while still producing a bright, long-lived infrared transient \citep[e.g.,][]{metzger17}. \citet{kashi17} have also suggested that dense, equatorial material may obscure transients driven by strong binary interactions, but that dust formation in polar outflows may still power IR emission in these events. As suggested by \citet{bally05}, dynamical interactions of massive, compact-multiple systems in star-forming regions may also produce luminous transients, the energy released likely emerging in the IR given their densely obscured environments (see also recent events in the Galactic star-forming regions Sh2-255 and NGC\,6334; \citealp{carattiogaratti17,hunter17,brogan18}). SPIRITS overcomes the selection biases of optical discovery and is sensitive to redder events that may lack optical counterparts altogether (e.g., the recently discovered eSPecially Red Intermediate-luminosity Transient Events [SPRITEs]; \citealp{kasliwal17}).

We have undertaken extensive follow-up, including optical/IR photometry, spectroscopy, and radio imaging to characterize the nature of each luminous IR transient presented in this work (Section~\ref{sec:obs}). We describe our analysis of the full data set, including host galaxy properties, progenitor constraints, photometric evolution, spectroscopic feature identification, and extinction estimates, in Section~\ref{sec:analysis}. In Section~\ref{sec:class}, we combine all available observational constraints in comparison to well-studied objects and attempt to classify each luminous SPIRITS transient. In Section~\ref{sec:redness}, we discuss the $A_V$ distribution of nearby, luminous IR transients and CCSNe in particular, and in Section~\ref{sec:rates}, we derive statistically robust estimates of the rate of CCSNe missed in nearby galaxies relative to the control sample of optically discovered CCSNe in SPIRITS galaxies. Finally, in Section~\ref{sec:summary}, we summarize the main results and conclusions of this work. 

\section{Survey design and sample selection}\label{sec:exp_design}

\subsection{Galaxy sample and imaging cadence}\label{sec:survey}
A full description of the SPIRITS survey design is given in \citet{kasliwal17}. It is a targeted search of nearby galaxies using the $5 \times 5\arcmin$ field of view of the IRAC camera. SPIRITS monitored a sample of 190 nearby galaxies for 3~yr from 2014 to 2016. The sample was selected based on the following criteria: (1) The 37 galaxies within 5\,Mpc including both early- and late-type galaxies, dwarf galaxies, and giant galaxies, (2) the 116 most luminous galaxies between 5 and 15\,Mpc, including 83\% of the $B$-band starlight within 15\,Mpc, and (3) the 37 most luminous and massive galaxies in the Virgo Cluster at 17\,Mpc.

In 2014, each galaxy was observed 3 times at ${\sim} 1$-month and ${\sim} 6$-month intervals. In 2015-2016, baselines of 1- and 3-week timescales were added. In 2017--2018, the galaxy sample was reduced to focus on only the 105 galaxies most likely to host new transients and SNe, including the 58 galaxies that had previously hosted at least one IR transient, and the 47 remaining most luminous and star-forming galaxies ($L > 2\times10^{10}~L_{\odot}$). Our cadence was also reduced to ${\sim} 6$-month intervals, typically with one observation per galaxy per visibility window. Each SPIRITS observation consists of seven dithered 100~s exposures in both IRAC bands. The nominal $5\sigma$ point-source limiting magnitudes of these observations are $20.0$ and $19.1$ at [3.6] and [4.5], respectively, using the zero-magnitude fluxes given in the IRAC instrument handbook\footnote{\url{http://irsa.ipac.caltech.edu/data/SPITZER/docs/irac/iracinstrumenthandbook/}} of $F_{\nu0} = 280.9$~Jy for [3.6] and $F_{\nu0} = 179.7$~Jy for [4.5].

\subsection{Image subtraction and transient identification}\label{sec:imsub}
For reference images, we make use of archival \textit{Spitzer} frames, including Super Mosaics\footnote{Super Mosaics are available as Spitzer Enhanced Imaging Products through the NASA/IPAC Infrared Science Archive: \url{https://irsa.ipac.caltech.edu/data/SPITZER/Enhanced/SEIP/overview.html}} or S4G (\textit{Spitzer} Survey of Stellar Structure in Galaxies; PID 61065; PI K. Sheth; \citealp{sheth10,munoz-mateos13,querejeta15}), or stacks of archival ``bcd'' images (where Super Mosaics or S4G images were not available). Further details of our image subtraction and transient identification pipelines are provided in \citet{kasliwal17}. Transient candidates automatically identified in by our pipeline in reference-subtracted images are vetted by human scanners, and sources passing human vetting are saved to our database and assigned a SPIRITS name. 

Photometry is performed at the location of SPIRITS transients in the reference-subtracted images using a 4 mosaicked pixel ($2\farcs4$) aperture and background annulus from 4 to 12 pixels ($2\farcs4$--$7\farcs2$). The extracted flux is multiplied by the aperture corrections of 1.215 for [3.6] and 1.233 for [4.5], as described in the IRAC instrument handbook, and converted to Vega system magnitudes using the handbook-defined zero-magnitude fluxes for each IRAC channel. 

\subsection{Luminous IR transient sample selection}\label{sec:obsc_sample}
\begin{figure*}
 \gridline{\fig{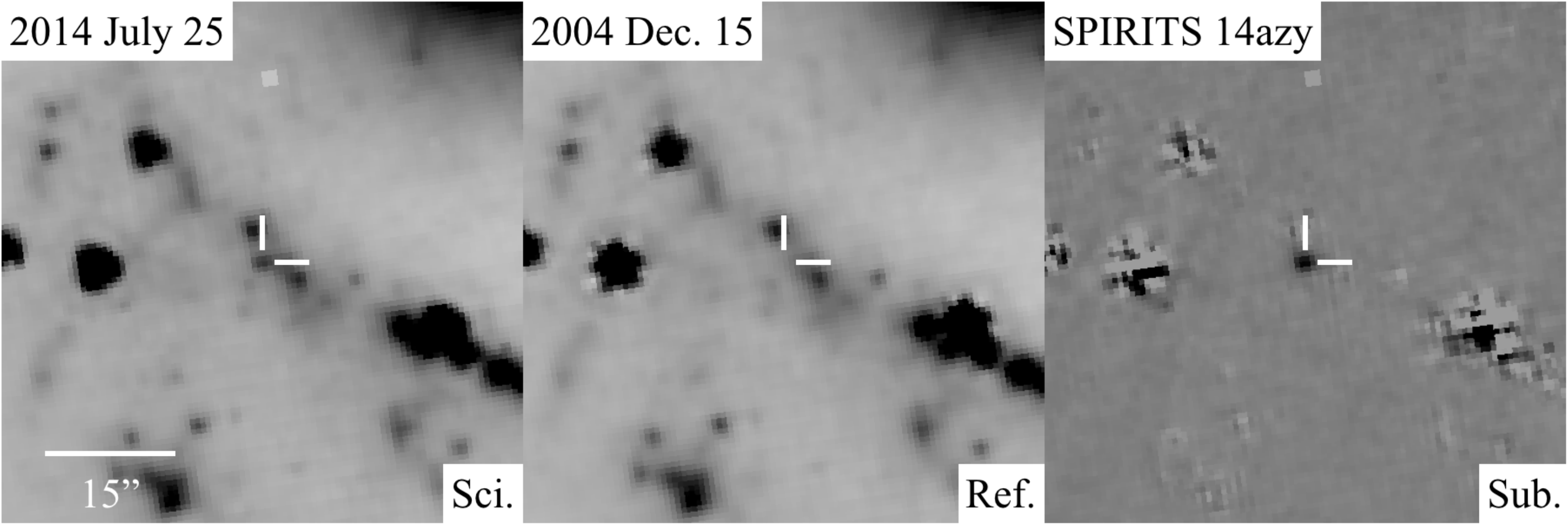}{0.49\textwidth}{} 
	\fig{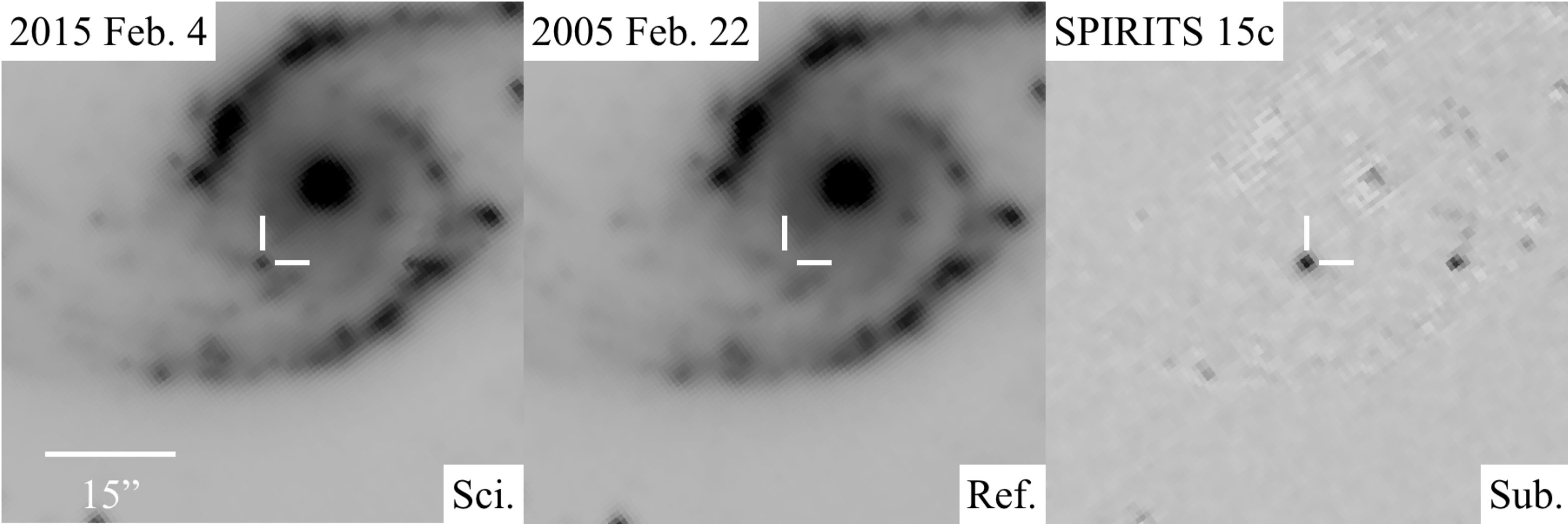}{0.49\textwidth}{}}
 \gridline{\fig{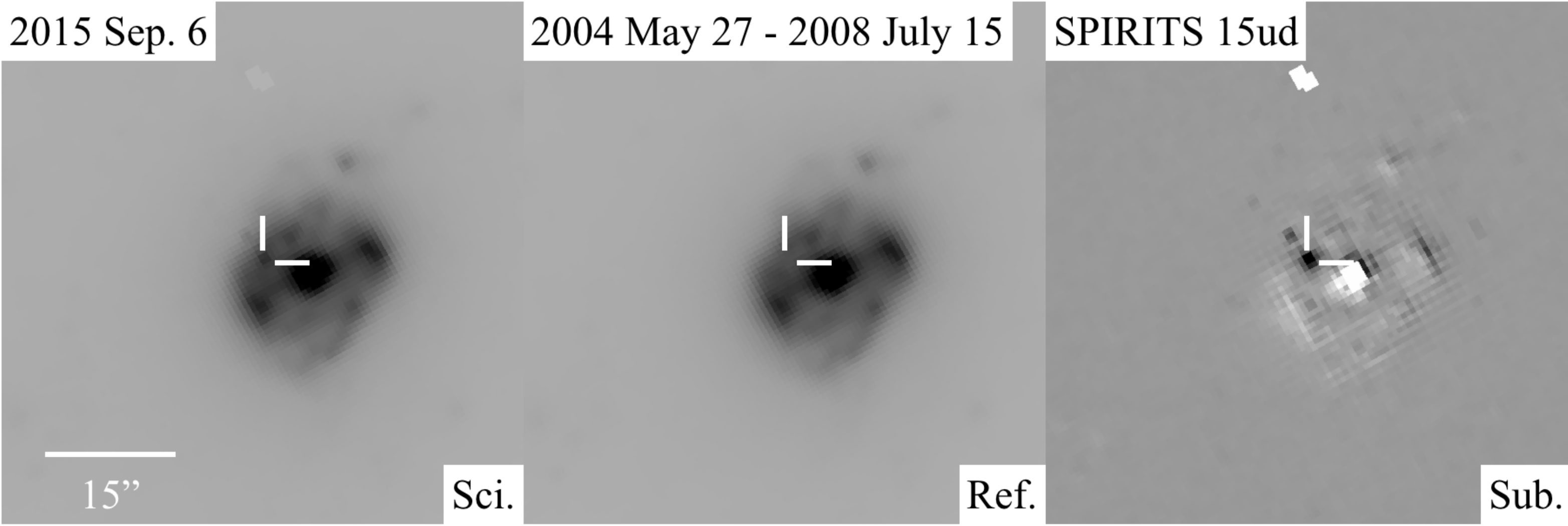}{0.49\textwidth}{} 
	\fig{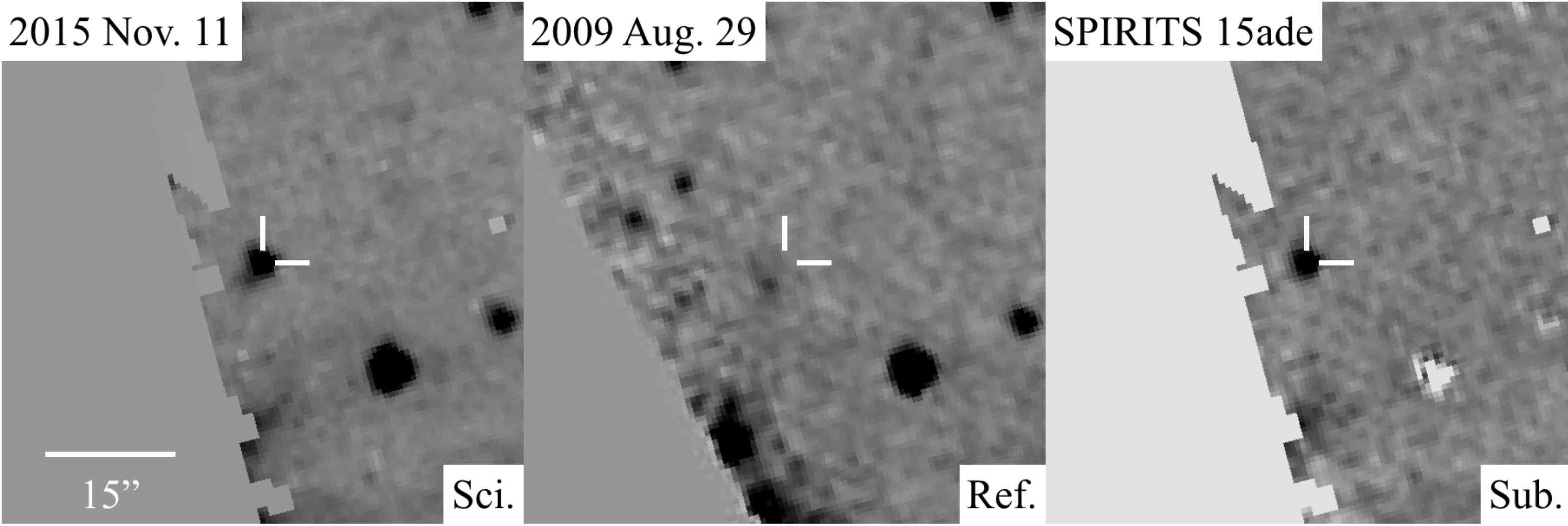}{0.49\textwidth}{}}
 \gridline{\fig{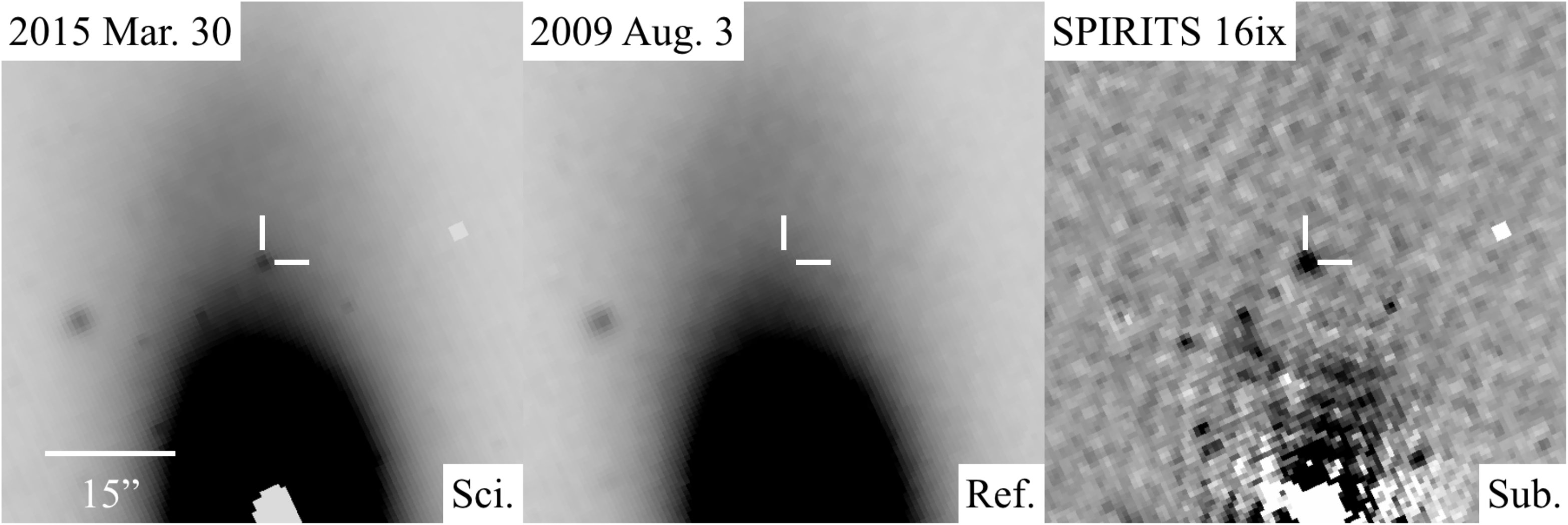}{0.49\textwidth}{} 
	\fig{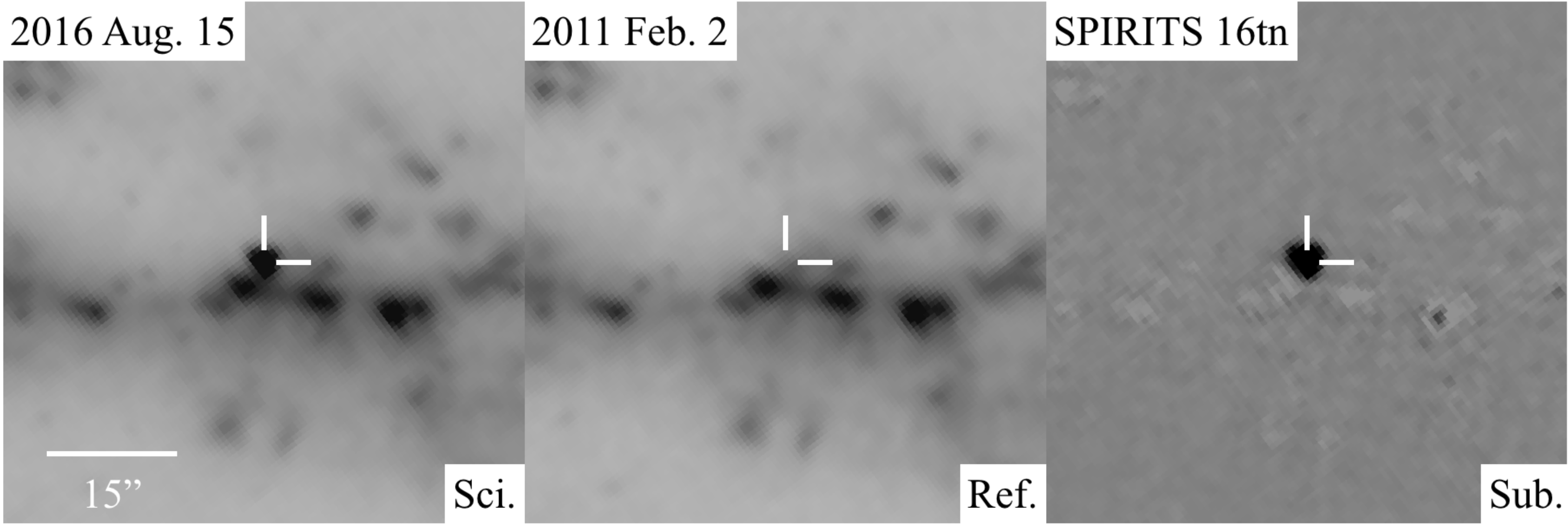}{0.49\textwidth}{}}
 \gridline{\fig{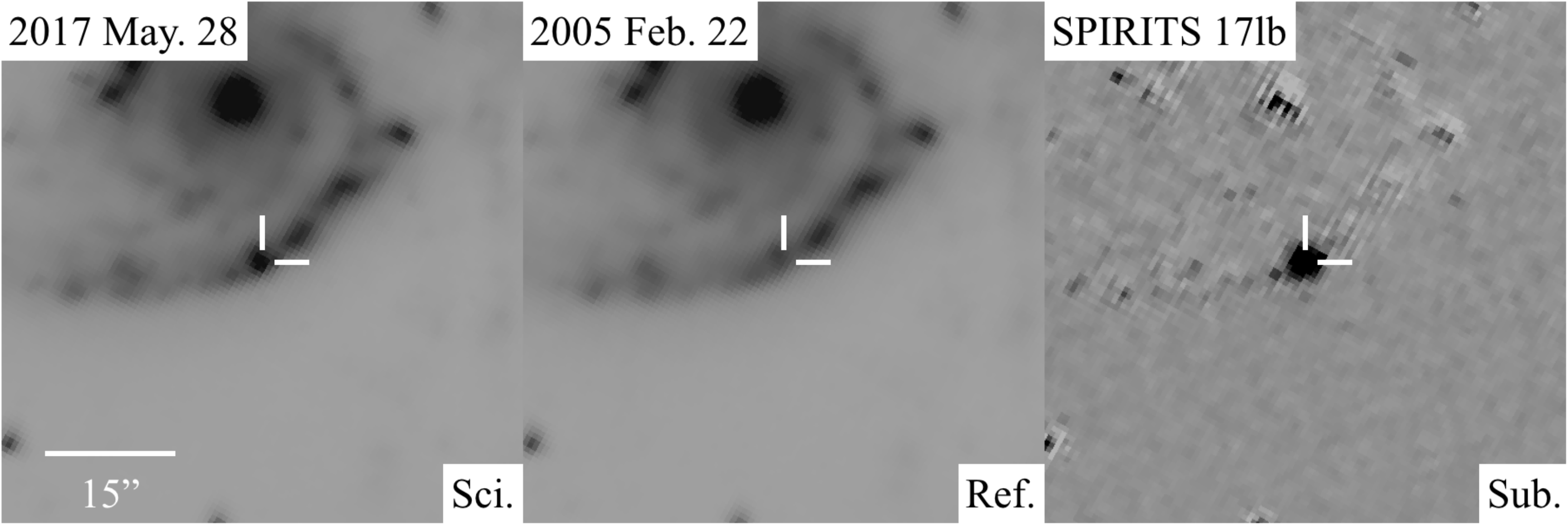}{0.49\textwidth}{} 
	\fig{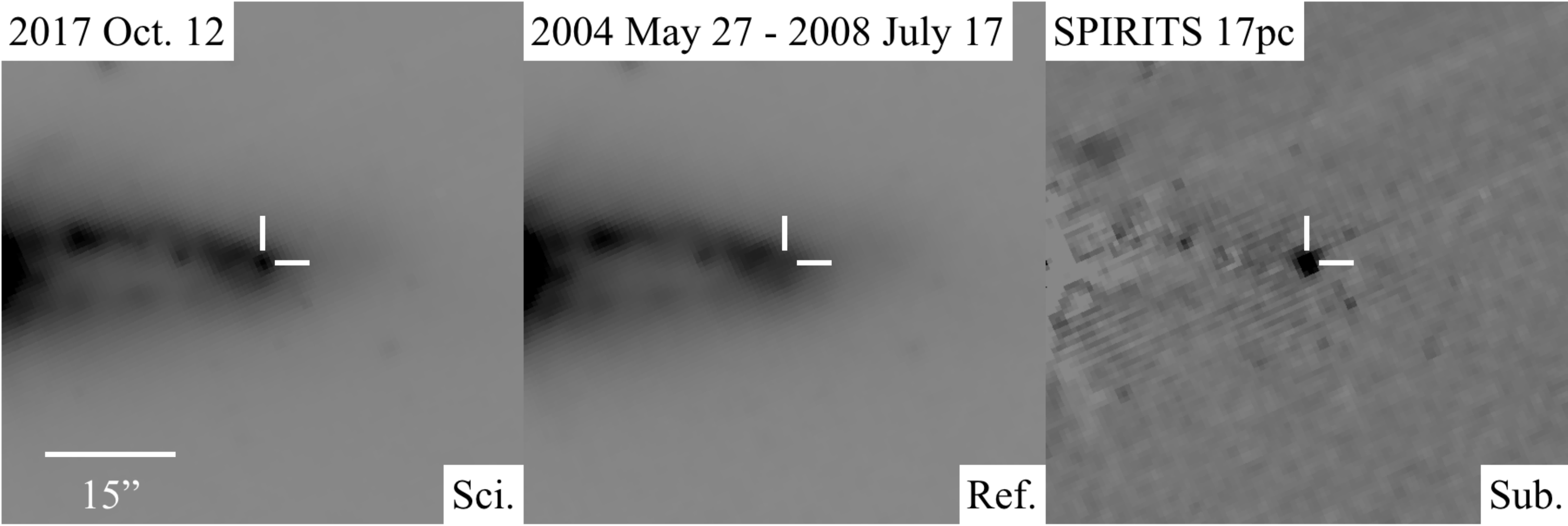}{0.49\textwidth}{}}
 \gridline{\fig{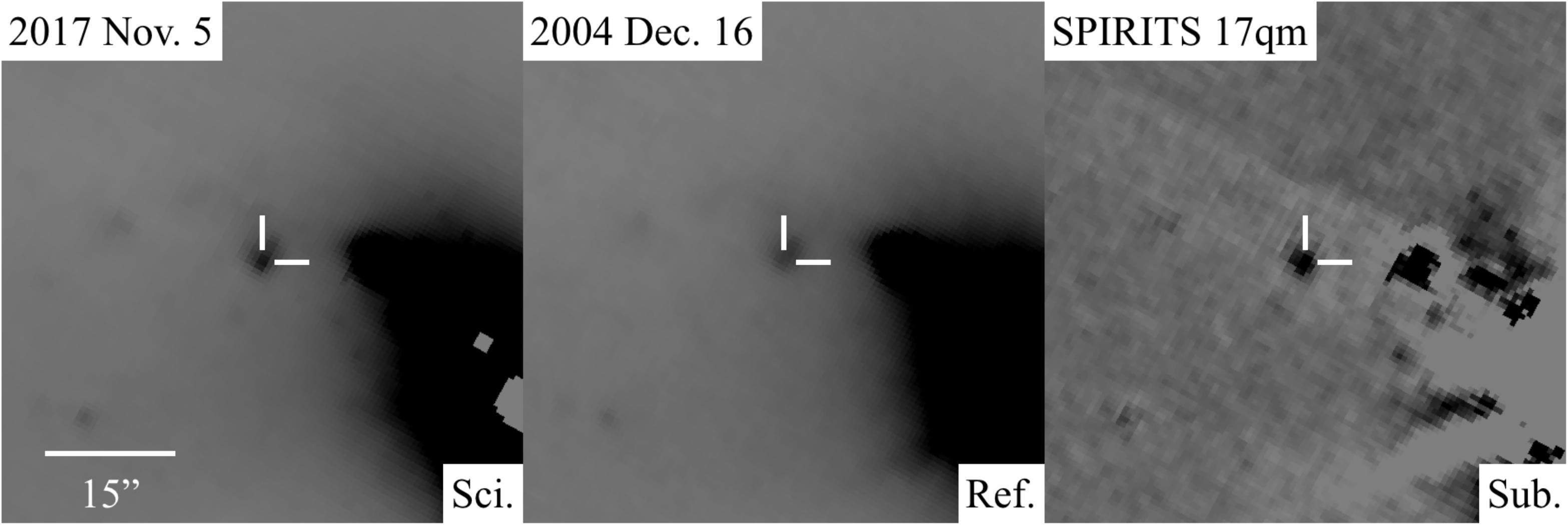}{0.49\textwidth}{}}
\caption{\label{fig:disc_images}
We show the \textit{Spitzer}/IRAC [4.5] discovery images for each luminous SPIRITS-discovered transient in our sample. In each panel, from left to right, we show the new science frame, reference frame, and science-minus-reference subtraction image. The dates of each image and labels for each object are shown along the top of each panel, and the locations of the transients are indicated by white crosshairs. Each image is $1\arcmin\times1\arcmin$, oriented with  N up and E left. 
}
\end{figure*}

We selected events that peaked at $M_{\mathrm{IR}}$ brighter than $-14$ ($\nu L_{\nu} > 1.5 \times 10^{5}~L_{\odot}$ at [4.5]) in either the [3.6] channel or [4.5] channel of \textit{Spitzer}/IRAC during SPIRITS observations\footnote{The full sample of SPIRITS transients includes many lower-luminosity events, including IR detections of classical novae (e.g., M81 possible nova AT~2018akh; \citealp{jencson18d}), and events belonging to the diverse and mysterious class of objects called SPRITEs, first described by \citet{kasliwal17}. These events are outside the scope of this work and will be studied in future publications.}. We further required at least two SPIRITS detections, to ensure that each event is astrophysically real and that the transients were not present in the first epoch of SPIRITS imaging, so that the age of the event is constrained by SPIRITS data. We list basic properties for these objects in Table~\ref{tab:IRsample}. SPIRITS\,14buu (first presented in \citealp{jencson17}) was identified in the first epoch of SPIRITS imaging of the galaxy IC\,2163. As such, we have no constraint on the age of this object from SPIRITS data, and thus we exclude it from the primary sample. The sample then consists of nine events discovered in SPIRITS between 2014 and 2018 that, to our knowledge, were not identified and spectroscopically classified by any other survey\footnote{An optical transient at the location of SPIRITS\,15ade was first discovered on 2015 September 11.5 and reported by M.\ Aoki as PSN\,J15220552+\allowbreak 0503160 through the Central Bureau for Astronomical Telegrams (CBAT; \url{http://www.cbat.eps.harvard.edu/unconf/followups/J15220552+0503160.html}). To our knowledge, no spectroscopy for classification or host confirmation was reported before this work.}. The \textit{Spitzer}/IRAC [4.5] discovery images for each object, including the new science frame, reference image, and science-minus-reference subtractions, are shown in Figure~\ref{fig:disc_images}.

\begin{deluxetable*}{lccccccccc}
\tablecaption{Luminous, IR-discovered SPIRITS transients\label{tab:IRsample}}
\tablehead{\colhead{Name} & \colhead{R.A.} & \colhead{Decl.} & \colhead{Host (Type)} & \colhead{$m-M$\tablenotemark{a}} & \colhead{Distance} & \colhead{ $E(B-V)_{\mathrm{MW}}$\tablenotemark{b} } & \colhead{UT Discovery} & \twocolhead{Host Offset} \\ 
\colhead{} & \colhead{(J2000)} & \colhead{(J2000)} & \colhead{} & \colhead{(mag)} & \colhead{(Mpc)} & \colhead{(mag)} & \colhead{} & \colhead{(arcsec)} & \colhead{(kpc)}}
\startdata
SPIRITS\,14buu & 06:16:27.2 & --21:22:29.2 & IC~2163 (SBc pec) & $32.75 \pm 0.4$ & 35.5 & 0.077 & 2014 Jan 13.9\tablenotemark{c} & 11.5 & 2.0 \\
\hline
SPIRITS\,14azy & 09:45:40.92 & --31:12:07.8 & NGC~2997 (SABc) & $30.43 \pm 0.16$ & 12.2 & 0.027 & 2014 Jul 26.0 & 48.4 & 2.9 \\
SPIRITS\,15c & 06:16:28.49 & --21:22:42.2 & IC~2163 (SBc pec) & $32.75 \pm 0.4$ & 35.5 & 0.077 & 2015 Feb 04.4 & 11.6 & 2.0 \\
SPIRITS\,15ud & 12:22:55.29 & +15:49:22.0 & M100 (SABbc) & $30.72 \pm 0.06$ & 13.9 & 0.023 & 2015 Sep 06.7 & 7.5 & 0.5 \\
SPIRITS\,15ade & 15:22:05.55 & +05:03:15.9 & NGC~5921 (SBbc) & $31.9 \pm 0.2$ & 24.0 & 0.036 & 2015 Nov 11.9 & 146.3 & 17.0 \\
SPIRITS\,16ix & 12:29:03.16 & +13:11:30.7 & NGC~4461 (SB0$^+$) & $31.48 \pm 0.25$ & 19.8 & 0.02 & 2016 Mar 30.9 & 29.2 & 2.8 \\
SPIRITS\,16tn & 11:11:20.40 & +55:40:17.3 & NGC~3556 (SBcd) & $29.7 \pm 0.4$ & 8.7 & 0.015 & 2016 Aug 15.0 & 89.9 & 3.8 \\
SPIRITS\,17lb & 06:16:27.78 & --21:22:51.7 & IC~2163 (SBc pec) & $32.75 \pm 0.4$ & 35.5 & 0.077 & 2017 May 28.7 & 18.8 & 3.2 \\
SPIRITS\,17pc & 12:25:44.43 & +12:39:44.5 & NGC~4388 (SAb) & $31.3 \pm 0.4$ & 18.2 & 0.029 & 2017 Oct 12.8 & 33.9 & 3.0 \\
SPIRITS\,17qm & 03:33:38.85 & --36:08:09.4 & NGC~1365 (SBb) & $31.31 \pm 0.06$ & 18.3 & 0.018 & 2017 Nov 05.2 & 34.1 & 3.0 \\
\enddata
\tablenotetext{a}{References for distance moduli: NGC~2997 \citep{hess09}, IC~2163 \citep{theureau07}, M100 and NGC~4461 \citep{tully13}, NGC~5921 \citep{rodriguez14}, NGC~3556 and NGC~4388 \citep{sorce14}, NGC~1365 \citep{riess16}.}
\tablenotetext{b}{Galactic extinction estimates taken from NED using the \citet{schlafly11} recalibration of the \citet{schlegel98} IR-based dust map assuming a \citet{fitzpatrick99} extinction law with $R_V = 3.1$.}
\tablenotetext{c}{Transient present in first 2014 SPIRITS epoch and therefore excluded from the primary sample.}
\end{deluxetable*}

\subsection{Optically discovered control sample}\label{sec:control_samp}
To place our sample of luminous IR transients in context, we define a control sample of optically discovered and classified transients recovered during normal operation of the SPIRITS survey, employing the same selection criteria as for the IR-selected sample. A total of 14 such transients hosted by SPIRITS galaxies and discovered at optical wavelengths have been reported since the start of the survey in 2014. We summarize basic properties of these events in Table~\ref{table:opt_sample}.

\begin{deluxetable*}{llccccccclccc}
\rotate
\tablecaption{Optically discovered and classified transients in SPIRITS \label{table:opt_sample}}
\tablehead{ \colhead{Name} & \colhead{SPIRITS Name\tablenotemark{a}} & \colhead{R.A.} & \colhead{Decl.} & \colhead{Host (Type)} & \colhead{$m-M$\tablenotemark{b}} & \colhead{$D$} & \colhead{$E(B-V)_{\mathrm{MW}}$\tablenotemark{c}} & \colhead{$A_{V}$\tablenotemark{d}} & \colhead{UT Discovery\tablenotemark{e}} & \twocolhead{Host Offset} & \colhead{Type\tablenotemark{f}} \\ 
\colhead{} & \colhead{} & \colhead{(J2000)} & \colhead{(J2000)} & \colhead{} & \colhead{(mag)} & \colhead{(Mpc)} & \colhead{(mag)} & \colhead{(mag)} & \colhead{} & \colhead{(arcsec)} & \colhead{(kpc)} & \colhead{} }
\startdata
SN~2014C & SPIRITS\,14aom & 22:37:05.60 & +34:24:31.9 & NGC~7331 (SAb) & $30.71 \pm 0.08$ & 13.9 & 0.08 & 0.22 & 2014 Jan 05.1\tablenotemark{g} & 31.0 & 2.1  & Ib/IIn \\
SN~2014J & SPIRITS\,14pw & 09:55:42.14 & +69:40:26.0 & M82 (I0) & $27.74 \pm 0.08$ & 3.5 & 0.14 & 0.98 & 2014 Jan 21.8\tablenotemark{g} & 58.6 & 1.0 & Ia \\
SN~2014L & SPIRITS\,14we & 12:18:48.68 & +14:24:43.5 & NGC~4254 (SAc) & $30.7 \pm 0.2$ & 13.8 & 0.11 & \nodata & 2014 Jan 26\tablenotemark{g} & 20.8 & 1.4 & Ic \\
\hline
SN~2014bc & \nodata & 12:18:57.71 & +47:18:11.3 & NGC~4258 (SABbc) & $29.32 \pm 0.05$ & 7.3 & 0.01 & \nodata & 2014 May 19.3 & 3.7 & 0.1 & II \\
SN~2014bi & SPIRITS\,15bx & 12:06:02.99 & +47:29:33.5 & NGC~4096 (SABc) & $30.4 \pm 0.4$ & 12.1 & 0.02 & $4.3$ & 2014 May 31.3 & 54.4 & 3.2 & IIP \\
SN~2014df & SPIRITS\,14bsc & 03:44:23.99 & --44:40:08.1 & NGC~1448 (SAcd) & $31.31 \pm 0.05$ & 18.3 & 0.01 & $\sim 0$ & 2014 Jun 03.2 & 121.1 & 10.7 & Ib \\
ASASSN-14ha & SPIRITS\,15yp & 04:20:01.41 & --54:56:17.0 & NGC~1566 (Sab pec?) & $31.1 \pm 0.4$ & 16.8 & 0.01 & $\sim 0$ & 2014 Sep 10.3 & 8.6 & 0.7 & II \\
SN~2014dt & SPIRITS\,15sd & 12:21:57.57 & +04:28:18.5 & M61 (SABbc) & $31.43 \pm 0.07$ & 19.3 & 0.02 & $\sim 0$ & 2014 Oct 29.8 & 40.5 & 3.8 & Iax \\
SN~2016C & SPIRITS\,16ot & 13:38:05.30 & --17:51:15.3 & NGC~5247 (SAbc) & $31.7 \pm 0.4$ & 21.9 & 0.07 & $\sim 0.2$ & 2016 Jan 03.8 & 111.9 & 11.9 & IIP \\
SN~2016adj & \nodata & 13:25:24.11 & --43:00:57.5 & Cen A (S0 pec) & $27.82 \pm 0.06$ & 3.7 & 0.10 & 0.23 & 2016 Feb 08.6 & 40.1 & 0.7 & Ib  \\
SN~2016bau & SPIRITS\,16is & 11:20:59.02 & +53:10:25.6 & NGC~3631 (SAc) & $31.2 \pm 0.4$ & 17.4 & 0.01 & $\sim 3.3$ & 2016 Mar 14.0 & 37.8 & 3.2 & Ib \\
SN~2016bkv & SPIRITS\,17eb & 10:18:19.31 & +41:25:39.3 & NGC~3184 (SABcd) & $30.79 \pm 0.05$ & 14.4 & 0.01 & 0.0 & 2016 Mar 21.7 & 30.3 & 2.1 & IIP \\
SN~2016cok & SPIRITS\,17ft & 11:20:19.10 & +12:58:56.0 & M66 (SABb) & $29.78 \pm 0.07$ & 9.0 & 0.03 & 0.16 & 2016 May 28.5 & 69.1 & 3.0 & IIP \\
SN~2017eaw & SPIRITS\,18k & 20:34:44.24 & +60:11:35.9 & NGC~6946 (SABcd) & $28.73 \pm 0.05$ & 5.6 & 0.30 & $\lesssim 0.13$ & 2017 May 14.2 & 154.1 & 4.2 & IIP \\
\enddata
\tablenotetext{a}{SN~2014bc was located near the saturated nucleus of NGC~4258 in the \textit{Spitzer}/IRAC images and not detected in SPIRITS. SN~2016adj itself was saturated in the \textit{Spitzer}/IRAC images of Centaurus~A and not recovered by the SPIRITS pipeline.}
\tablenotetext{b}{References for distance moduli: NGC~7331, M82, NGC~4254, NGC~4258, Centaurus~A, M66 \citep{tully13}, NGC~4096, \citep{sorce14}, NGC~1448 \citep{riess16}, NGC~1566 \citep{nasonova11}, M61 and NGC~6946 \citep{rodriguez14}, NGC~5247 \citep{tully88}, NGC~3631 \citep{theureau07}, NGC~3184 \citep{ferrarese00}.}
\tablenotetext{c}{Galactic extinction estimates taken from NED using the \citet{schlafly11} recalibration of the \citet{schlegel98} IR-based dust map assuming a \citet{fitzpatrick99} extinction law with $R_V = 3.1$.}
\tablenotetext{d}{References for host extinction estimates: SN~2014C \citep{milisavljevic15}, SN~2014J \citep[with $R_V = 1.4$;][]{amanullah14}, SN~2014bi (J. Johansson et al.\ 2019, in preparation), SN~2014dt \citep{foley14}, SN~2016adj \citep[with $R_V = 2.57$;][]{banerjee18}, SN~2016bkv \citep{hosseinzadeh18}, SN~2016cok \citep{kochanek17a}, SN~2017eaw \citep{kilpatrick18}.}
\tablenotetext{e}{References for SN discovery: SN~2014C \citep{kim14}, SN~2014J \citep{fossey14}, SN~2014L \citep{zhang14}, SN~2014bc \citep{smartt14}, SN~2014bi \citep{kumar14}, SN~2014df \citep{monard14}, ASASSN-14ha \citep{kiyota14}, SN~2014dt \citep{nakano14}, SN~2016C \citep{aoki16}, SN~2016ajd \citep{marples16}, SN~2016bau \citep{arbour16}, SN~2016bkv \citep{itagaki16}, SN~2016cok \citep{bock16}, and SN~2017eaw \citep{wiggins17}.}
\tablenotetext{f}{References for SN classification: SN~2014C \citep{kim14,milisavljevic15}, SN~2014J \citep{cao14}, SN~2014L \citep{li14}, SN~2014bc \citep{cortini14}; SN~2014bi \citep{kumar14}, SN~2014df \citep{monard14}, ASASSN-14ha \citep{arcavi14}, SN~2014dt \citep{ochner14}, SN~2016C \citep{sahu16}, SN~2016adj \citep{stritzinger14}, SN~2016bau \citep{granata16}, SN~2016bkv \citep{hosseinzadeh16}, SN~2016cok \citep{zhang16}, and SN~2017eaw \citep{cheng17}.} 
\tablenotetext{g}{SN present in first 2014 SPIRITS epoch and therefore excluded from the control sample.}
\end{deluxetable*}

Three events, SN~2014C (Type Ib/IIn), SN~2014J (Type Ia), and SN~2014L (Type Ic), were present in the first epoch of SPIRITS observations such that we would not have a meaningful constraint on the explosion epoch from SPIRITS data alone. Two additional events were not recovered in SPIRITS by our automated image subtraction and transient identification pipeline owing to saturation of the IRAC detector at the location of the transient. SN~2014bc (Type II) was located only $3\farcs4$ from the bright nucleus of NGC~4258, which is saturated in all epochs of SPIRITS imaging. SN~2016adj (Type Ib), the nearest object in the optically discovered sample at only 3.7\,Mpc, is itself saturated in all epochs where the SN is present in SPIRITS imaging and was not recovered by the SPIRITS pipeline. 

Thus, we define our primary comparison sample as the nine objects that were recovered by SPIRITS and have constraints on their explosion dates from SPIRITS data. By optical spectroscopic subtype this sample includes six SNe~II, two SNe~Ib, and one SN~Iax. 

\begin{figure*}
 \gridline{\fig{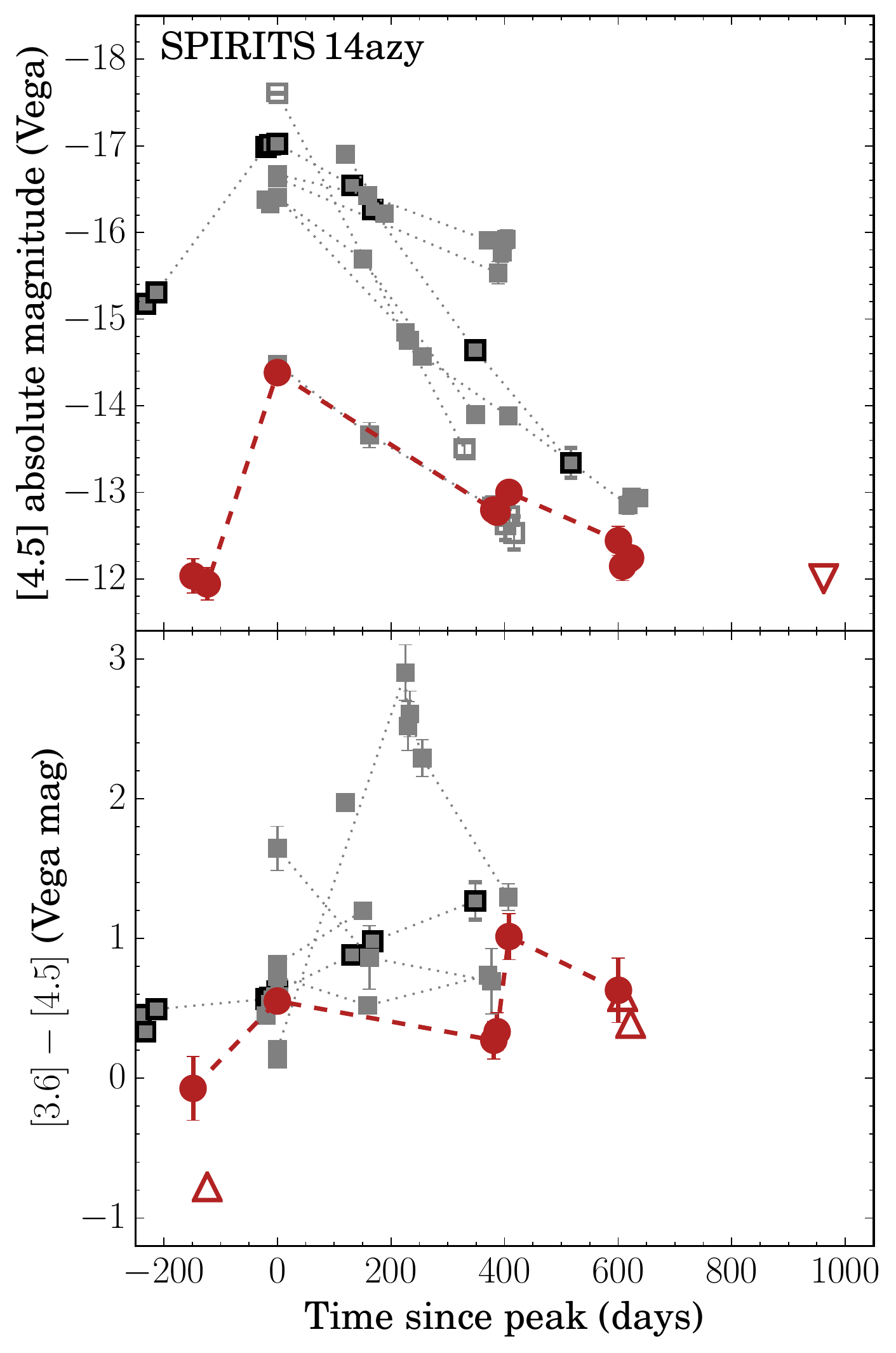}{0.3\textwidth}{} 
	\fig{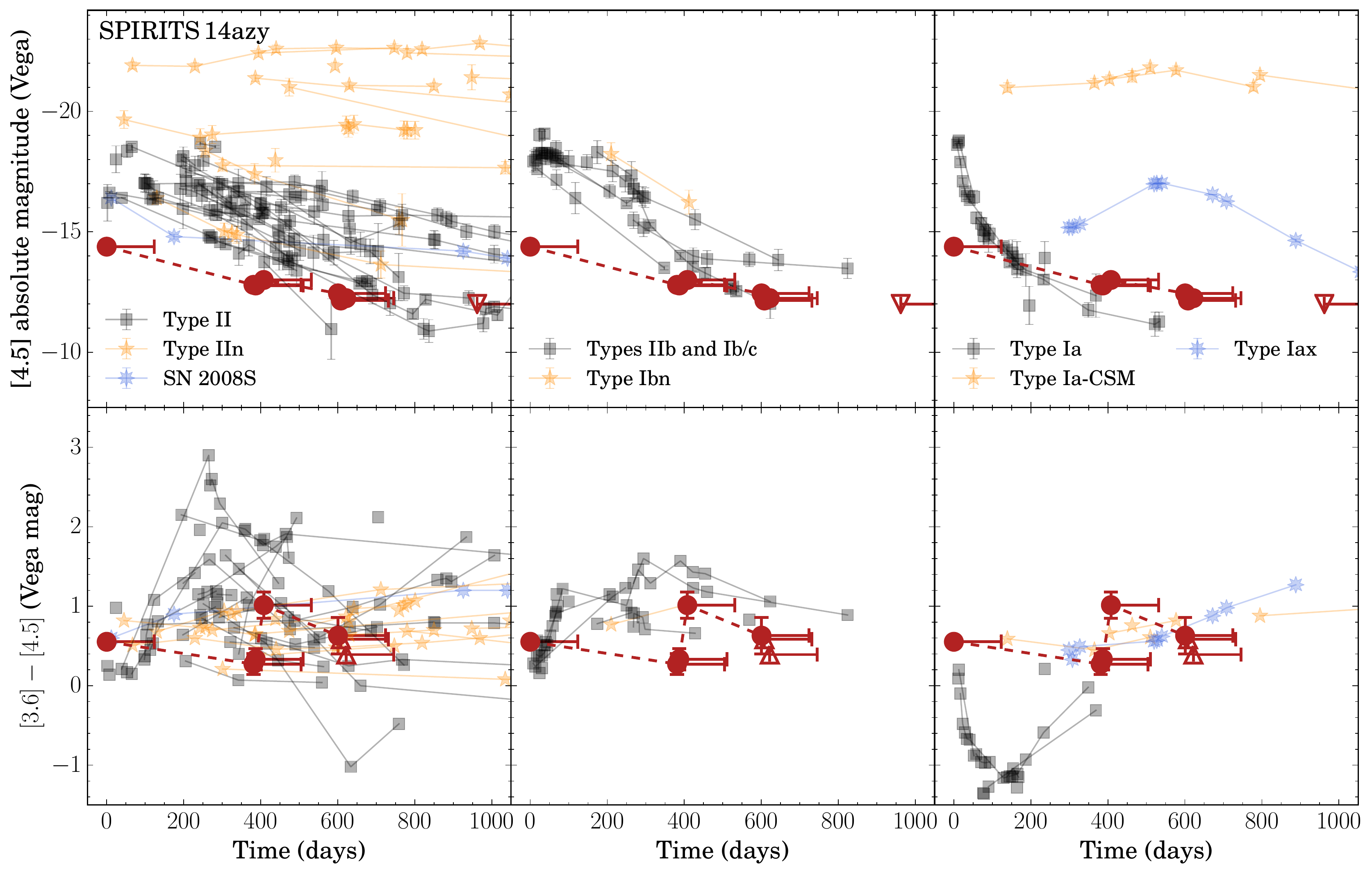}{0.7\textwidth}{}}
\gridline{\fig{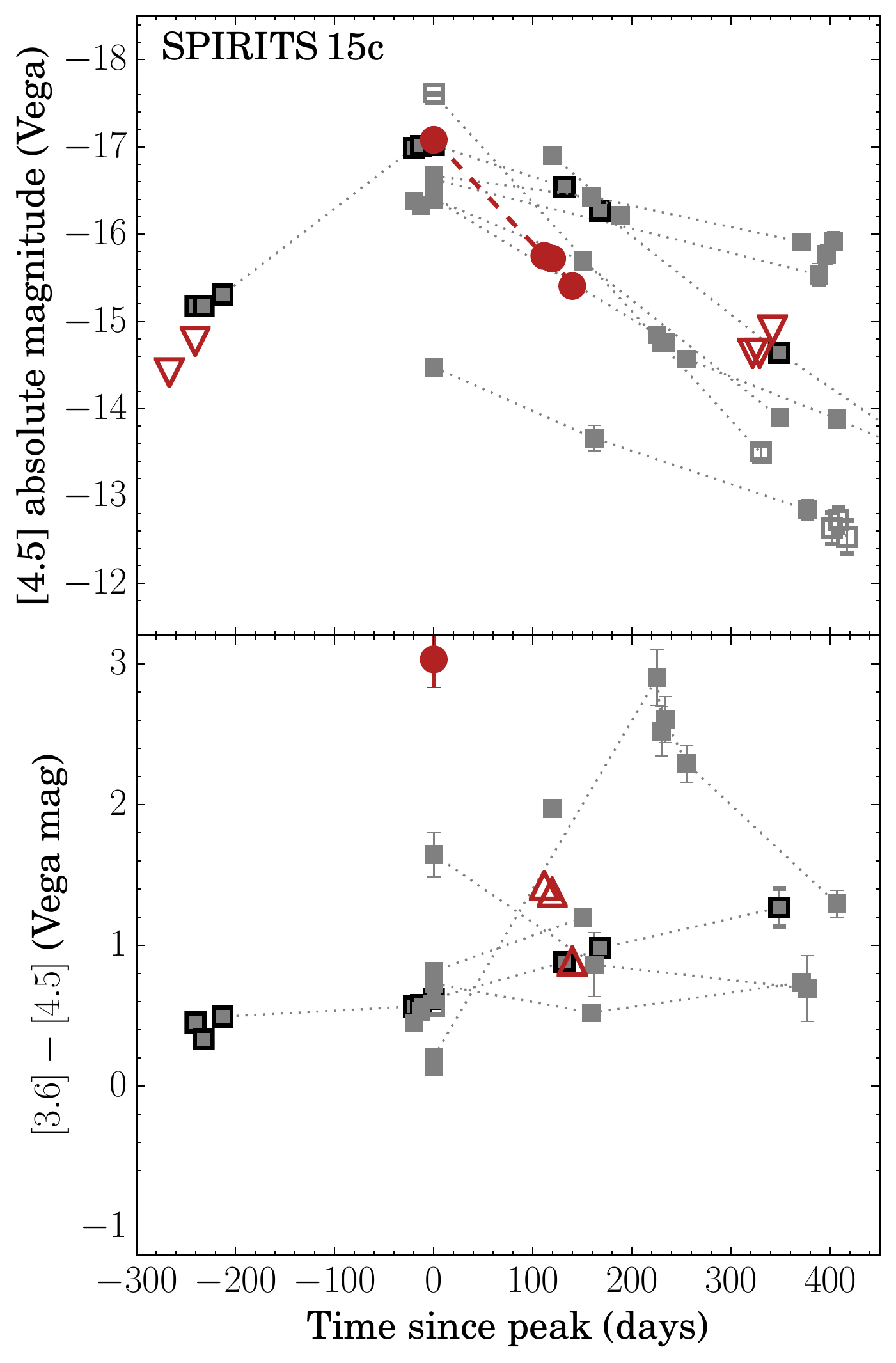}{0.3\textwidth}{}
	\fig{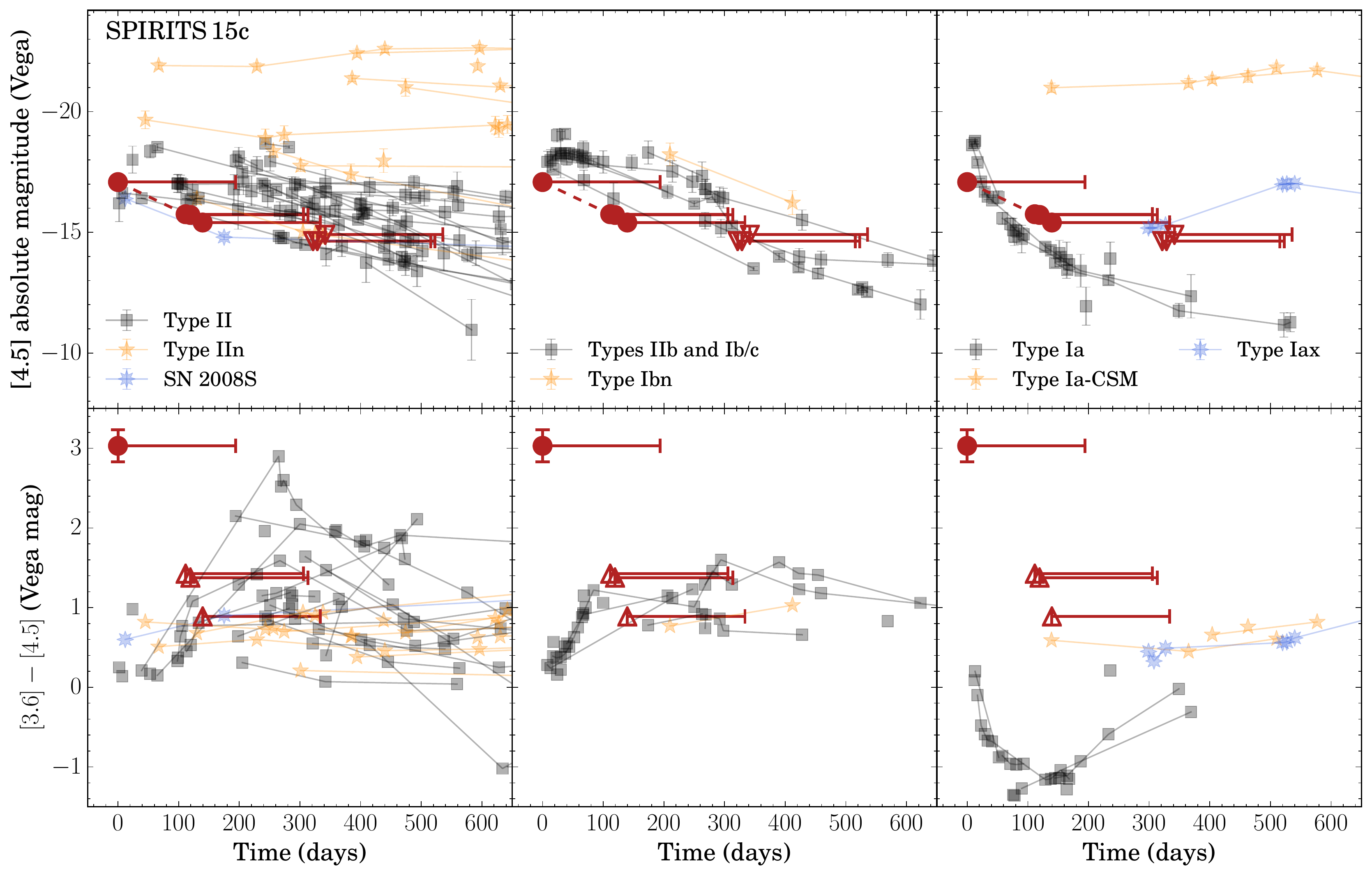}{0.7\textwidth}{}}   
\caption{\label{fig:IR_lcs}
For each luminous IR SPIRITS transient, we show the [4.5] light curve (top row) and $[3.6]{-}[4.5]$ color curve (bottom row) as the large red circles. In the left panel for each object, the phase is measured in days since the observed IR peak, and we compare to the control sample of optically discovered SNe recovered in SPIRITS as described in Section~\ref{sec:control_samp}, including SNe~II (light-gray squares), stripped-envelope SNe~Ib (open squares), and the Type Iax SN~2014dt (black outlined squares). In the right panel for each object, we compare to the full sample of SNe detected by \textit{Spitzer} compiled by \citet{szalai19}, including previous compilations by \citet{szalai13}, \citet{tinyanont16}, and \citet{johansson17}, and references therein. Additionally, we include the SNe observed by SPIRITS since 2014 and presented in this work. From left to right, we compare to SNe~II (IIn), stripped-envelope SNe~IIb and Ib/c (Ibn), and SNe~Ia (Ia-CSM) as light-gray squares (orange stars). The blue, eight-pointed stars indicate the prototypical for its class SN\,2008S, and the unusual, dusty Type Iax SN~2014dt. The phase for comparison SNe is measured as days since discovery, and for obscured SN candidates we represent our uncertainty in the phase of the primary outburst by the horizontal error bars. Upper (lower) limits from nondetections are indicated by open, downward-pointing (upward-pointing) triangles. 
}
\end{figure*}

Our photometry from \textit{Spitzer}/IRAC imaging for both the IR-discovered sample of luminous IR transients and the optically discovered control sample in SPIRITS is given in Table~\ref{table:Sp_phot_IR} and shown in Figure~\ref{fig:IR_lcs} (see Section~\ref{sec:IR_lcs} for further discussion).

\begin{figure*}
\figurenum{\ref{fig:IR_lcs}, continued}
 \gridline{\fig{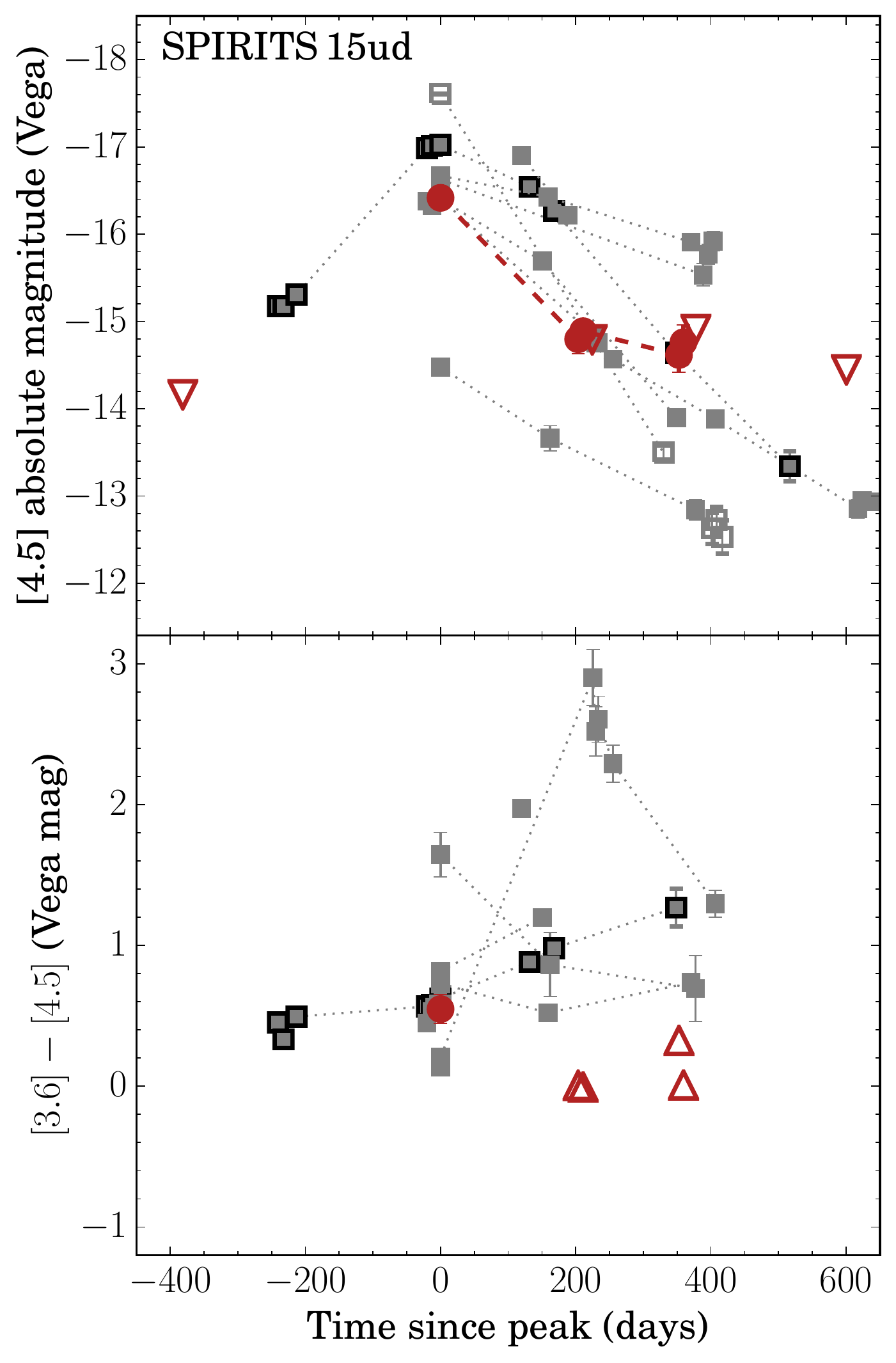}{0.3\textwidth}{}
	\fig{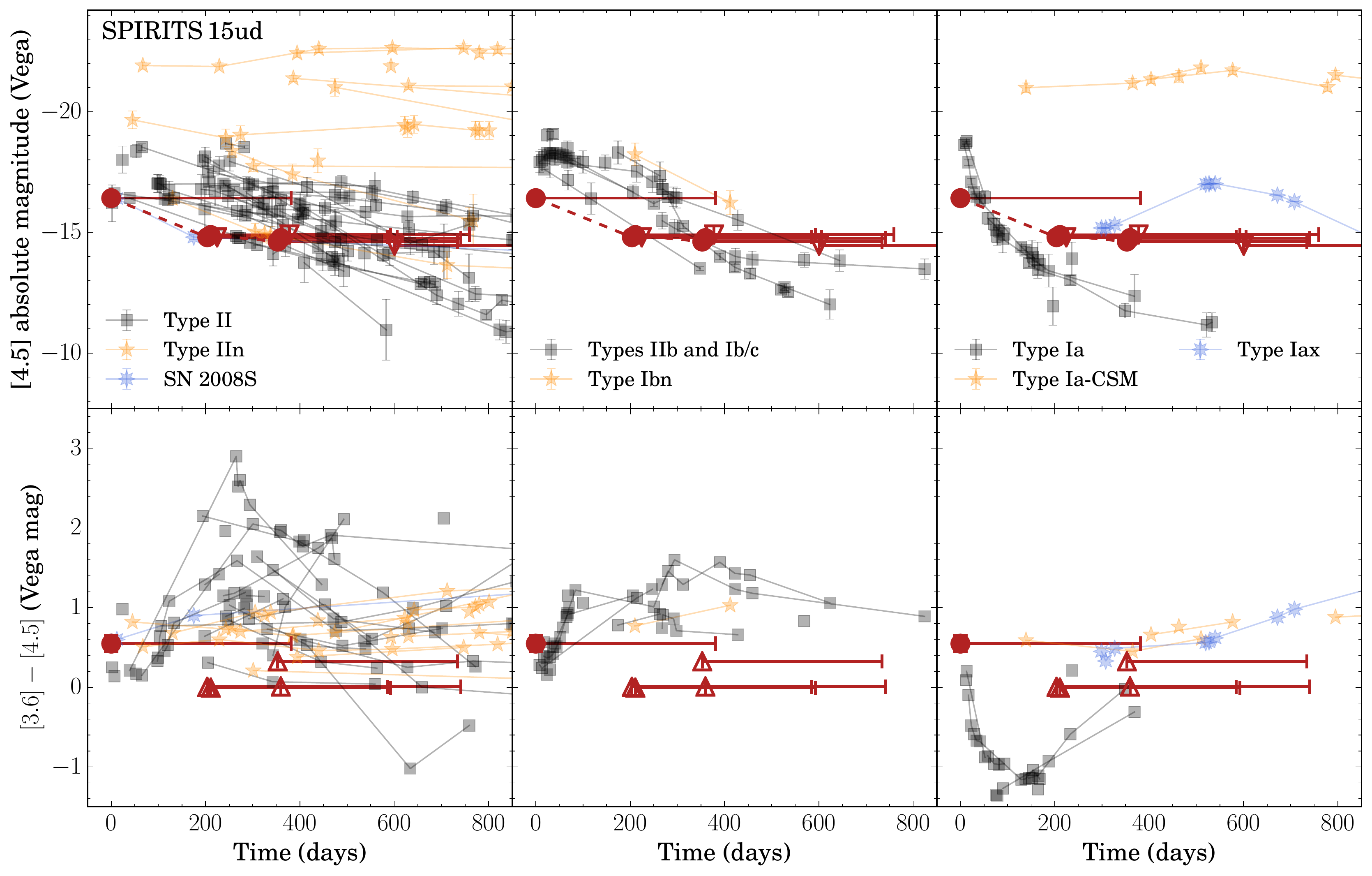}{0.7\textwidth}{}}
\gridline{\fig{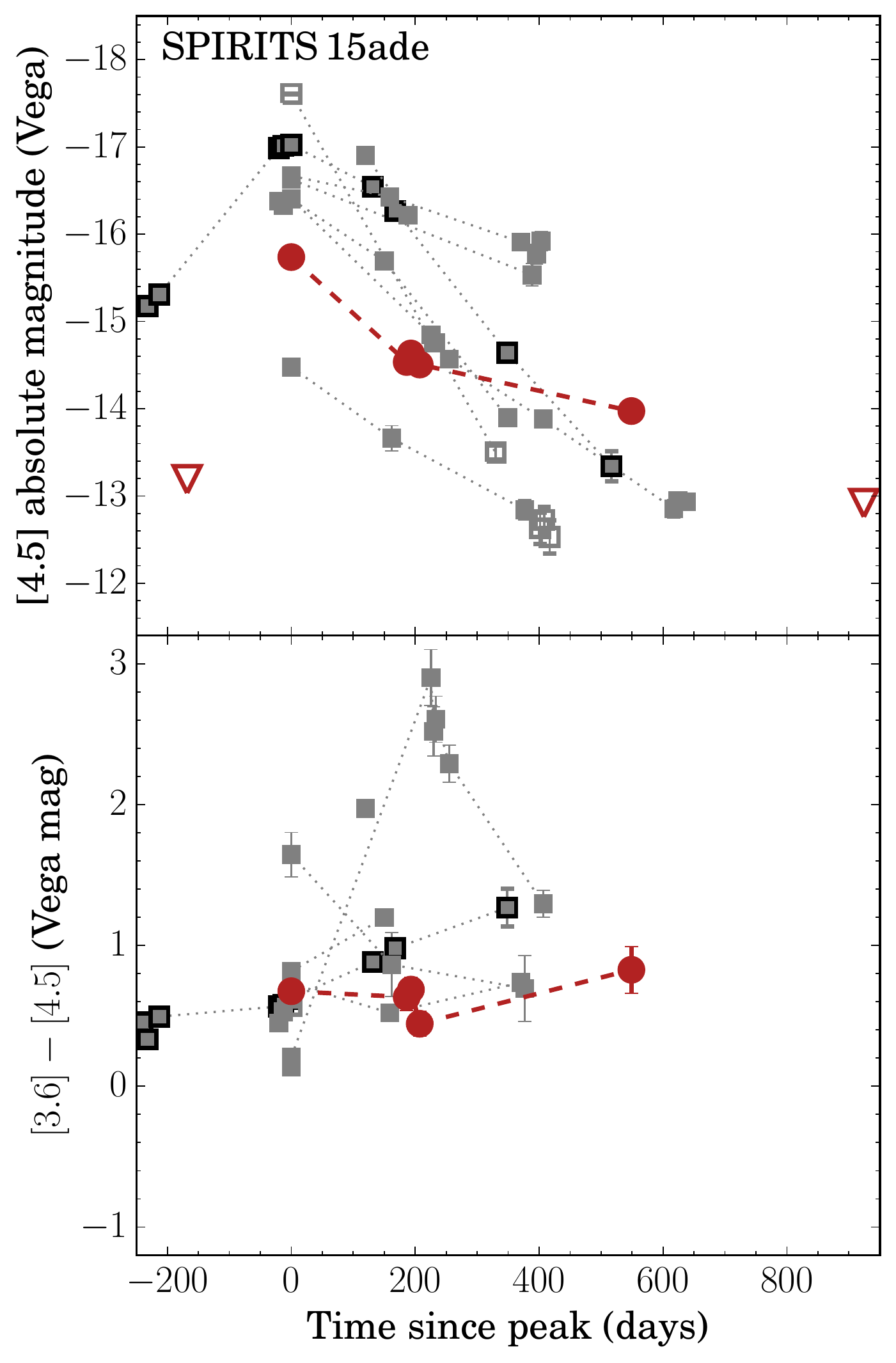}{0.3\textwidth}{}
	\fig{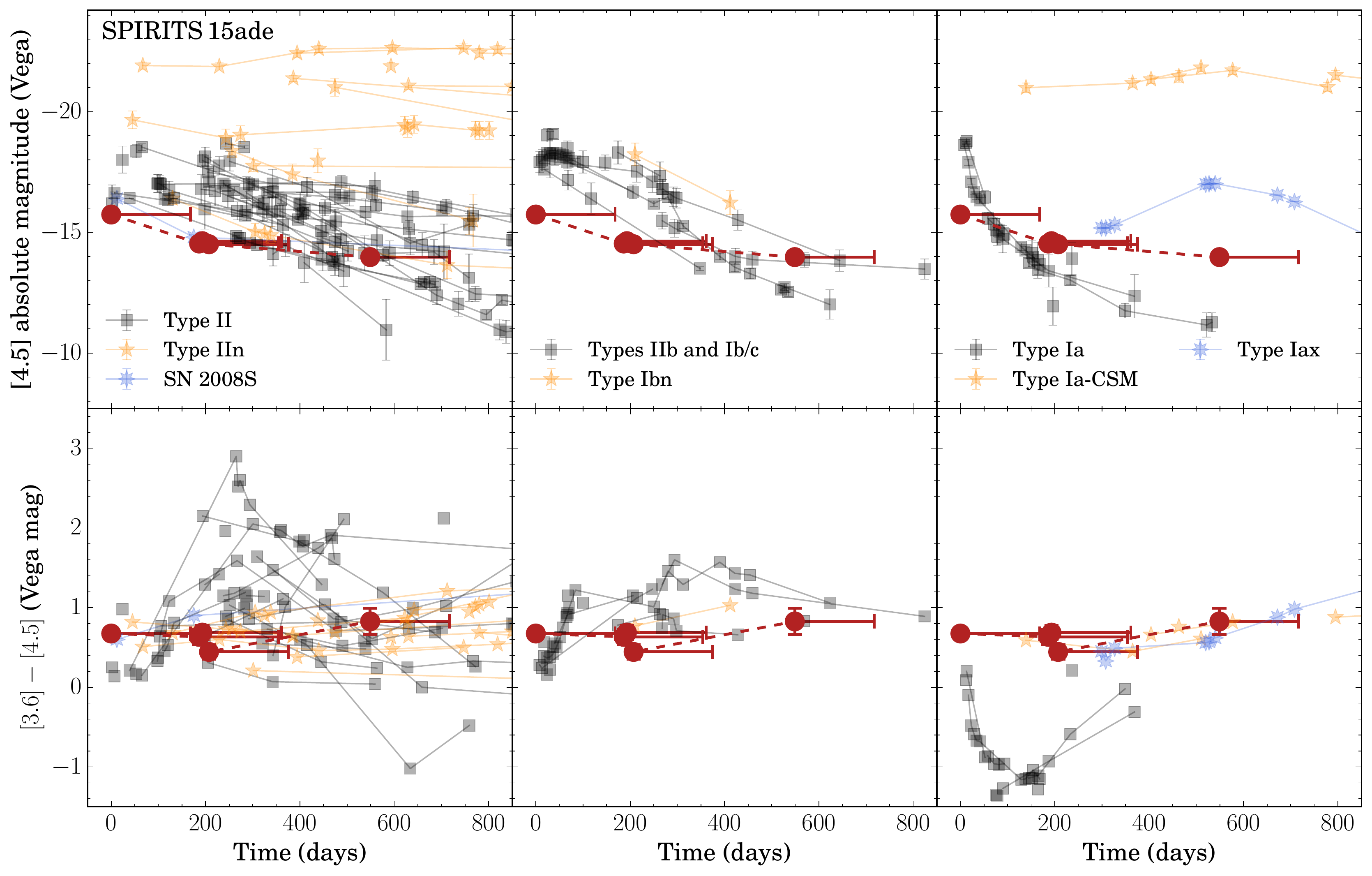}{0.7\textwidth}{}}
\caption{}
\end{figure*}

\begin{figure*}
\figurenum{\ref{fig:IR_lcs}, continued}
 \gridline{\fig{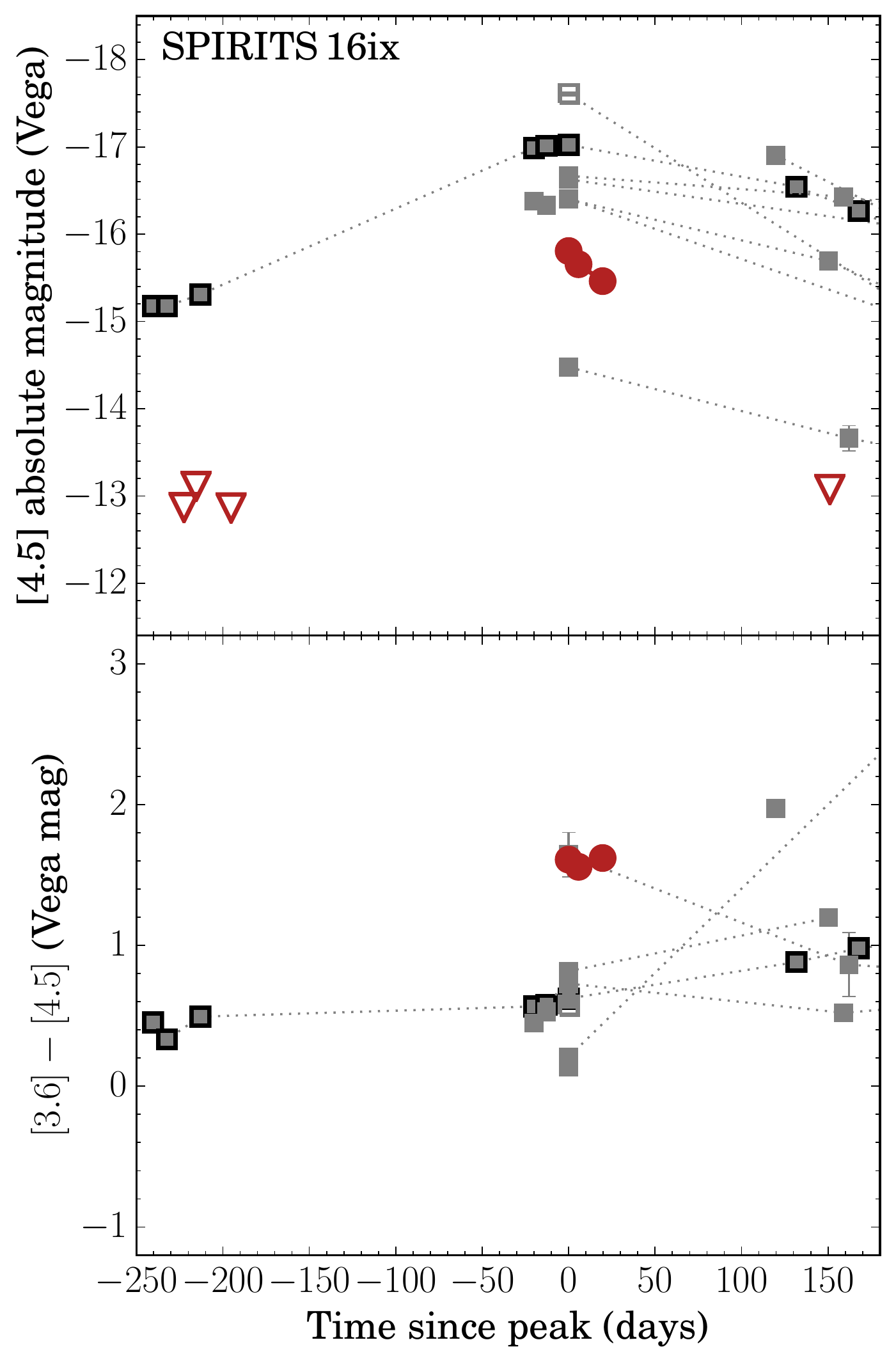}{0.3\textwidth}{}
	\fig{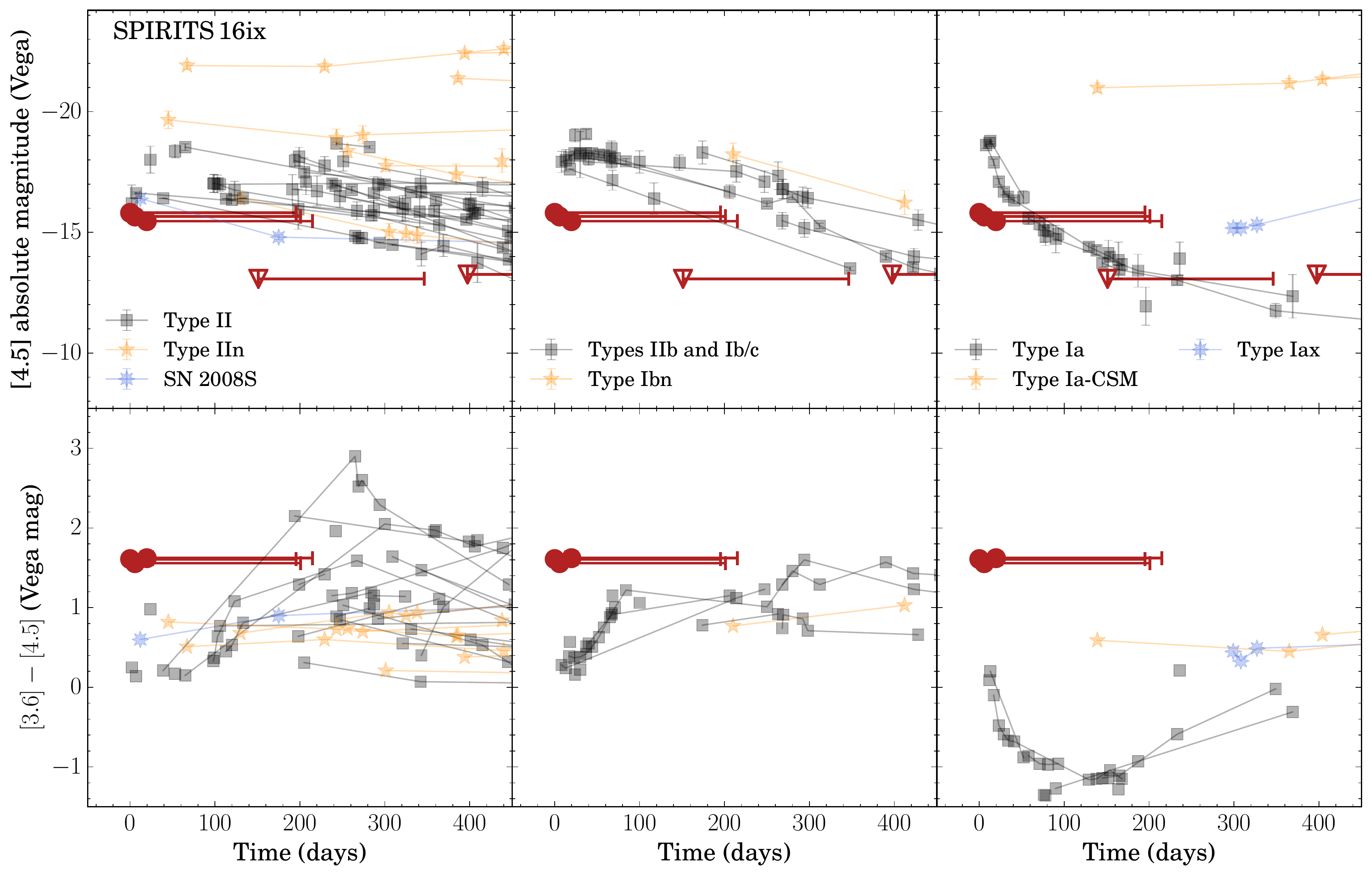}{0.7\textwidth}{}}
\gridline{\fig{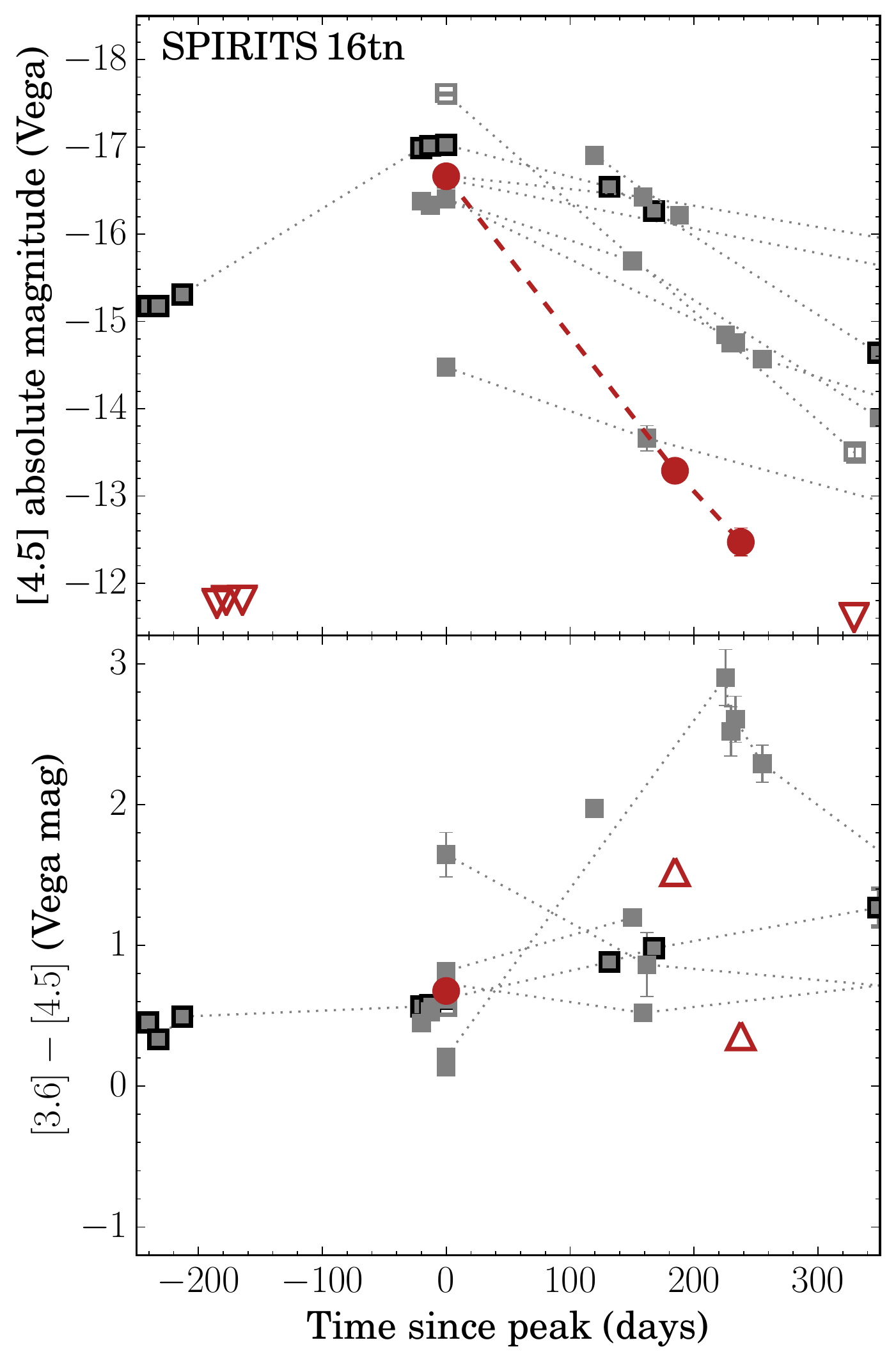}{0.3\textwidth}{}
	\fig{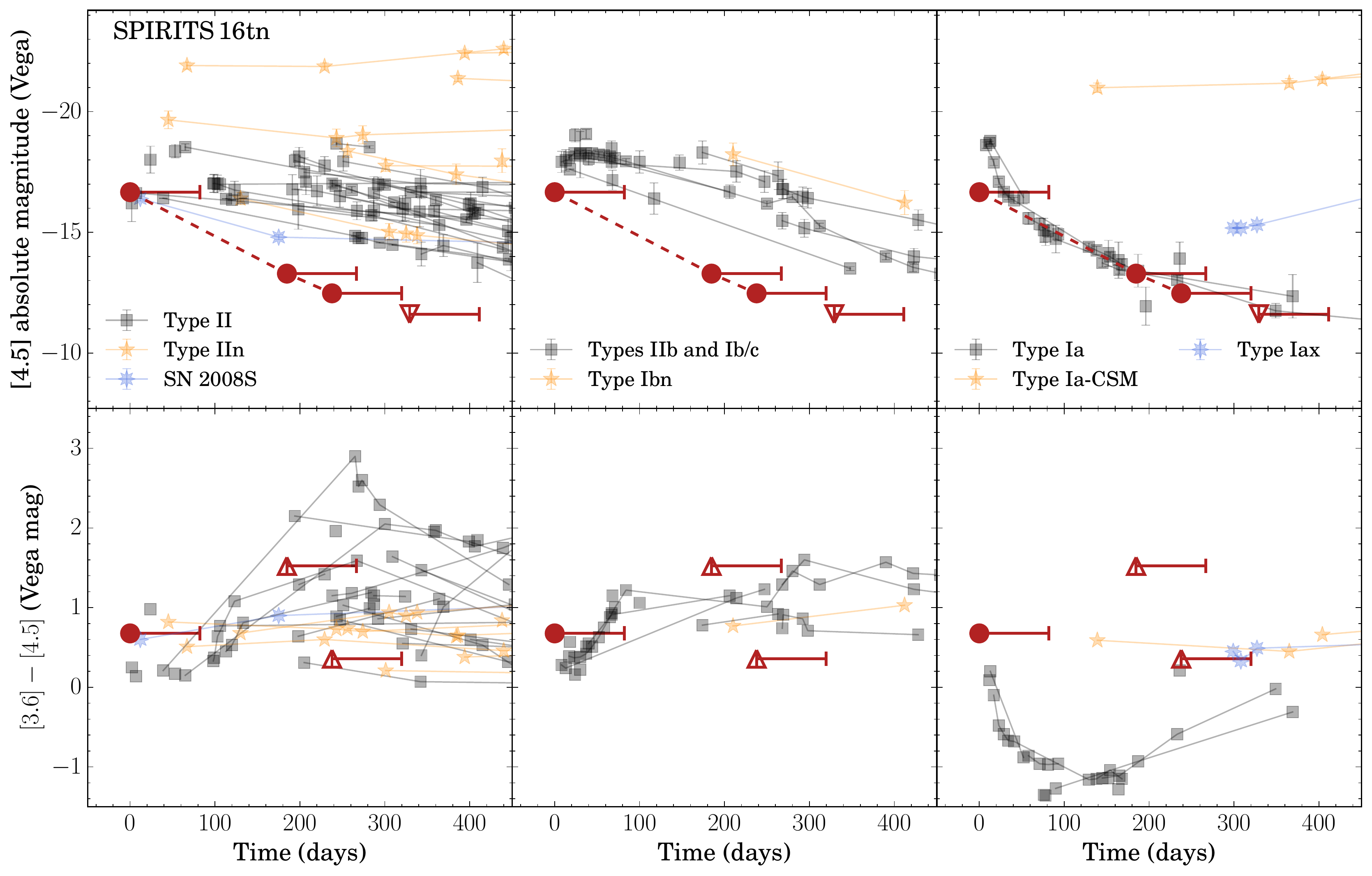}{0.7\textwidth}{}}
\caption{}
\end{figure*}

\begin{figure*}
\figurenum{\ref{fig:IR_lcs}, continued}
 \gridline{\fig{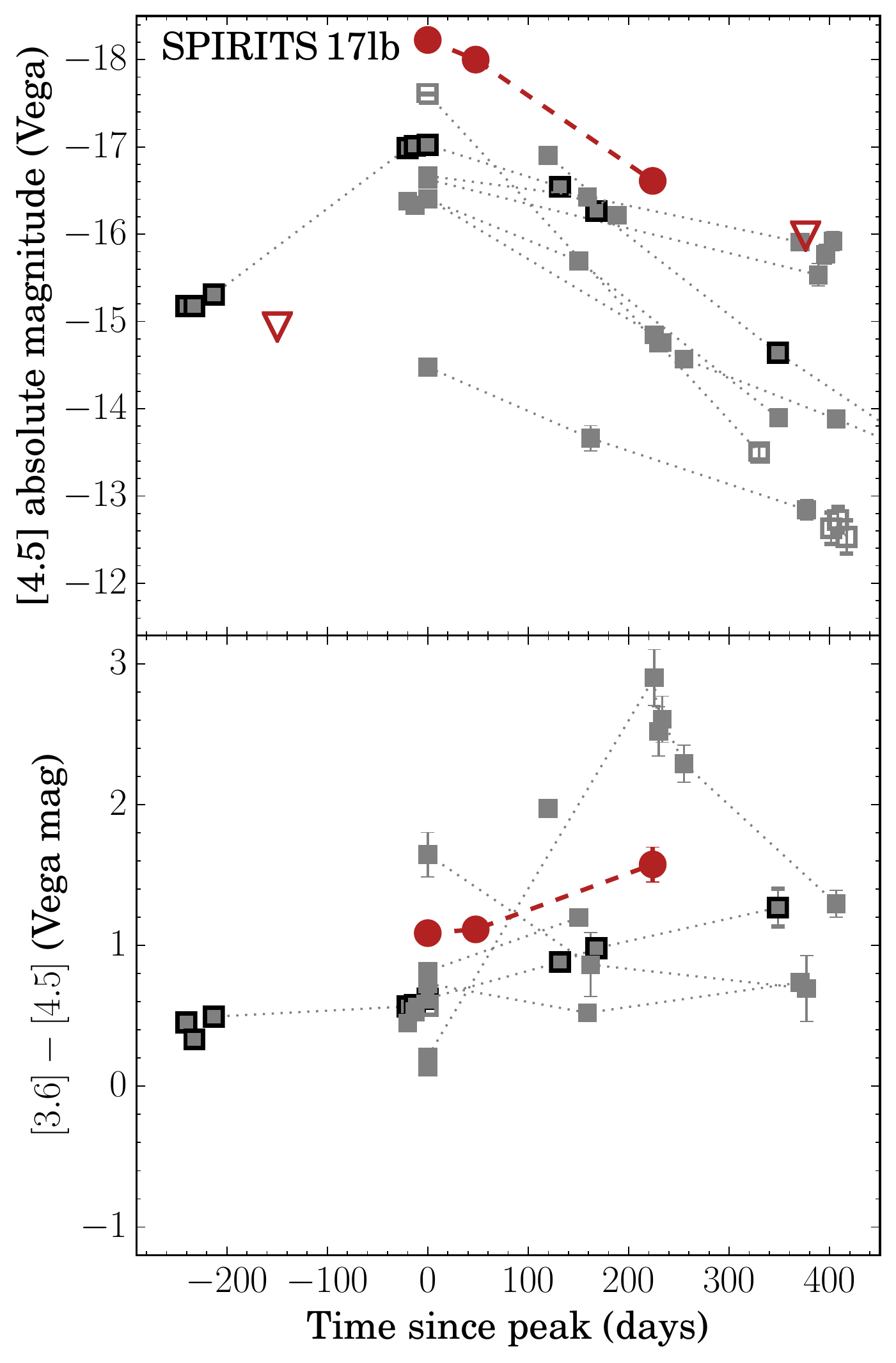}{0.3\textwidth}{}
	\fig{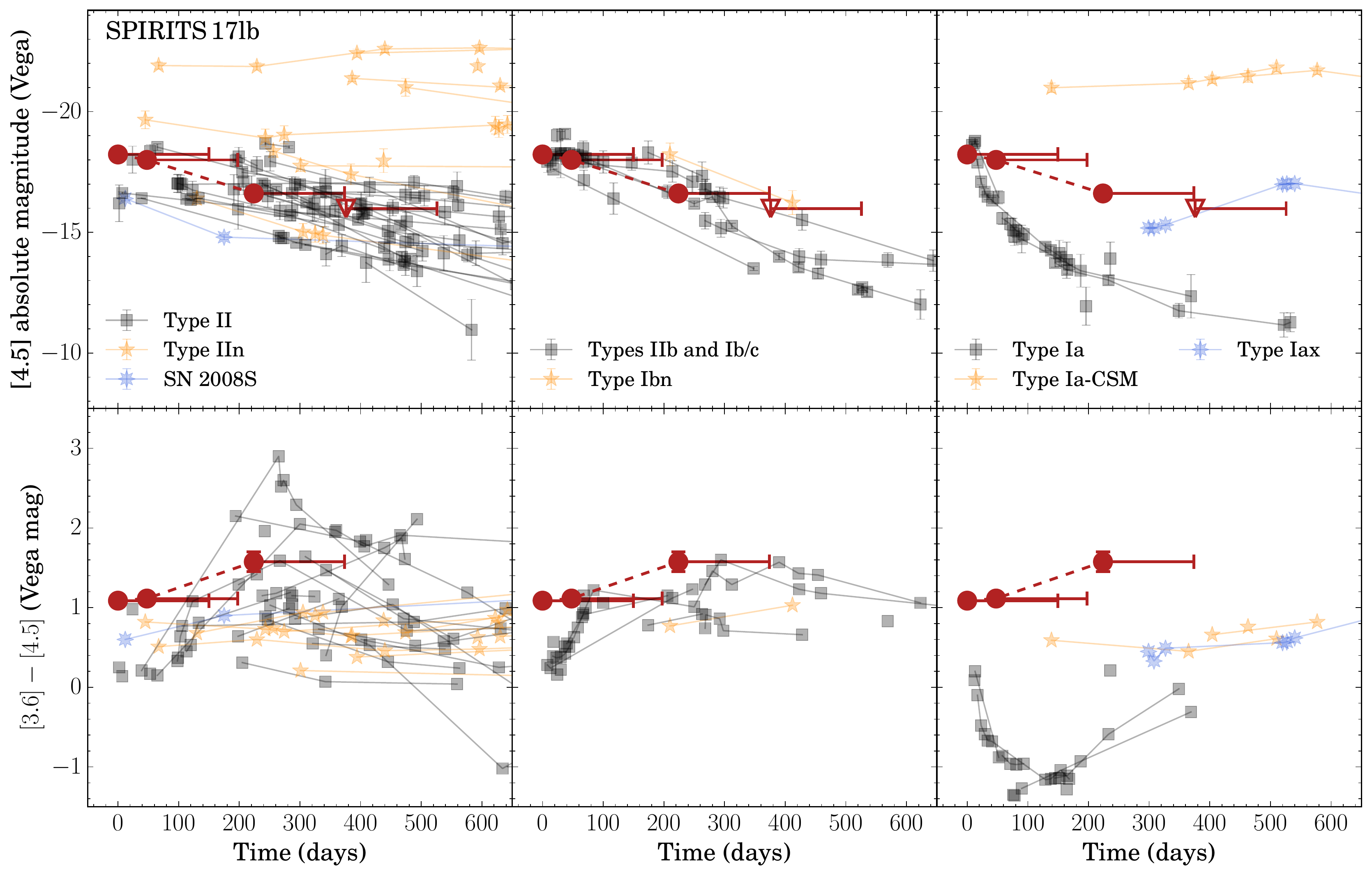}{0.7\textwidth}{}}
\gridline{\fig{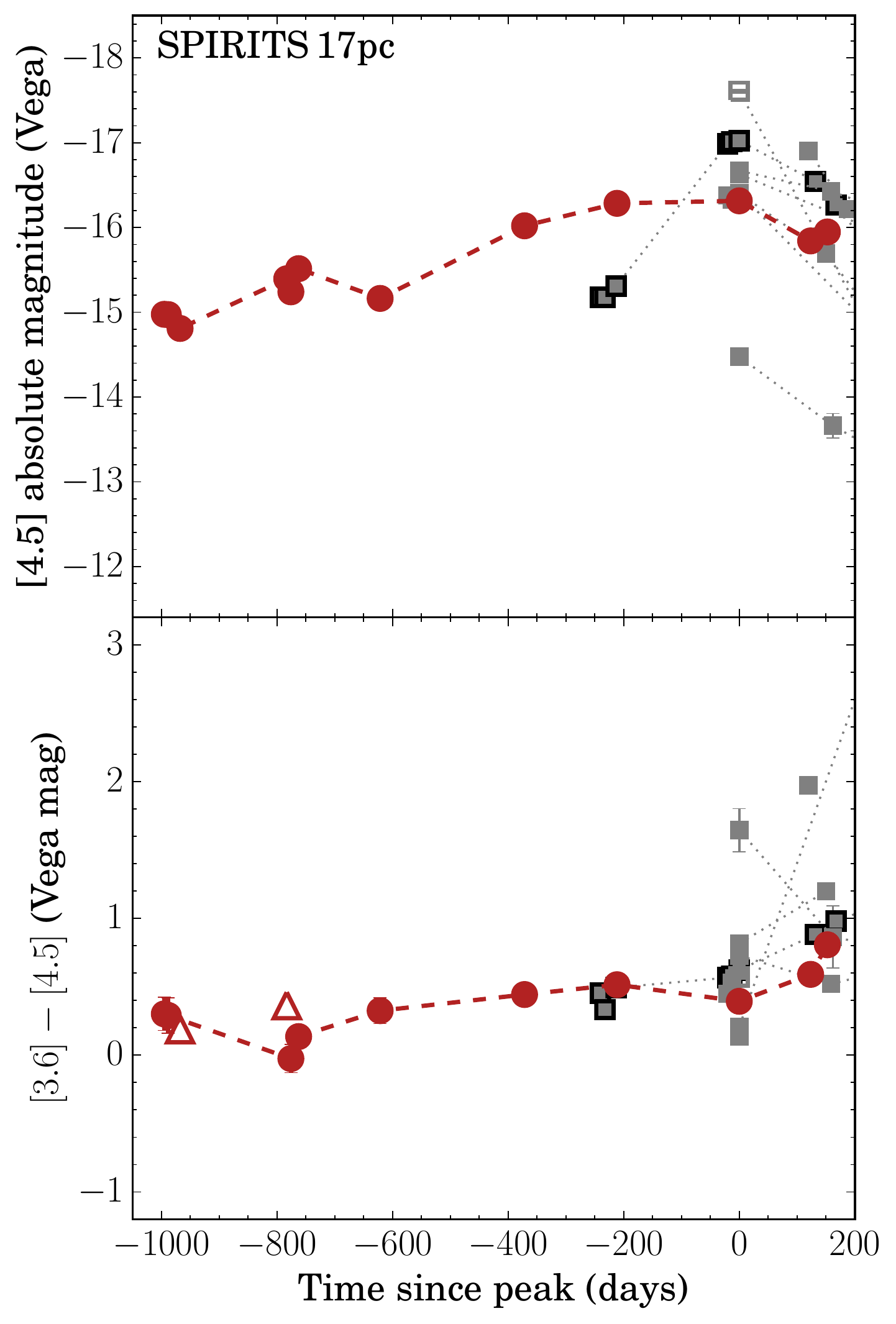}{0.3\textwidth}{}
	\fig{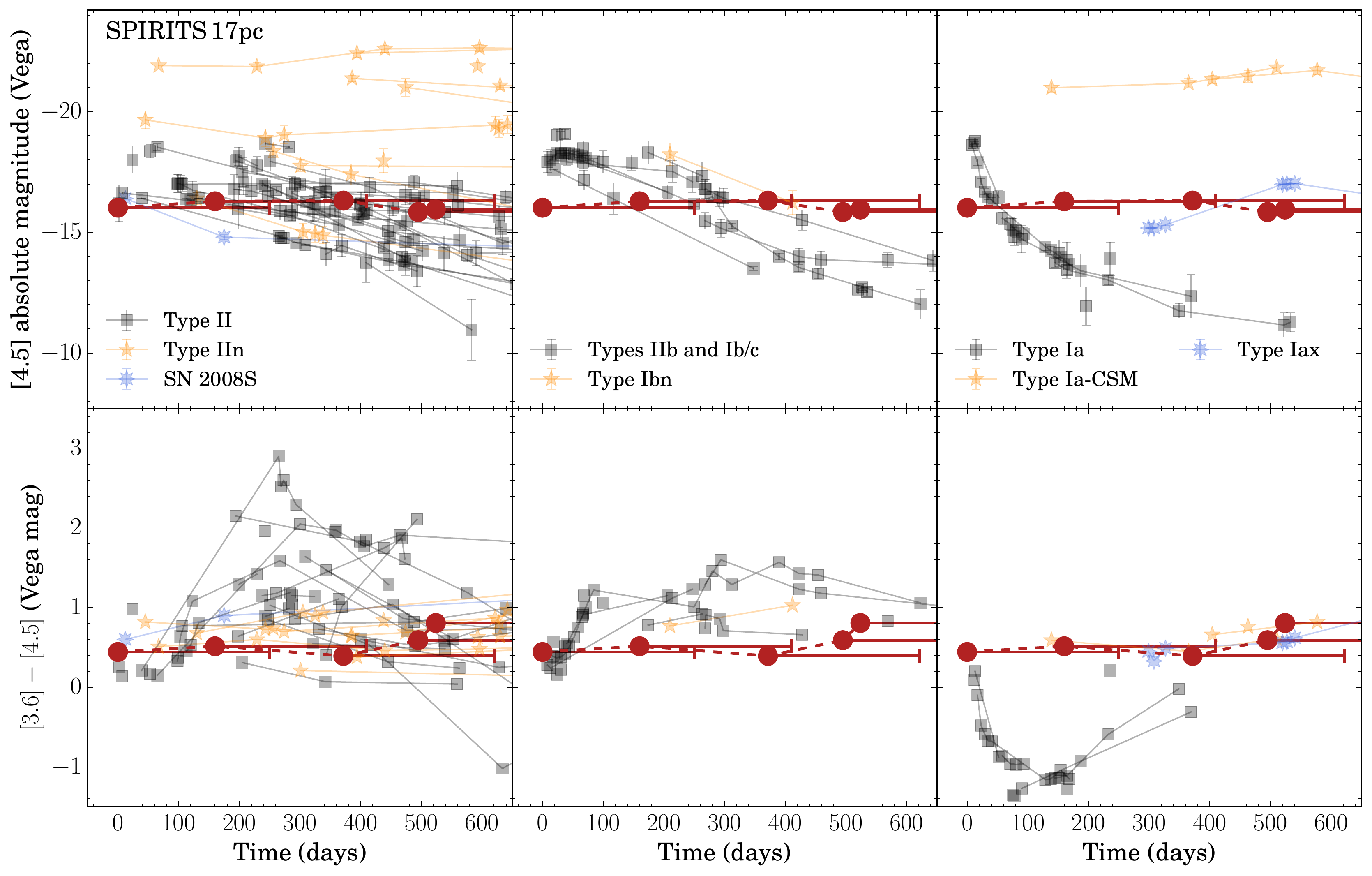}{0.7\textwidth}{}}
\caption{}
\end{figure*}

\begin{figure*}
\figurenum{\ref{fig:IR_lcs}, continued}
 \gridline{\fig{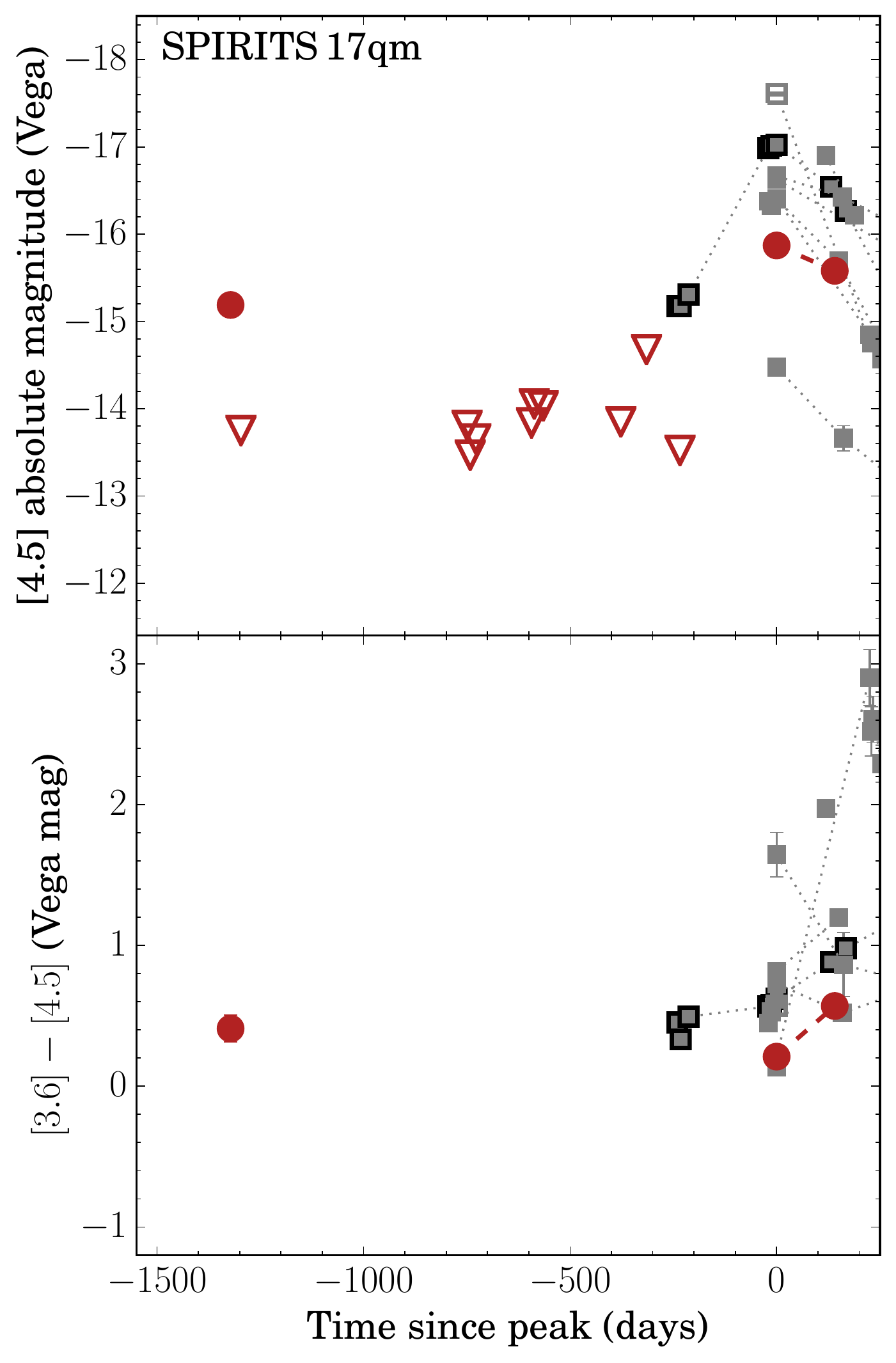}{0.3\textwidth}{}
	\fig{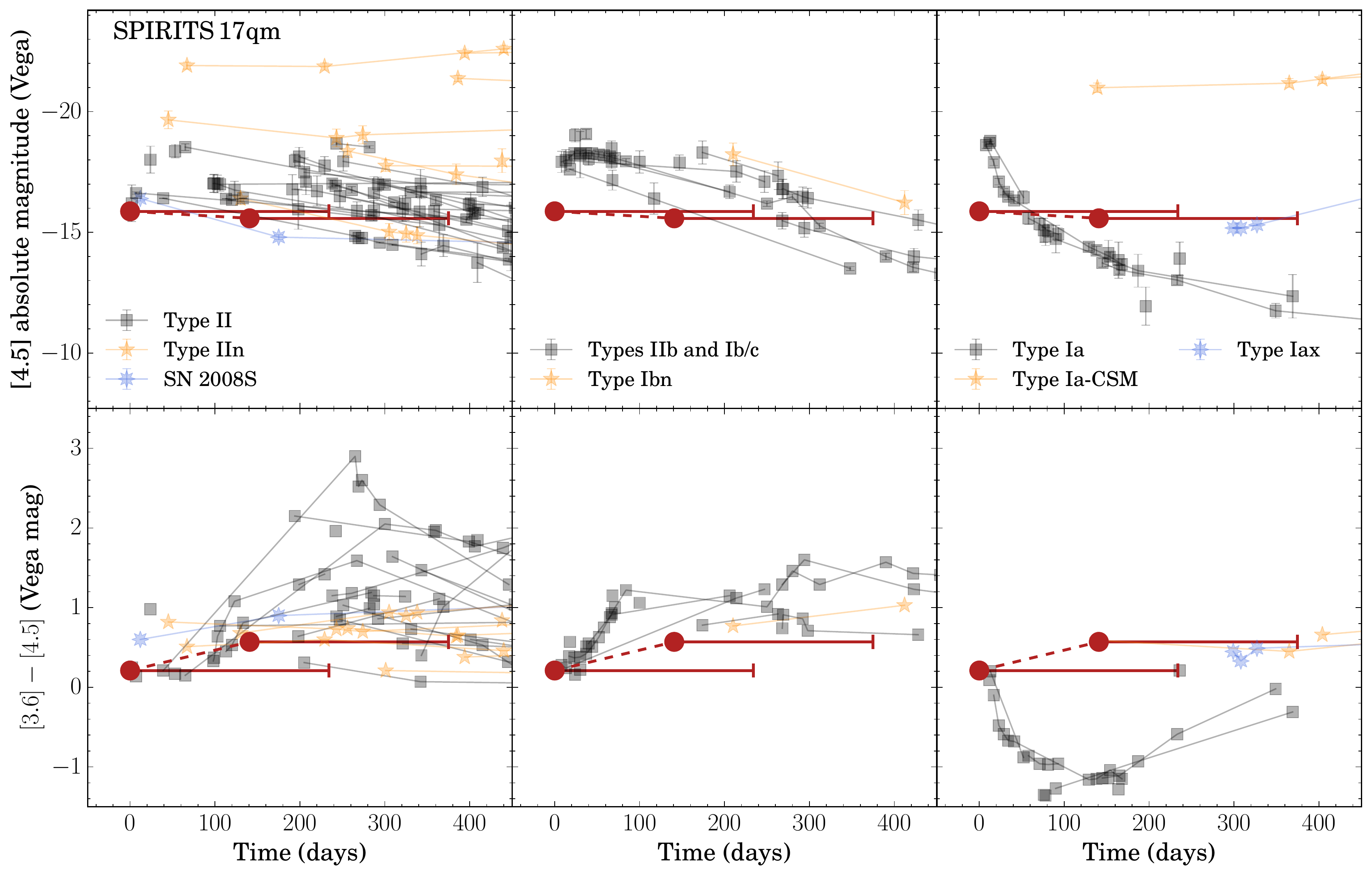}{0.7\textwidth}{}}
\caption{}
\end{figure*}

\begin{deluxetable*}{lrrrrrrrr}
\tablecaption{\textit{\textit{Spitzer}/IRAC Photometry of Luminous, IR Transients in SPIRITS\tablenotemark{a}} \label{table:Sp_phot_IR}}
\tablehead{ \colhead{Name} & \colhead{MJD} & \colhead{Phase\tablenotemark{b}} & \colhead{$f_{\nu,[3.6]}$} & \colhead{$f_{\nu,[4.5]}$} & \colhead{[3.6]} & \colhead{[4.5]} & \colhead{$M_{[3.6]}$} & \colhead{$M_{[4.5]}$} \\ 
\colhead{} & \colhead{} & \colhead{(days)} & \colhead{($\mu$Jy)} & \colhead{($\mu$Jy)} & \colhead{(Vega mag.)} & \colhead{(Vega mag.)} & \colhead{} & \colhead{} }
\startdata
SPIRITS\,14azy & 56,715.5 &   -56.5 &  $13.5 \pm 1.4$ &   $8.1 \pm 1.5$ & $18.29 \pm 0.11$ & $18.37 \pm 0.20$ &   $-12.1$ &   $-12.1$ \\
               & 56,740.5 &   -31.5 &        $< 23.8$ &   $7.4 \pm 1.3$ &         $> 17.7$ & $18.46 \pm 0.19$ & $> -12.8$ &   $-12.0$ \\
               & 56,864.0 &    92.0 &  $66.1 \pm 1.8$ &  $70.3 \pm 1.4$ & $16.57 \pm 0.03$ & $16.02 \pm 0.02$ &   $-13.9$ &   $-14.4$ \\
               & 57,245.1 &   473.1 &  $19.9 \pm 1.7$ &  $16.3 \pm 1.4$ & $17.88 \pm 0.09$ & $17.60 \pm 0.09$ &   $-12.6$ &   $-12.8$ \\
               & 57,251.0 &   479.0 &  $18.3 \pm 1.6$ &  $15.9 \pm 1.4$ & $17.96 \pm 0.09$ & $17.63 \pm 0.10$ &   $-12.5$ &   $-12.8$ \\
               & 57,271.8 &   499.8 &  $12.1 \pm 1.6$ &  $19.7 \pm 1.5$ & $18.41 \pm 0.14$ & $17.40 \pm 0.08$ &   $-12.0$ &   $-13.0$ \\
               & 57,464.6 &   692.6 &  $10.3 \pm 1.5$ &  $11.8 \pm 1.8$ & $18.59 \pm 0.16$ & $17.96 \pm 0.17$ &   $-11.8$ &   $-12.5$ \\
               & 57,472.0 &   700.0 &         $< 8.2$ &   $9.0 \pm 1.4$ &         $> 18.8$ & $18.25 \pm 0.17$ & $> -11.6$ &   $-12.2$ \\
               & 57,486.0 &   714.0 &        $< 10.7$ &   $9.8 \pm 1.3$ &         $> 18.5$ & $18.16 \pm 0.14$ & $> -11.9$ &   $-12.3$ \\
               & 57,826.1 &  1054.1 &         $< 7.3$ &         $< 7.9$ &         $> 19.0$ &         $> 18.4$ & $> -11.5$ & $> -12.0$ \\
               & 57,974.2 &  1202.2 &         $< 8.2$ &         $< 6.6$ &         $> 18.8$ &         $> 18.6$ & $> -11.6$ & $> -11.9$ \\
\hline
SN\,2014bi\tablenotemark{c} & 56737.5 &   -70.8 &          $< 8.9$ &          $< 9.8$ &         $> 18.8$ &         $> 18.2$ & $> -11.6$ & $> -12.2$ \\
 (SPIRITS\,15bx) & 56,847.1 &    38.8 & $585.3 \pm 10.8$ &  $453.7 \pm 8.4$ & $14.20 \pm 0.02$ & $13.99 \pm 0.02$ &   $-16.2$ &   $-16.4$ \\
                 & 57,072.5 &   264.2 &   $11.7 \pm 2.1$ &  $108.1 \pm 2.0$ & $18.45 \pm 0.20$ & $15.55 \pm 0.02$ &   $-11.9$ &   $-14.8$ \\
                 & 57,076.8 &   268.5 &   $15.2 \pm 2.4$ &   $99.1 \pm 2.0$ & $18.17 \pm 0.17$ & $15.65 \pm 0.02$ &   $-12.2$ &   $-14.8$ \\
                 & 57,080.5 &   272.2 &   $14.1 \pm 2.1$ &   $99.6 \pm 1.8$ & $18.25 \pm 0.16$ & $15.64 \pm 0.02$ &   $-12.2$ &   $-14.8$ \\
                 & 57,102.2 &   293.9 &   $15.8 \pm 1.9$ &   $83.6 \pm 1.8$ & $18.12 \pm 0.13$ & $15.83 \pm 0.02$ &   $-12.3$ &   $-14.6$ \\
                 & 57,253.6 &   445.3 &   $21.0 \pm 1.6$ &   $44.4 \pm 1.9$ & $17.81 \pm 0.08$ & $16.52 \pm 0.05$ &   $-12.6$ &   $-13.9$ \\
                 & 57,464.4 &   656.1 &          $< 8.4$ &   $17.2 \pm 1.7$ &         $> 18.8$ & $17.55 \pm 0.11$ & $> -11.6$ &   $-12.9$ \\
                 & 57,470.9 &   662.6 &          $< 8.8$ &   $18.7 \pm 1.5$ &         $> 18.8$ & $17.46 \pm 0.09$ & $> -11.6$ &   $-12.9$ \\
                 & 57,484.4 &   676.1 &          $< 8.8$ &   $18.5 \pm 1.4$ &         $> 18.8$ & $17.47 \pm 0.08$ & $> -11.6$ &   $-12.9$ \\
                 & 57,842.9 &  1034.6 &          $< 8.8$ &         $< 11.0$ &         $> 18.8$ &         $> 18.0$ & $> -11.6$ & $> -12.4$ \\
                 & 57,963.7 &  1155.4 &          $< 7.5$ &          $< 9.1$ &         $> 18.9$ &         $> 18.2$ & $> -11.5$ & $> -12.2$ \\
\enddata
\tablenotetext{a}{We show measurements for SPIRITS\,14azy and SN\,2014bi to illustrate the form and content of this table. A full version including all nine objects in each of IR-discovered and optical control samples is published in machine-readable format.}
\tablenotetext{b}{Time since $t_0$ as reported in Table~\ref{table:lc_prop} for the IR-discovered sample, and time since discovery as reported in Table~\ref{table:opt_sample} for the optical control sample.}
\tablenotetext{c}{Measurements through $\mathrm{MJD} = 57,253.6$ were previously reported in \citet{tinyanont16}.}
\end{deluxetable*}

\section{Follow-up and supplementary observations}\label{sec:obs}

\subsection{Space-based imaging}
For transients that were inaccessible to ground-based observing at the time of discovery with \textit{Spitzer}, we attempted to trigger Target of Opportunity (ToO) observations with the \textit{Neil Gehrels Swift Observatory} UV/Optical Telescope (\textit{Swift}/UVOT; \citealp{gehrels04,nousek04,roming05}) to detect an optical counterpart or obtain limits on the contemporaneous optical flux. For SPIRITS\,16tn we triggered a 2000~s observation on 2016 August 29.1 split between the $U$, $B$, and $V$ bands, with limits reported in \citet{adams16b} and \citet{jencson18c}. For SPIRITS\,17lb we obtained a 1200~s integration in the $V$ band on 2017 June 9.4 and derived a $5\sigma$ limiting magnitude of $V > 17.7$. For SPIRITS\,17pc we obtained a 2000~s observation on 2017 November 9.1 split between $U$, $B$, and $V$. NGC~4388 was previously observed by \textit{Swift}/UVOT 1\,yr prior on 2016 November 8.8, but with shorter integrations of only ${\approx}60$~s in each band. We find no evidence of significant variability at the location of SPIRITS\,17pc between the two epochs. We derive limits for the earlier (later) epochs of $U > 18.0~(18.8)$, $B > 18.2~(18.8)$, and $V > 17.2~(18.0)$. 

We also executed \textit{Hubble Space Telescope} (\textit{HST}) ToO observations of SPIRITS\,16tn using the Wide Field Camera 3 (WFC3) in the UVIS channel with the F814W filter and the IR channel with the F110W and F160W filters as part of our program to follow up SPIRITS transients (GO-14258; PI: H. Bond) on 2016 September 25 as described in \citet{jencson18c}. 

\subsection{Ground-based imaging}
SPIRITS galaxies were regularly monitored from the ground in the optical and near-IR with several telescopes. For the SPIRITS transient host galaxies discussed here, sequences of optical $g^{\prime}r^{\prime}i^{\prime}$-band images of IC\,2163, M100, NGC~4461, NGC~3556, and NGC~4388 were obtained with the CCD camera on the fully automated Palomar 60-inch telescope \citep[P60;][]{cenko06} throughout 2014--2018 and reduced by a fully automated pipeline. Where available, we used Sloan Digital Sky Survey (SDSS) images as templates for image subtraction to remove host galaxy background emission and obtain deeper limits on optical emission from the transients. 

Similarly for SPIRITS galaxies located in the southern hemisphere, namely, NGC~2997, IC~2163, and NGC~1365, sequences of optical $gri$-band images were obtained with the CCD camera on the 1\,m Swope Telescope at Las Campanas Observatory (LCO) throughout 2014--2015. Near-IR $YJH$-band images were also obtained throughout 2014-2015 with the RetroCam IR camera \citep{morgan05} on the 2.5\,m du Pont Telescope at LCO. Photometry was performed at the locations of the transients by fitting the point-spread function (PSF) of the images, measured using stars in the field, simultaneously with the background emission, modeled using low-order polynomials. 

We also obtained and performed PSF photometry on the images of NGC~5921 reported at CBAT for PSN\,J15220552+\allowbreak 0503160 (SPIRITS\,15ade), including unfiltered CCD images obtained by M.\ Aoki, G.\ Masi, and T.\ Noguchi, along with post-discovery images of this galaxy in the $R$ band from the Hiroshima One-shot Wide-field Polarimeter (HOWpol; \citealp{kawabata08}) on the Kanata 1.5\,m telescope at Higashi-Hiroshima Observatory and in the $r$ band with the Dark Energy Camera (DECam) on the Blanco 4-m telescope at Cerro Tololo Inter-American Observatory (CTIO). 

Post-discovery, ground-based follow-up imaging in the optical and near-IR was also obtained for several SPIRITS transients. Near-IR $JHK_s$-band images were obtained with the Multi-object Spectrometer for Infra-red Exploration \citep[MOSFIRE;][]{mosfire10,mosfire12} on the 10\,m Keck I Telescope of the W. M. Keck Observatory on the summit of Maunakea, the Wide Field Infrared Camera \citep[WIRC;][]{wilson03} on the 200-inch telescope at Palomar Observatory (P200), and the FourStar IR camera \citep{persson13} on the Magellan Baade Telescope at LCO. Flat-fielding, background subtraction, astrometric alignment, and final stacking of images in each filter were performed using a custom pipeline. Additional near-IR imaging for SPIRITS\,17lb was obtained with the FLAMINGOS-2 imaging spectrograph on the 8.1\,m Gemini S Telescope (PID GS-2017B-Q-15;  PI J.\ Jencson) and reduced using the Gemini \textsc{iraf} package following procedures online in the FLAMINGOS-2 Data Reduction Cookbook\footnote{Data reduction procedures for FLAMINGOS-2 are found at \url{https://gemini-iraf-flamingos-2-cookbook.readthedocs.io/en/latest/}}. Optical imaging was obtained for SPIRITS\,17pc using the DEep Imaging Multi-Object Spectrograph (DEIMOS; \citealp{faber03}) on the 10\,m Keck II Telescope and reduced using standard tasks in \textsc{iraf}. 

Additional near-IR imaging was obtained with the Wide Field Camera \citep[WFCAM;][]{casali07} on
the United Kingdom Infrared Telescope (UKIRT) at Maunakea Observatories. Simultaneous optical/near-IR
$rizYJH$ were obtained with the Reionization and Transients InfraRed camera \citep[RATIR;][]{butler12} on the 1.5\,m Johnson Telescope at the Mexican Observatorio Astronomico Nacional on the Sierra San Pedro Martir in Baja California, Mexico \citep{watson12}. 

For our follow-up imaging, we performed simple aperture photometry at the location of the transients with the aperture size defined by the seeing in each image. Our photometry (including PSF photometry above) was calibrated using $\gtrsim 10$ isolated stars in SDSS for the optical images and in the Two Micron All Sky Survey (2MASS) for the near-IR images. Where necessary, we adopt the conversions of \citet{jordi06} to convert from the Sloan $ugriz$ system to $UBVRI$ magnitudes. For $Y$-band images, we adopt the conversion from 2MASS used for the UKIRT/WFCAM from \citet{hodgkin09}. Unfiltered CCD images of NGC~5921 were calibrated to the SDSS $r$ band. Our supplementary photometry is included in Table~\ref{table:Sup_phot} and shown for each object in Figure~\ref{fig:multiband_lcs}.

\begin{deluxetable*}{lrrccrr}
\tablecaption{Supplementary photometry\tablenotemark{a} \label{table:Sup_phot}}
\tablehead{ \colhead{Name} & \colhead{MJD} & \colhead{Phase\tablenotemark{b}} & \colhead{Tel./Inst.} & \colhead{Band} & \colhead{Apparent Mag.} & \colhead{Absolute Mag.\tablenotemark{c}} \\ 
\colhead{} & \colhead{} & \colhead{(days)} & \colhead{} & \colhead{} & \colhead{} & \colhead{} }
\startdata
SPIRITS\,14azy & 56,772.0 &     0.0 &        Swope/CCD &   $g$ & $21.94 \pm 0.47$ &    $-8.8$ \\
               & 56,772.0 &     0.0 &        Swope/CCD &   $r$ & $20.49 \pm 0.18$ &   $-10.2$ \\
               & 56,772.0 &     0.0 &        Swope/CCD &   $i$ & $19.81 \pm 0.18$ &   $-10.8$ \\
               & 56,804.0 &    32.0 &        Swope/CCD &   $g$ & $20.04 \pm 0.10$ &   $-10.7$ \\
               & 56,804.0 &    32.0 &        Swope/CCD &   $r$ & $19.02 \pm 0.06$ &   $-11.7$ \\
               & 56,804.0 &    32.0 &        Swope/CCD &   $i$ & $18.73 \pm 0.06$ &   $-11.9$ \\
               & 56,828.0 &    56.0 &        Swope/CCD &   $g$ & $20.79 \pm 0.27$ &   $-10.0$ \\
               & 56,828.0 &    56.0 &        Swope/CCD &   $r$ & $19.63 \pm 0.10$ &   $-11.0$ \\
               & 56,828.0 &    56.0 &        Swope/CCD &   $i$ & $19.20 \pm 0.10$ &   $-11.4$ \\
               & 56,829.0 &    57.0 &        Swope/CCD &   $g$ & $21.09 \pm 0.19$ &    $-9.7$ \\
               & 56,829.0 &    57.0 &        Swope/CCD &   $r$ & $19.69 \pm 0.07$ &   $-11.0$ \\
               & 56,829.0 &    57.0 &        Swope/CCD &   $i$ & $19.17 \pm 0.07$ &   $-11.4$ \\
               & 56,996.8 &   224.8 & du~Pont/RetroCam &   $H$ & $19.83 \pm 0.05$ &   $-10.6$ \\
               & 57,053.7 &   281.7 & du~Pont/RetroCam &   $H$ & $20.32 \pm 0.19$ &   $-10.2$ \\
               & 57,066.7 &   294.7 & du~Pont/RetroCam &   $H$ & $20.11 \pm 0.19$ &   $-10.4$ \\
               & 57,069.7 &   297.7 &   Baade/FourStar &   $J$ & $21.11 \pm 0.27$ &    $-9.4$ \\
               & 57,069.7 &   297.7 &   Baade/FourStar & $K_s$ & $18.58 \pm 0.27$ &   $-11.9$ \\
               & 57,089.7 &   317.7 & du~Pont/RetroCam &   $H$ & $20.42 \pm 0.22$ &   $-10.1$ \\
               & 57,115.6 &   343.6 & du~Pont/RetroCam &   $H$ & $20.41 \pm 0.11$ &   $-10.1$ \\
\enddata
\tablenotetext{a}{We show measurements for SPIRITS\,14azy to illustrate the form and content of this table. A full version including all of our supplementary photometry is published in machine-readable format.}
\tablenotetext{b}{Time since $t_0$ as reported in Table~\ref{table:lc_prop} for the IR-discovered sample.}
\tablenotetext{c}{Absolute magnitudes corrected for Galactic extinction to the host from NED.}
\end{deluxetable*}

\subsection{Spectroscopy}
We obtained near-IR spectroscopy of the luminous IR SPIRITS transients at several epochs using the Folded-port InfraRed Echellette spectrograph \citep[FIRE;][]{simcoe08,simcoe13} on the Magellan Baade Telescope at LCO, MOSFIRE on the Keck I Telescope, the Near-Infrared Echellette Spectrometer\footnote{\url{https://www2.keck.hawaii.edu/inst/nires/}} (NIRES) on the Keck II Telescope, the Gemini Near-InfraRed Spectrograph (GNIRS; \citealp{elias06}) on the 8.1 m Gemini N Telescope (PIDs GN-2016B-FT-25, GN-2017B-Q-14; PI J.\ Jencson), and the FLAMINGOS-2 spectrograph \citep{eikenberry06} on the 8.1 m Gemini S Telescope (PID GS-2017B-Q-15; PI J. Jencson). Optical spectroscopy was obtained for SPIRITS\,16tn, SPIRITS\,17pc, and SPIRITS\,17qm with the Low Resolution Imaging Spectrometer (LRIS; \citealp{goodrich03}) on the Keck I telescope. Our spectroscopic observations, including the integration times used, are summarized in Table~\ref{table:spec}.

\begin{deluxetable*}{llcrlccc}
\tablecaption{Spectroscopic Observations \label{table:spec}}
\tablehead{\colhead{Name} & \colhead{UT Date} & \colhead{MJD} & \colhead{Phase} & \colhead{Tel./Instr.} & \colhead{Range} & \colhead{Resolution} & \colhead{Integration} \\ 
\colhead{} & \colhead{} & \colhead{} & \colhead{(days)} & \colhead{} & \colhead{(\AA)} & \colhead{($\lambda / \delta \lambda$)} & \colhead{} }
\startdata
SPIRITS\,15c   & 2015 Mar 14.1 & 57,095.1 & 204.7 & Baade/FIRE           & 8000--24000  & 300--500 & $4 \times 120$~s  \\
               & 2015 Mar 31.2 & 57,112.2 & 221.8 & Keck I/MOSFIRE       & 9700--11100  & 3400     & $8 \times 180$~s  \\
               & 2015 Mar 31.3 & 57,112.3 & 221.9 & Keck I/MOSFIRE       & 11400--13100 & 3300     & $6 \times 120$~s  \\
               & 2015 Mar 31.3 & 57,112.3 & 221.9 & Keck I/MOSFIRE       & 14500--17500 & 3700     & $10 \times 120$~s \\
               & 2015 Mar 31.3 & 57,112.3 & 221.9 & Keck I/MOSFIRE       & 19500--23500 & 3600     & $6 \times 180$~s  \\
               & 2015 Sep 19.3 & 57,284.6 & 394.2 & Keck I/MOSFIRE       & 9700--11100  & 3400     & $10 \times 180$~s \\
SPIRITS\,15ud  & 2016 Jan 23.6 & 57,410.6 & 138.9 & Keck I/MOSFIRE       & 19500--23500 & 3600     & $12 \times 180$~s \\
SPIRITS\,15ade & 2016 Jan 23.6 & 57,410.6 & 134.1 & Keck I/MOSFIRE       & 19500--23500 & 3600     & $30 \times 180$~s \\
               & 2016 Apr 16.6 & 57,494.6 & 218.1 & Keck I/MOSFIRE       & 9700--11100  & 3400     & $2 \times 180$~s  \\
               & 2016 Apr 16.6 & 57,494.6 & 218.1 & Keck I/MOSFIRE       & 11400--13100 & 3300     & $10 \times 120$~s \\
               & 2016 May 30.5 & 57,538.5 & 262.0 & Keck I/MOSFIRE       & 14500--17500 & 3700     & $6 \times 120$~s  \\
SPIRITS\,16ix  & 2016 Apr 16.5 & 57,494.5 & 16.6  & Keck I/MOSFIRE       & 19500--23500 & 3600     & $10 \times 180$~s \\
SPIRITS\,16tn  & 2016 Nov 2.6  & 57,694.6 & 79.5  & Keck I/LRIS          & 3500--6000   & 600      & $1800$~s          \\
               & 2016 Nov 2.6  & 57,694.6 & 79.5  & Keck I/LRIS          & 5500--10300  & 1000     & $2 \times 860$~s  \\
               & 2016 Dec 29.5 & 57,751.5 & 136.4 & Gemini N/GNIRS       & 8500--25000  & 1200     & $14 \times 300$~s \\
               & 2017 Jan 9.6  & 57,762.6 & 147.5 & Gemini N/GNIRS       & 8500--25000  & 1200     & $10 \times 300$~s \\
SPIRITS\,17lb  & 2017 Sep 28.6 & 58,024.6 & 122.9 & Keck I/MOSFIRE       & 19500--23500 & 3600     & $24 \times 180$~s \\
               & 2017 Nov 1.3  & 58,058.3 & 156.6 & Gemini S/FLAMINGOS-2 & 13500--24000 & 600      & $24 \times 150$~s \\
               & 2017 Nov 20.6 & 58,077.6 & 175.9 & Keck I/MOSFIRE       & 11400--13100 & 3300     & $6  \times 120$~s \\
               & 2017 Nov 20.6 & 58,077.6 & 175.9 & Keck I/MOSFIRE       & 19500--23500 & 3600     & $8  \times 180$~s \\
SPIRITS\,17pc  & 2017 Nov 11.6 & 58,071.6 & 192.8 & Keck I/LRIS          & 3500--6000   & 600      & $1200$~s          \\
               & 2017 Nov 11.6 & 58,071.6 & 192.8 & Keck I/LRIS          & 5500--10300  & 1000     & $2 \times 560$~s  \\
               & 2017 Nov 20.6 & 58,077.6 & 198.8 & Keck I/MOSFIRE       & 9700--11100  & 3400     & $6 \times 180$~s  \\
               & 2017 Nov 20.7 & 58,077.7 & 198.9 & Keck I/MOSFIRE       & 19500--23500 & 3600     & $4 \times 180$~s  \\
               & 2017 Dec 7.6  & 58,094.6 & 215.8 & Gemini N/GNIRS       & 8500--25000  & 1200     & $4 \times 180$~s  \\
               & 2017 Dec 8.6  & 58,095.6 & 216.8 & Gemini N/GNIRS       & 8500--25000  & 1200     & $8 \times 180$~s  \\
               & 2017 Dec 11.6 & 58,098.6 & 219.8 & Gemini N/GNIRS       & 8500--25000  & 1200     & $12 \times 180$~s \\
               & 2018 Jan 8.5  & 58,126.5 & 247.7 & Keck I/MOSFIRE       & 9700--11100  & 3400     & $6 \times 180$~s  \\
               & 2018 Jan 8.6  & 58,126.6 & 247.8 & Keck I/MOSFIRE       & 11400--13100 & 3300     & $6 \times 120$~s  \\
               & 2018 Jan 8.6  & 58,126.6 & 247.8 & Keck I/MOSFIRE       & 14500--17500 & 3700     & $6 \times 120$~s  \\
               & 2018 Jan 8.6  & 58,126.6 & 247.8 & Keck I/MOSFIRE       & 19500--23500 & 3600     & $6 \times 180$~s  \\
               & 2018 May 4.4  & 58,242.4 & 363.6 & Keck II/NIRES        & 9500--24600  & 2700     & $6 \times 300$~s  \\
SPIRITS\,17qm  & 2017 Nov 11.5 & 58,071.5 & 9.3   & Keck I/LRIS          & 3500--6000   & 600      & $2 \times 1200$~s \\
               & 2017 Nov 11.5 & 58,071.5 & 9.3   & Keck I/LRIS          & 5500--10300  & 1000     & $4 \times 560$~s  \\
               & 2017 Nov 20.5 & 58,077.5 & 15.3  & Keck I/MOSFIRE       & 9700--11100  & 3400     & $14 \times 180$~s \\
               & 2018 Jan 8.2  & 58,126.2 & 64.0  & Keck I/MOSFIRE       & 9700--11100  & 3400     & $8 \times 180$~s  \\
               & 2018 Jan 8.3  & 58,126.3 & 64.1  & Keck I/MOSFIRE       & 11400--13100 & 3300     & $12 \times 120$~s \\
               & 2018 Jan 8.3  & 58,126.3 & 64.1  & Keck I/MOSFIRE       & 14500--17500 & 3700     & $12 \times 120$~s \\
               & 2018 Jan 8.3  & 58,126.3 & 64.1  & Keck I/MOSFIRE       & 19500--23500 & 3600     & $10 \times 180$~s \\
\enddata
\end{deluxetable*}

The FIRE and MOSFIRE spectra of SPIRITS\,15c and the GNIRS and LRIS spectra of SPIRITS\,16tn, as well as details of the data reduction procedures, were previously published in \citet{jencson17,jencson18c}.

\begin{figure*}
\plotone{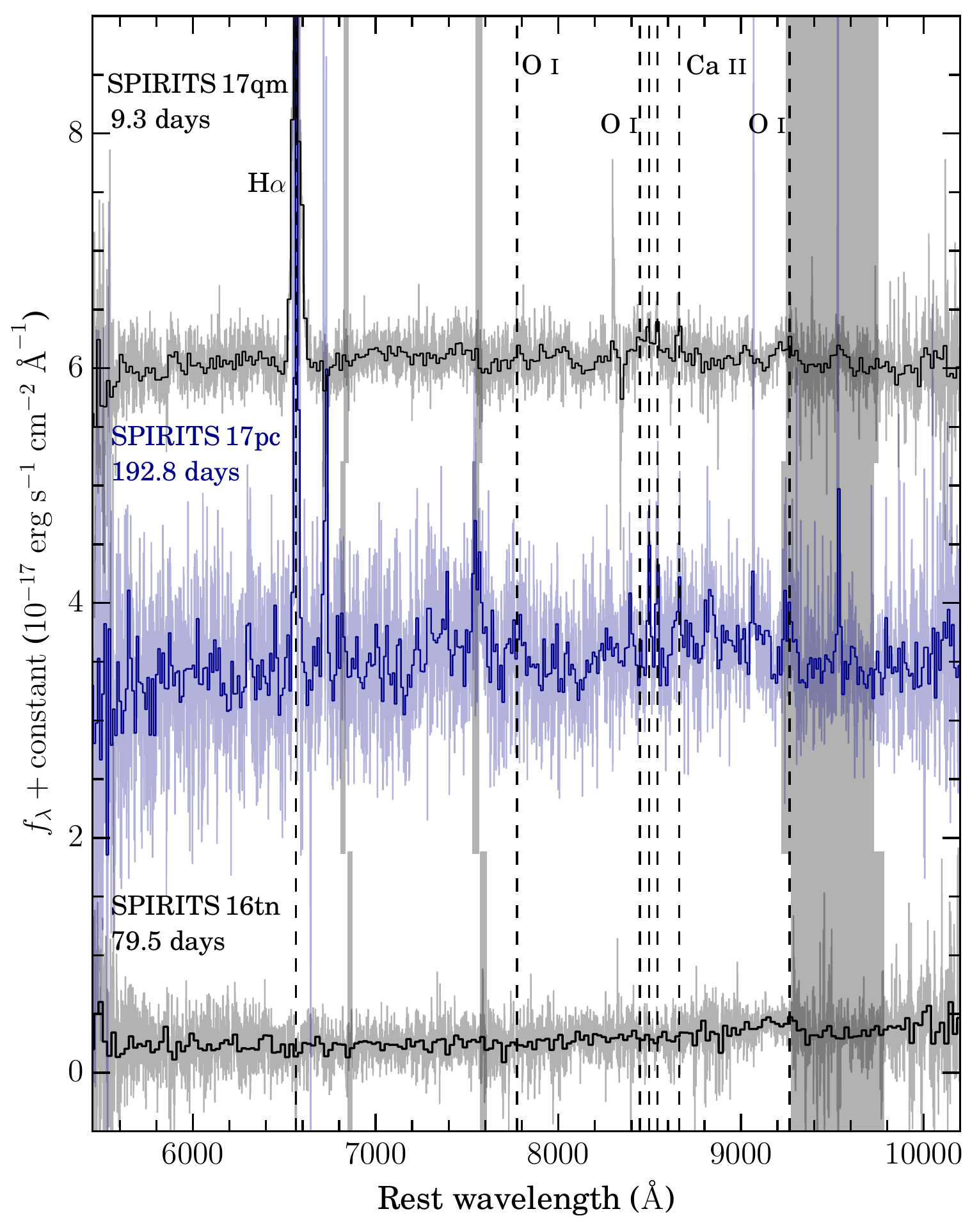}
\caption{\label{fig:opt_spec}
Optical spectroscopy from Keck I/LRIS of SPIRITS\,16tn, SPIRITS\,17pc, and SPIRITS\,17qm in alternating colors and offset in flux by an arbitrary constant so that spectra for each object may be easily distinguished. Spectra are labeled for each object along with their respective phases along the left side of the figure. Data are shown in light gray or light blue, with binned data for each object overplotted in black or dark blue, respectively. Regions contaminated by telluric absorption are indicated by the light-gray vertical bands. Atomic emission features detected in the spectra, including broad ${\approx}1000$\,km~s$^{-1}$ H$\alpha$ and lines of O~\textsc{i} in SPIRITS\,17qm and the Ca~\textsc{ii} IR triplet in SPIRITS\,17qm and SPIRITS\,17pc, are indicated by the dashed black vertical lines and labeled near the top of the figure. 
}
\end{figure*}

For the LRIS observations of SPIRITS\,17pc and SPIRITS\,17qm, we used the D560 dichroic to split the light between the red and blue sides, and we used a $1\arcsec$ long slit with the 400/8500 grating on the red side and the 400/3400 grism on the blue side providing the resolution and wavelength coverage given in Table~\ref{table:spec}. Spectroscopic reductions for LRIS were performed using the analysis pipeline LPIPE\footnote{Software available at \url{http://www.astro.caltech.edu/~dperley/programs/lpipe.html}.} \citep{perley19}. For SPIRITS\,17pc, a weak trace is visible at the position of the transient on the red-side camera. For SPIRITS\,17qm, there is no obvious continuum trace; however we detect several emission lines associated with the transient on the red side. No emission was detected on the blue side for either object. The 1D spectra were extracted at these positions along the slit and flux-calibrated using observations of the standard stars G191-B2B and BD\,$+75^{\circ}325$. Our optical spectra of SPIRITS\,16tn, SPIRITS\,17pc, and SPIRITS\,17qm are shown in Figure~\ref{fig:opt_spec}. 

For observations with the near-IR spectrographs the target was nodded along the slit between exposures to allow for accurate subtraction of the sky background. Observations of an A0V telluric standard star near the target location were also taken immediately before or after each science target observation for flux calibration and correction of the strong near-IR telluric absorption features. For MOSFIRE, we used a $0\farcs7$ slit with the standard grating/filter setups for each of the $YJHK$ spectral regions, providing the wavelength coverage and resolution listed for each observation in Table~\ref{table:spec}. For GNIRS, we used the cross-dispersed (XD), multiorder mode providing coverage of the full near-IR spectral region at once, with a $0\farcs45$ slit, the 32~line~mm$^{-1}$ grating, and the short blue camera with its XD prism, providing an average spectral resolution of $R = 1200$. For FLAMINGOS-2, we use the long-slit mode and the low-resolution $HK$ grism with a 3 pixel, $0\farcs54$ slit providing wavelength coverage from 13,500 to 24,000~\AA\ and an average spectral resolution of $R = 600$. NIRES employs a $0\farcs55$ slit and provides wavelength coverage from 9500 to 24,600~\AA\ across five spectral orders at a mean resolution of $R = 2700$.

Reductions for MOSFIRE, including flat-fielding, the wavelength solution, background subtraction, and frame stacking for each object on a given night, were performed with the MOSFIRE Data Reduction Pipeline\footnote{\url{https://keck-datareductionpipelines.github.io/MosfireDRP/}}. 1D extractions, where the continuum trace of the target and/or emission lines were visible in the reduced 2D spectra, were performed using standard tasks in \textsc{iraf}\footnote{\textsc{iraf} is distributed by the National Optical Astronomy Observatory, which is operated by the Association of Universities for Research in Astronomy (AURA) under a cooperative agreement with the National Science Foundation.}. For GNIRS and FLAMINGOS-2 reductions, including detector pattern noise cleaning and radiation event removal (GNIRS only), flat-fielding, background subtraction, spatial distortion corrections (GNIRS only), wavelength calibration, and 1D extractions, we used standard tasks in the Gemini \textsc{iraf} package following procedures outlined on the Gemini webpage\footnote{Procedures for reducing GNIRS XD spectra are found at \url{http://www.gemini.edu/sciops/instruments/gnirs/data-format-and-reduction/reducing-xd-spectra}}. NIRES data were reduced, including flat-fielding, wavelength calibration, background subtraction, and 1D spectral extractions steps, using a version of the IDL-based data reduction package Spextool developed by \citet{cushing04}, updated by M.\ Cushing specifically for NIRES. Corrections for the strong near-IR telluric absorption features and flux calibrations for spectra from all instruments were performed with the A0V standard-star observations using the method developed by \citet{vacca03} implemented with the IDL tools \textsc{xtellcor} or \textsc{xtellcor\_general} developed by \citet{cushing04} as part of Spextool. 

\begin{figure*}
\plotone{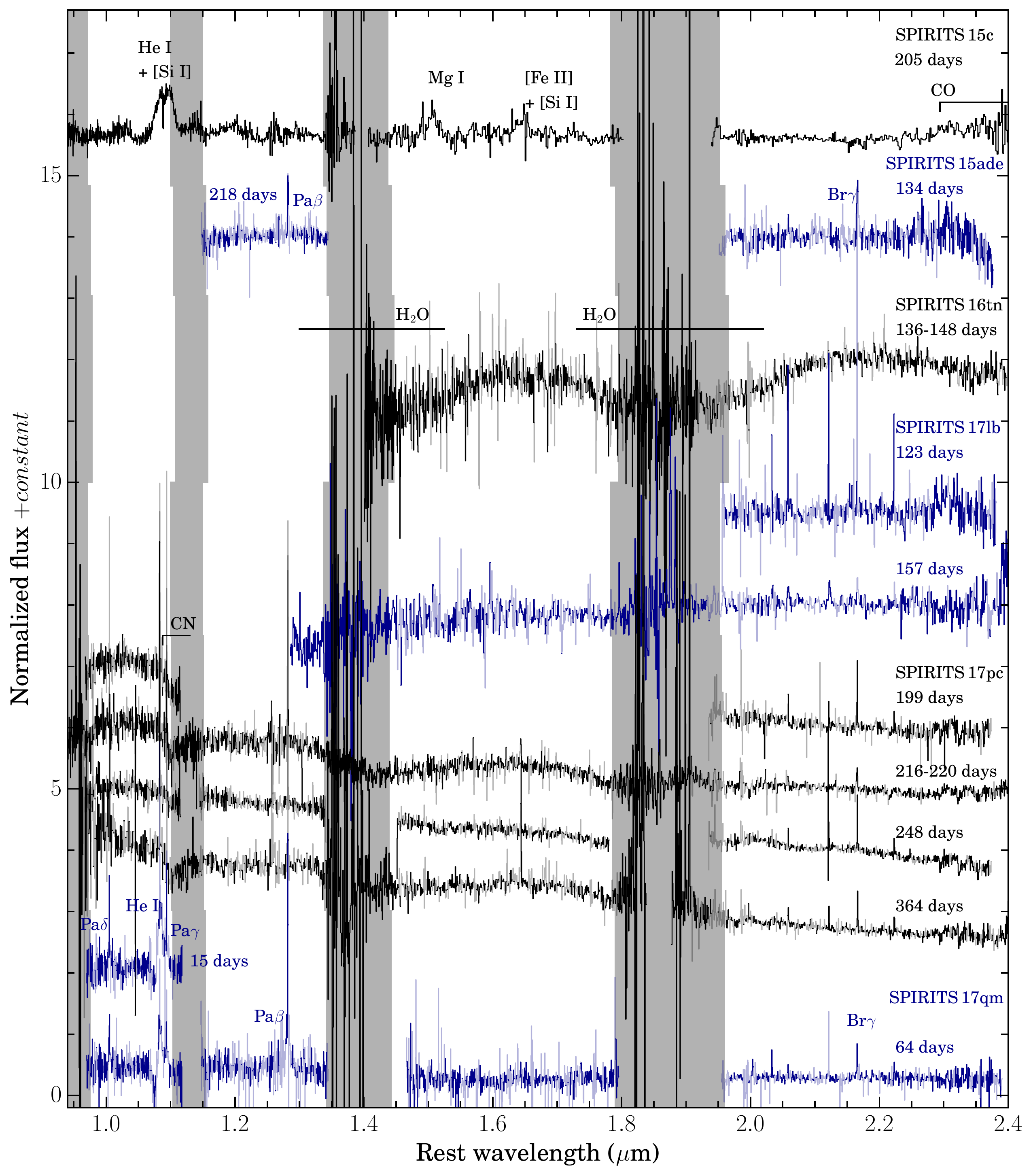}
\caption{\label{fig:nir_spec}
Full sequence of near-IR spectra obtained for the SPIRITS luminous IR transient sample. Spectra correspond to the objects and phases (measured from $t_0$ as in Table~\ref{table:lc_prop}) listed along the right side of the figure, shown in alternating colors so that spectra for separate objects may be easily distinguished. The spectra for each object are shifted to the rest frame of their respective host galaxies, and regions of low S/N due to coincidence with OH airglow emission lines are shown in lighter colors. The spectra have been scaled in flux and shifted by arbitrary constants for clarity. The major features identified in each spectrum and discussed in the text are labeled, including the SN~Ib/IIb emission features in the spectrum of SPIRITS\,15c, the ${\sim}200$\,km~s$^{-1}$ width emission lines of H~\textsc{i} in the spectra of SPIRITS\,15ade, the possible broad H$_2$O absorption features in the spectrum of SPIRITS\,16tn, the broader ${\sim}2000$\,km~s$^{-1}$ features of H~\textsc{i} and He~\textsc{i} in the spectrum of SPIRITS\,17qm, and the CO $\delta v = 2$ vibrational transition band heads detected in the spectra of SPIRITS\,15c, SPIRITS\,17lb, and SPIRITS\,17pc. 
}
\end{figure*}

We did not detect any line emission or a discernible continuum trace from the transients in the MOSFIRE spectrum of SPIRITS\,15ud, the MOSFIRE 2016 April 16.6 $Y$-band and 2016 May 30.5 $H$-band spectra of SPIRITS\,15ade, the MOSFIRE spectrum of SPIRITS\,16ix, or the 2017 November 20.6 MOSFIRE spectrum of SPIRITS\,17lb. The full sequence of near-IR spectra for which we detected emission from the transient, either a continuum trace or specific emission lines, is shown in Figure~\ref{fig:nir_spec}.

\subsection{Radio Observations}\label{radio}
We obtained radio continuum imaging observations of each of our luminous IR transients, primarily using the Karl G. Jansky Very Large Array (VLA) between 2017 June and 2018 April in the C, B, and A configurations with 3.4, 11.1, and 36.4\,km maximum baselines, respectively (PIDs 16B-388, 17A-365, 17B-331, 18A-418; PI J.\ Jencson). The bulk of our observations were carried out in 1\,hr blocks in $C$ band (6\,GHz central frequency) using the full, wide-band capabilities of the upgraded VLA with 3-bit samplers offering 4\,GHz bandwidth. For some objects and epochs, we split the block between multiple bands including the $S$ (3 GHz central frequency, 2\,GHz bandwidth with 8\,bit samplers), $X$ (10 GHz central frequency, 4\,GHz bandwidth), and $Ku$ bands (15.5\,GHz central frequency, 4\,GHz bandwidth). Each observing block included observations of the VLA standard calibrators 3C\,286 or 3C\,138 for calibration of the flux density scale and instrument bandpass. For complex gain calibrations, we took observations of a nearby calibrator source, cycling between the calibrator and science target at sufficient intervals for the given array configuration and observing frequency based on recommendations in the VLA observing guide\footnote{VLA calibration information available here: \url{https://science.nrao.edu/facilities/vla/docs/manuals/obsguide/calibration}}.

We also analyzed the 2016 February 19 observations of IC\,2163 targeting the unrelated event SN~2010jp, but fortuitously covering the sites of SPIRITS\,14buu, SPIRITS\,15c, and SPIRITS\,17lb (PID 16A-101; PI C.\ Kilpatrick). These observations used the 8-bit sampler setup providing $2\times1$\,GHz bandwidth tuned to the ranges 4.5--5.5\,GHz and 6.9--7.9\,GHz in $C$ band and 8.0--9.0\,GHz and 10.5--11.5~\,GHz in $X$ band. 

The data for each observation were run through the VLA CASA calibration pipeline,\footnote{\url{https://science.nrao.edu/facilities/vla/data-processing/pipeline/scripted-pipeline}} suitable for automated flagging and calibration of Stokes $I$ continuum data sets. The calibrated data were carefully inspected, and additional flagging and recalibration were performed as necessary. We imaged our data using the standard \textsc{clean} task in CASA. First-pass images of SPIRITS\,17pc in NGC~4388 were limited in sensitivity owing to artifacts from residual phase errors of the bright nucleus of the host, a known Seyfert~2 active galactic nucleus. We performed self-calibration of the visibility phases on the nucleus, which significantly improved the final images, reaching near the theoretical thermal noise sensitivity for our observations. 

We inspected the location of the transients in each image for the presence of a coincident point source and report our flux measurements with $1\sigma$ errors, estimated as the rms noise in a relatively clean region of the image, in Table~\ref{table:radio}. In several cases, the location of the transients suffered significant contamination from extended emission from the host galaxy or nearby star-forming regions, particularly in the lower-resolution $C$-configuration observations. For nondetections or when possible emission from the transient cannot be distinguished from background contamination, we report upper limits on the transient flux as either $5\sigma$ ($5\times \mathrm{image~rms}$), or the level of the contaminating flux at the location plus $2\times \mathrm{image~rms}$, whichever is larger. 

Specifically for the 2016 February 19.1 $C$-band measurement of SPIRITS\,15c, a source is clearly detected at the position and has faded significantly in the subsequent image on 2017 June 10.8 (taken in the same configuration with similar resolution), confirming its association with the transient. We note, however, that possible residual emission from the transient in the later epoch is blended with a nearby, somewhat extended contaminating source. We adopt the flux at the transient location as an estimate of the maximum contamination for our previous measurement and adopt this as a lower bound on the flux in Table~\ref{table:radio}. In the subsequent epochs in the larger B and A configurations, the contaminating emission appears resolved out, and the radio counterpart to SPIRITS\,15c is clearly detected as a relatively isolated, fading point source. 

Radio observations of SPIRITS\,16tn were also obtained with the Arcminute Microkelvin Image Large Array (AMI-LA), which were previously reported in \citet{jencson18c}. Additional observations of IC~2163 containing SPIRITS\,14buu, SPIRITS\,15c, and SPIRITS\,17lb were obtained with the Australia Telescope Compact Array (ATCA) over 6\,hr on 2017 September 4 at 9.0 and 5.5\,GHz simultaneously. The ATCA primary flux calibrator, PKS~B1934\==638, was used to set the absolute flux scale, as well as define the bandpass calibration in each 2\,GHz band. Frequent observations of the nearby source PKS~0606\==223 allowed us to monitor and correct for variations in gain and phase throughout each run. The data were processed using the Miriad package \citep{sault95} and using procedures outlined in Section~4.3 of the ATCA User Guide\footnote{\url{http://www.narrabri.atnf.csiro.au/observing/users_guide/html/atug.html}}. After editing and calibrating the data, images at each frequency were made using robust weighting ($\mathrm{robust}=0.5$) and then cleaned down to 3$\times$ the rms noise level. No point sources exactly matching the locations of these three SNe were evident at either frequency within the clumpy disk structure. Fitting a Gaussian point source to the nearest peak of emission in each case yielded the upper limits on flux densities shown in Table~\ref{table:radio}.

\begin{deluxetable*}{llccccccc}
\tablecaption{Radio observations \label{table:radio}}
\tablehead{\colhead{Name} & \colhead{UT Date} & \colhead{MJD} & \colhead{Phase} & \colhead{Inst.} & \colhead{Max. Baseline} & \colhead{Frequency} & \colhead{Flux} & \colhead{Luminosity} \\ 
\colhead{} & \colhead{} & \colhead{} & \colhead{(days)} & \colhead{} & \colhead{(km)} & \colhead{(GHz)} & \colhead{(mJy)} & \colhead{(erg~s$^{-1}$~Hz$^{-1}$)} } 
\startdata
SPIRITS\,14buu & 2016 Feb 19.1  & 57,437.1 & 786.8  & VLA    & 3.4 & 10.0  & $<0.04$         & $<6.0\times10^{25}$ \\
               & 2016 Feb 19.1  & 57,437.1 & 786.8  & VLA    & 3.4 & 6.0   & $<0.13$         & $<2.0\times10^{26}$ \\
               & 2017 Jun 10.8  & 57,914.8 & 1264.5 & VLA    & 3.4 & 6.0   & $<0.12$         & $<1.8\times10^{26}$ \\
               & 2017 Sep 4     & 58,000   & 1349.7 & ATCA   & 4.5 & 9.0   & $<22$           & $<3.3\times10^{28}$ \\
               & 2017 Sep 4     & 58,000   & 1349.7 & ATCA   & 4.5 & 5.5   & $<25$           & $<3.8\times10^{28}$ \\
               & 2018 Jan 6.2   & 58,124.2 & 1473.9 & VLA    & 11.1 & 6.0  & $<0.039$        & $<5.9\times10^{25}$ \\
               & 2018 May 9.0   & 58,247.0 & 1596.7 & VLA    & 36.4 & 6.0  & $<0.035$        & $<5.3\times10^{25}$ \\
               & 2018 May 9.0   & 58,247.0 & 1596.7 & VLA    & 36.4 & 6.0  & $<0.055$        & $<8.3\times10^{25}$ \\
\hline
SPIRITS\,14azy & 2017 Jun 12.9  & 57,916.9 & 1144.9 & VLA    & 3.4  & 6.0  & $<0.075$        & $<1.3\times10^{25}$ \\
SPIRITS\,15c   & 2016 Feb 19.1  & 57,437.1 & 546.7  & VLA    & 3.4  & 10.0 & $0.12\pm0.01$ & $1.8\times10^{26}$    \\
               & 2016 Feb 19.1  & 57,437.1 & 546.7  & VLA    & 3.4  & 6.0  & $0.23^{+0.01}_{-0.09}$ & $3.5\times10^{26}$ \\
               & 2017 Jun 10.8  & 57,914.8 & 1024.4 & VLA    & 3.4  & 6.0  & $<0.11$         & $<1.7\times10^{26}$ \\
               & 2017 Sep 4     & 58,000   & 1110   & ATCA   & 4.5  & 9.0  & $< 1.7$         & $< 2.6\times10^{27}$ \\
               & 2017 Sep 4     & 58,000   & 1110   & ATCA   & 4.5  & 5.5  & $< 7.6$         & $< 1.1\times10^{28}$ \\
               & 2018 Jan 6.2   & 58,124.2 & 1233.8 & VLA    & 11.1 & 6.0  & $0.055\pm0.007$ & $8.3\times10^{25}$  \\
               & 2018 May 9.0   & 58,247.0 & 1356.6 & VLA    & 36.4 & 6.0  & $0.051\pm0.008$ & $7.7\times10^{25}$  \\
               & 2018 May 9.0   & 58,247.0 & 1356.6 & VLA    & 36.4 & 3.0  & $0.082\pm0.014$ & $1.2\times10^{26}$  \\
SPIRITS\,15ud  & 2017 Jun 14.1  & 57,918.1 & 646.4  & VLA    & 3.4  & 6.0  & $<1.1$          & $<2.5\times10^{26}$ \\
SPIRITS\,15ade & 2017 Jun 16.0  & 57,920.0 & 643.5  & VLA    & 3.4  & 6.0  & $<0.045$        & $<3.1\times10^{25}$ \\
SPIRITS\,16ix  & 2017 Jun 16.0  & 57,920.0 & 442.1  & VLA    & 3.4  & 6.0  & $<0.025$        & $<1.2\times10^{25}$ \\
SPIRITS\,16tn  & 2016 Sep 3     & 57,634   & 19     & AMI-LA & 0.11 & 15.0 & $<0.3$          & $<2.8\times10^{25}$ \\
               & 2016 Sep 4.0   & 57,635.0 & 20.0   & VLA    & 11.1 & 10.0 & $<0.047$        & $<4.4\times10^{24}$ \\
               & 2016 Sep 4.0   & 57,635.0 & 20.0   & VLA    & 11.1 & 6.0  & $<0.075$        & $<7.0\times10^{24}$ \\
               & 2016 Sep 4.0   & 57,635.0 & 20.0   & VLA    & 11.1 & 3.0  & $<0.10$         & $<9.3\times10^{24}$ \\
               & 2017 Jan 12.4  & 57,765.4 & 150.4  & VLA    & 36.4 & 15.5 & $<0.029$        & $<2.7\times10^{24}$ \\
               & 2017 Jan 12.4  & 57,765.4 & 150.4  & VLA    & 36.4 & 6.0  & $<0.029$        & $<2.7\times10^{24}$ \\
               & 2018 Jan 23.6  & 58,141.6 & 526.6  & VLA    & 11.1 & 15.5 & $<0.030$        & $<2.8\times10^{24}$ \\
               & 2018 Jan 23.6  & 58,141.6 & 526.6  & VLA    & 11.1 & 6.0  & $<0.030$        & $<2.8\times10^{24}$ \\
SPIRITS\,17lb  & 2017 Jun 10.8  & 57,914.8 & 13.1   & VLA    & 3.4  & 6.0  & $<0.25$         & $<3.8\times10^{26}$ \\
               & 2017 Sep 4     & 58,000   & 99     & ATCA   & 4.5  & 9.0  & $< 1.7$         & $<2.6\times10^{27}$ \\
               & 2017 Sep 4     & 58,000   & 99     & ATCA   & 4.5  & 5.5  & $< 6.5$         & $<9.8\times10^{27}$ \\
               & 2018 Jan 6.2   & 58,124.2 & 222.5  & VLA    & 11.1 & 6.0  & $0.059\pm0.006$ & $8.9\times10^{25}$  \\
               & 2018 May 9.0   & 58,247.0 & 345.3  & VLA    & 36.4 & 6.0  & $0.039\pm0.009$ & $5.9\times10^{25}$  \\
               & 2018 May 9.0   & 58,247.0 & 345.3  & VLA    & 36.4 & 3.0  & $0.067\pm0.013$ & $1.0\times10^{26}$  \\
SPIRITS\,17pc  & 2017 Oct 30.5  & 58,056.5 & 177.7  & VLA    & 11.1 & 6.0  & $<0.090$        & $>3.6\times10^{25}$ \\
               & 2018 Jan 21.7  & 58,139.7 & 260.9  & VLA    & 11.1 & 6.0  & $<0.13$         & $>5.2\times10^{25}$ \\
               & 2018 Apr 14.2  & 58,222.1 & 343.3  & VLA    & 36.4 & 6.0  & $<0.025$        & $>9.9\times10^{24}$ \\
SPIRITS\,17qm  & 2017 Nov 23.3  & 58,080.3 & 18.1   & VLA    & 11.1 & 6.0  & $<0.05$         & $>2.0\times10^{25}$ \\
               & 2018 Apr 16.8  & 58,224.8 & 162.6  & VLA    & 36.4 & 6.0  & $<0.025$        & $>1.0\times10^{25}$ \\ 
\enddata
\end{deluxetable*}

\subsection{Limits from wide-field optical surveys}\label{opt_limits}
We obtained limits on the optical emission from SPIRITS transients that were fortuitously covered by the intermediate Palomar Transient Factory (iPTF) survey during their outbursts in the $g$, $R$, and/or $i$ filters. The iPTF operated from 2013 January 1 to 2017 March 2 using a wide-field camera with a 7.26~deg$^2$ field of view on the 48-inch Samuel Oschin Schmidt Telescope at Palomar Observatory (P48). During its operation, iPTF images were processed through a real-time image subtraction pipeline \citep{cao16} to search for transients and variables. We ran forced PSF photometry at the locations of our SPIRITS transients on iPTF reference-subtracted images using the PTF IPAC/iPTF Discovery Engine (PTFIDE) tool \citep{masci17}. To obtain deeper constraints, we stacked limits from individual observations within 10-day windows. Constraints on the optical emission from our sample transients from the iPTF are included in Figure~\ref{fig:multiband_lcs}.

\section{Analysis}\label{sec:analysis}

\subsection{Host environments and archival imaging}\label{sec:archival_imaging}
\begin{figure*}
\plotone{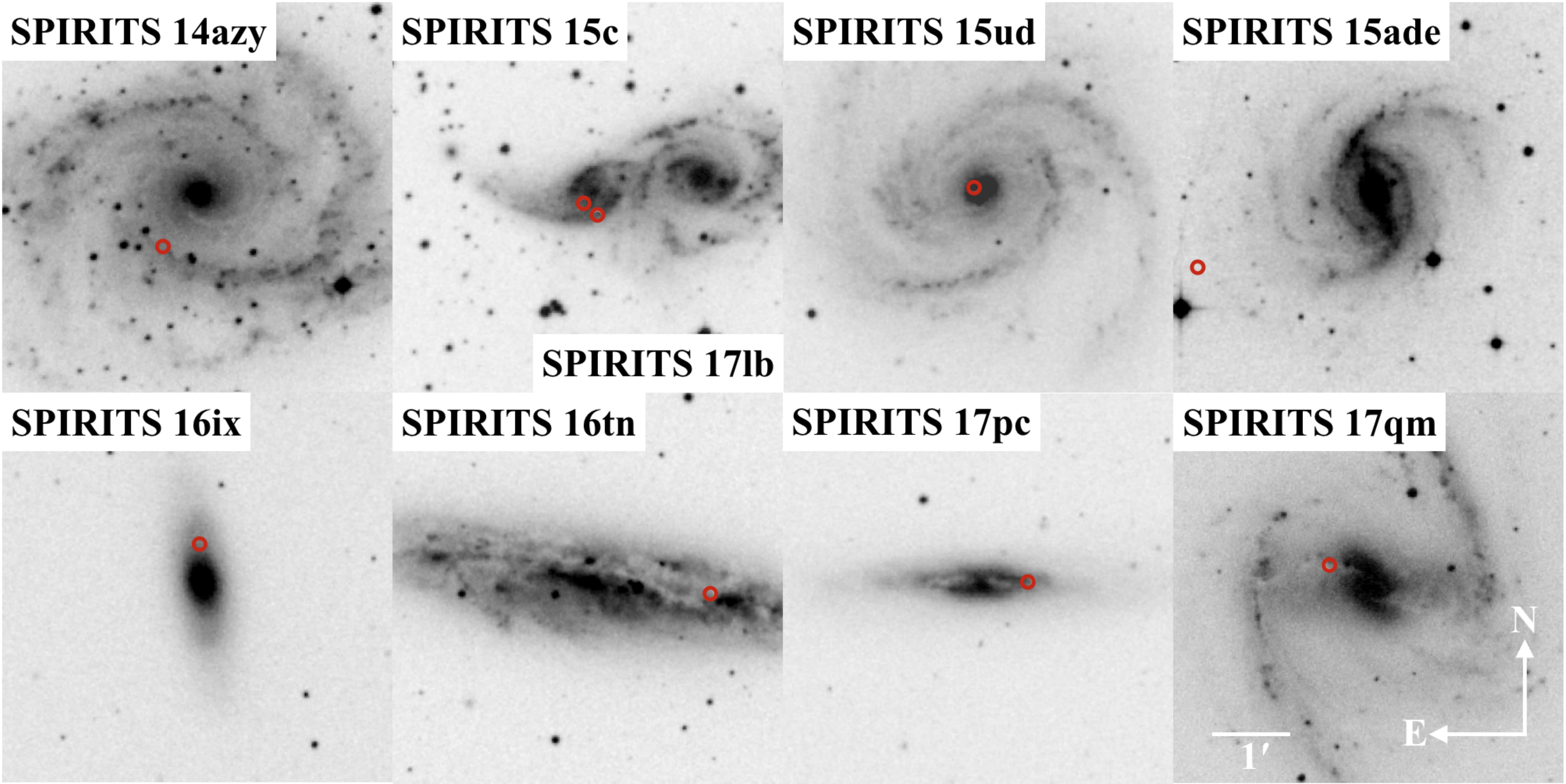}
\caption{\label{fig:DSS_hosts}
We show the DSS2-red images of the host galaxies of the IR-selected sample of SPIRITS transients. The location of the transient(s) in each panel is indicated by the red circle. The orientation and scale are the same in each panel as indicated in the bottom right panel.
}
\end{figure*}

The host galaxy morphological types (obtained from NED) of our sample of IR-selected transients are given in Table~\ref{tab:IRsample}. With the exception of SPIRITS\,16ix in the lenticular SB0$^+$ galaxy NGC~4461, these events were found in late-type, star-forming galaxies. Similarly, the entire control sample of optically discovered SNe was also found in star-forming galaxies (Table~\ref{table:opt_sample}). In Figure~\ref{fig:DSS_hosts}, we indicate the locations of the IR-selected transients in the Digitized Sky Survey 2 (DSS2) images of their host galaxies, showing a clear trend of positional associations with the active star-forming regions of the hosts' spiral arms. This indicates a likely physical association of the bulk of the sample to ongoing star formation and young, massive stars. Of particular note, SPIRITS\,15ade was discovered in the outskirts of NGC~5921 at a projected distance from the galaxy center of $146\farcs3$ ($17.0$~kpc), while SPIRITS\,15ud was found very near the nucleus of M100 at a projected distance of only $7\farcs5$ ($0.5$~kpc).  

We examined the location of each source in the IR-selected sample in the archival \textit{Spitzer}/IRAC [3.6] and [4.5] reference images for the presence of possible IR progenitor stars. The regions near the objects in our sample were typically crowded by several sources or dominated by the bright, spatially variable background emission from the host galaxy. To identify possible progenitors, we constructed source catalogs and performed PSF photometry for each \textit{Spitzer}/IRAC reference image using the \textsc{daophot}/\textsc{allstar} package \citep{stetson87}, where a model of the PSF was constructed using isolated stars in the image. The PSF-fitting and photometry procedure, including corrections for the finite radius of the PSF (using the method of \citealp{khan17}), is described in \citet{karambelkar19}. 

With the exception of SPIRITS\,17qm, there are no sources in our catalogs consistent with the transient positions in \textit{Spitzer}/IRAC reference images. We estimate upper limits on the progenitor flux as 5 times the standard deviation in a $25 \times 25$ pixel box at the transient position. The limits at [3.6] and [4.5] on the progenitor flux of each object in our sample are given in Table~\ref{tab:prog_lims}. 

For SPIRITS\,17qm, we identified a source in the [3.6] and [4.5] reference image PSF catalogs of NGC~1365, separated from the location of the transient by only $0\farcs12$ and $0\farcs24$, respectively, less than one IRAC mosaicked pixel. However, we note that this source is heavily blended with several other sources in our catalogs, and its centroid is slightly offset from the optical progenitor star identified in archival \textit{HST} imaging below. While the coincident IR PSF catalog sources likely contain flux from the progenitor, we consider the PSF magnitudes given in Table~\ref{tab:prog_lims} as upper limits given the possibility of significant contamination from nearby sources. 

\begin{deluxetable*}{llccccc}
\tablecaption{Archival imaging progenitor constraints\label{tab:prog_lims}}
\tablehead{\colhead{Name} & \colhead{UT Date} & \colhead{Tel./Inst.} & \colhead{Program/PI} & \colhead{Band} & \colhead{Vega Mag.} & \colhead{Abs. Mag.}\tablenotemark{a} }
\startdata
SPIRITS\,14azy & 2004 Dec 16.7 & \textit{Spitzer}/IRAC  & PID 3333/M.\ Barlow   & [3.6] & $>16.7$ & $>-13.8$ \\
               & 2004 Dec 16.7 & \textit{Spitzer}/IRAC  & PID 3333/M.\ Barlow   & [4.5] & $>16.5$ & $>-14.0$ \\
               & 2010 Oct 28.8 & \textit{HST}/WFC3 UVIS & SNAP-12229/L.\ Smith  & F336W & $>25.5$ & $>-5.4$  \\
SPIRITS\,15c   & 2005 Feb 22.7 & \textit{Spitzer}/IRAC  & Super Mosaic          & [3.6] & $>15.5$ & $>-17.3$ \\
               & 2005 Feb 22.7 & \textit{Spitzer}/IRAC  & Super Mosaic          & [4.5] & $>15.5$ & $>-17.3$ \\
               & 1998 Nov 11.7 & \textit{HST}/WFPC2 WFC & GO-6483/D.\ Elmegreen & F555W & $>25.1$ & $>-7.9$  \\
               & 1998 Nov 11.7 & \textit{HST}/WFPC2 WFC & GO-6483/D.\ Elmegreen & F814W & $>24.0$ & $>-8.9$  \\
SPIRITS\,15ud  & 2004 May 27.5--2008 Jul 15.7 & \textit{Spitzer}/IRAC & Super Mosaic & [3.6] & $>12.8$ & $>-17.9$ \\
               & 2004 May 27.5--2008 Jul 15.7 & \textit{Spitzer}/IRAC & Super Mosaic & [4.5] & $>12.7$ & $>-18.0$ \\
               & 2001 Nov 12.1 & \textit{HST}/WFC3 UVIS & GO-11646/A.\ Crotts   & F775W & $>24.5$ & $>-6.3$  \\
               & 2005 May 31.0 & \textit{HST}/ACS HRC   & GO-9776/D.\ Richstone & F814W & $>24.4$ & $>-6.4$  \\
SPIRITS\,15ade & 2009 Aug 29.4 & \textit{Spitzer}/IRAC  & S4G/K.\ Sheth         & [3.6] & $>19.7$ & $>-12.2$ \\
               & 2009 Aug 29.4 & \textit{Spitzer}/IRAC  & S4G/K.\ Sheth         & [4.5] & $>19.1$ & $>-12.8$ \\
SPIRITS\,16ix  & 2010 Aug 3.5  & \textit{Spitzer}/IRAC  & S4G/K.\ Sheth         & [3.6] & $>16.3$ & $>-15.2$ \\
               & 2010 Aug 3.5  & \textit{Spitzer}/IRAC  & S4G/K.\ Sheth         & [4.5] & $>16.3$ & $>-15.2$ \\
SPIRITS\,16tn  & 2011 Feb 7.6  & \textit{Spitzer}/IRAC  & S4G/K.\ Sheth         & [3.6] & $>15.0$ & $>-14.7$ \\
               & 2011 Feb 7.6  & \textit{Spitzer}/IRAC  & S4G/K.\ Sheth         & [4.5] & $>14.8$ & $>-14.9$ \\
               & 1994 Jul 4.8  & \textit{HST}/WFPC2 WFC & SNAP-5446/G.\ Illingworth & F606W & $>24.5$ & $>-5.2$ \\
SPIRITS\,17lb  & 2005 Feb 22.7 & \textit{Spitzer}/IRAC  & Super Mosaic          & [3.6] & $>14.9$ & $>-17.9$ \\
               & 2005 Feb 22.7 & \textit{Spitzer}/IRAC  & Super Mosaic          & [4.5] & $>14.8$ & $>-18.0$ \\
               & 2012 Dec 4.5  & \textit{HST}/WFC3 UVIS & SNAP-13029/A.\ Filippenko & F625W & $>26.8$ & $>-9.4$ \\
               & 2012 Dec 4.5  & \textit{HST}/WFC3 UVIS & SNAP-13029/A.\ Filippenko & F814W & $>26.2$ & $>-8.9$ \\
SPIRITS\,17pc  & 2004 May 27.6--2008 Jul 17.7 & \textit{Spitzer}/IRAC & Super Mosaic & [3.6] & $>14.6$ & $>-16.7$ \\
               & 2004 May 27.6--2008 Jul 17.7 & \textit{Spitzer}/IRAC & Super Mosaic & [4.5] & $>14.4$ & $>-16.9$ \\
               & 2011 Jun 8.3  & \textit{HST}/WFC3 UVIS & GO-12185/J.\ Greene   & F336W & $>26.06$        & $>-5.4$ \\
               & 2011 Jun 8.4  & \textit{HST}/WFC3 UVIS & GO-12185/J.\ Greene   & F438W & $>26.2$        & $>-5.2$ \\
               & 2011 Jun 8.4  & \textit{HST}/WFC3 UVIS & GO-12185/J.\ Greene   & F814W & $24.21\pm0.05$ & $-7.1$  \\
               & 2011 Jun 8.2  & \textit{HST}/WFC3 IR   & GO-12185/J.\ Greene   & F110W & $21.76\pm0.04$ & $-9.6$  \\
               & 2011 Jun 8.2  & \textit{HST}/WFC3 IR   & GO-12185/J.\ Greene   & F160W & $20.60\pm0.02$ & $-10.7$ \\
SPIRITS\,17qm  & 2004 Dec 16.4 & \textit{Spitzer}/IRAC  & Super Mosaic          & [3.6] & $>15.3$ & $>-16.0$ \\
               & 2004 Dec 16.4 & \textit{Spitzer}/IRAC  & Super Mosaic          & [4.5] & $>14.9$ & $>-16.5$ \\
               & 2001 Mar 8.3  & \textit{HST}/WFPC2 PC  & SNAP-8597/M.\ Regan   & F606W & $22.45$ & $-9.3$ \\
               & 2001 Jun 9.1  & \textit{HST}/WFPC2 WFC & SNAP-8597/M.\ Regan   & F606W & $20.77$ & $-11.0$ \\
\enddata
\tablenotetext{a}{Corrected for Galactic extinction only.}
\end{deluxetable*}

We also examined the available archival \textit{HST} imaging for each source in our sample. To determine the precise locations of our transients in the archival \textit{HST} frames, we registered the archival images with a detection image of the transient, usually a \textit{Spitzer}/IRAC image where the transient is strongly detected. We used higher-resolution images of the active transients where they were available and where registration with \textit{Spitzer} frames was insufficient to determine a precise location of the transient. To perform the registrations, we used centroid measurements of several (at least 10) relatively isolated, bright stars detected in both frames. We then determined the geometric transformations from the archival \textit{HST} frame to the frame containing the transient using the Space Telescope Science Data Analysis System (STSDAS)\footnote{STSDAS is a product of STScI, which is operated by AURA for NASA.} \textsc{geomap} task. By applying the \textsc{geotran} task to the archival frames and blinking these transformed images against the transient detection images, we verified the quality of the registrations. We then examined the precise location of each transient in the available archival frames to search for the presence of a possible progenitor star.

For SPIRITS\,15c and SPIRITS\,16tn, we discussed the limits we obtained on their progenitors from this procedure in \citet{jencson17} and \citet{jencson18c}. We obtained precise registrations of the \textit{HST}/WFPC2 F555W and F814W images taken 1998 November 11.7 with program GO-6483 (PI: D. Elmegreen) using a Baade/IMACS WB6226-7171 image of the SPIRITS\,15c from 2015 January 20.0. We derived $5\sigma$ limiting magnitudes on the progenitor of  $V > 25.1$ and $I > 24.0$, corresponding to limits on the absolute magnitude of the progenitor of $M_V > -7.9$ and $M_I > -8.9$, correcting for Galactic extinction to IC~2163 only. We constrained the flux from the progenitor of SPIRITS\,16tn to $V\gtrsim 24.5$ in an archival WFPC2/WFC F606W frame from 1994 July 4.8 (PID SNAP-5446; PI: G. Illingworth), corresponding to limits on the absolute magnitude of $M_V > -5.2$ (correcting for Galactic extinction only; $L < 2.9\times10^{4}~L_{\odot}$ for a red supergiant [RSG] progenitor of spectral type M0); however the limit is not constraining for an SN~II progenitor if one assumes heavy extinction of $A_V \sim 8$ as was inferred for SPIRITS\,16tn in \citet{jencson18c} based on the optical/near-IR SED of the transient. 

\begin{figure*}
 \gridline{\fig{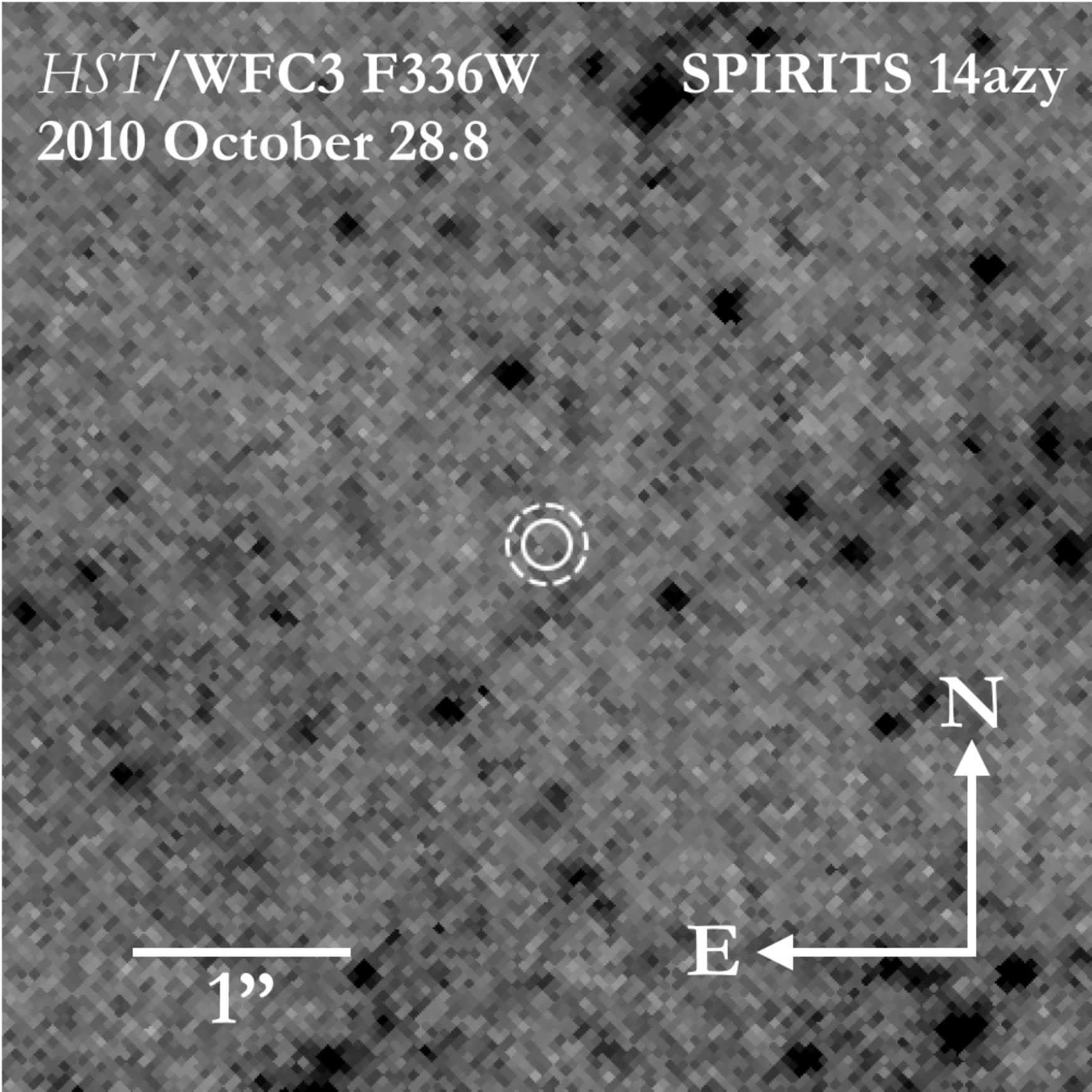}{0.33\textwidth}{}
           \fig{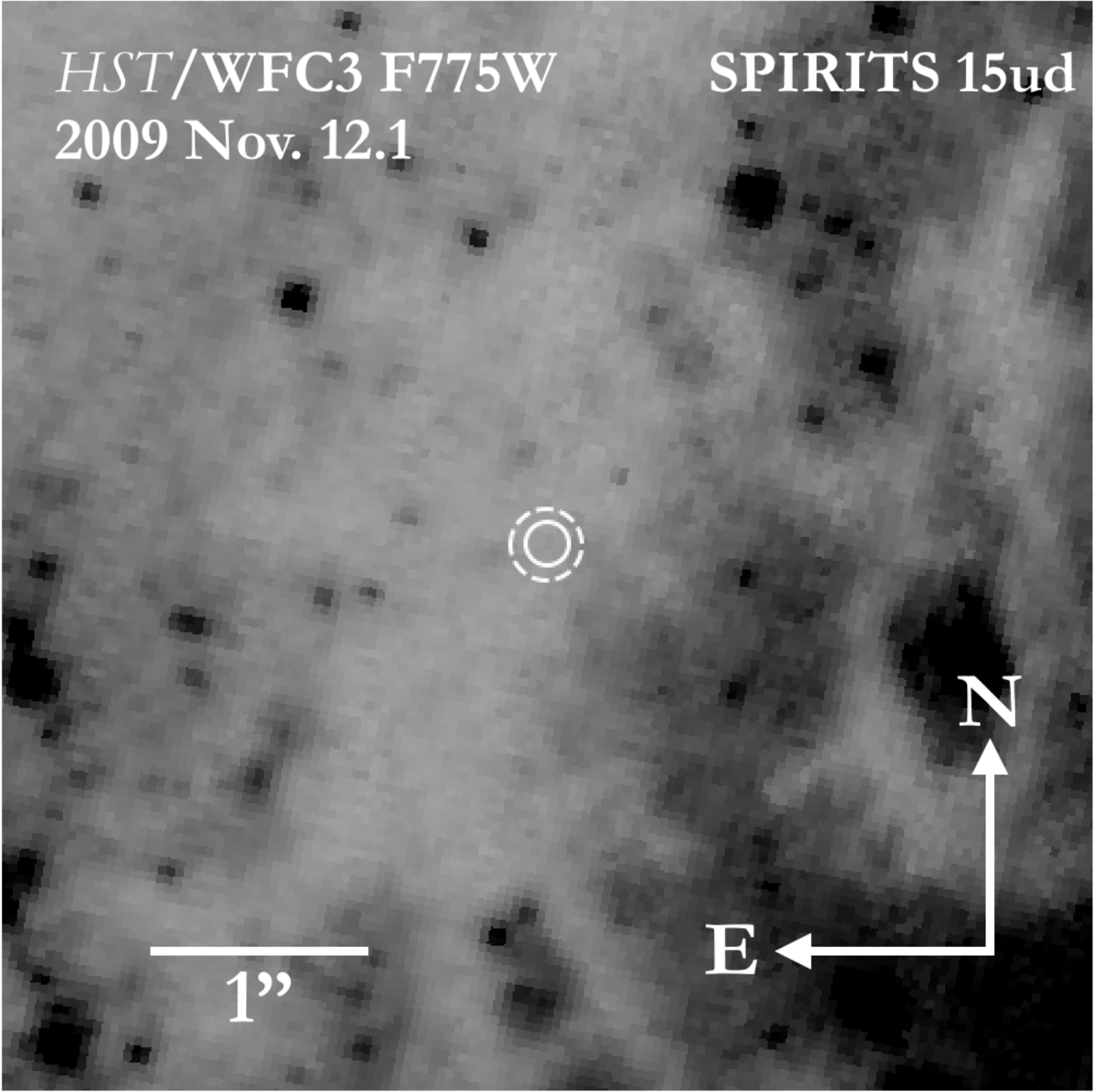}{0.33\textwidth}{}
           \fig{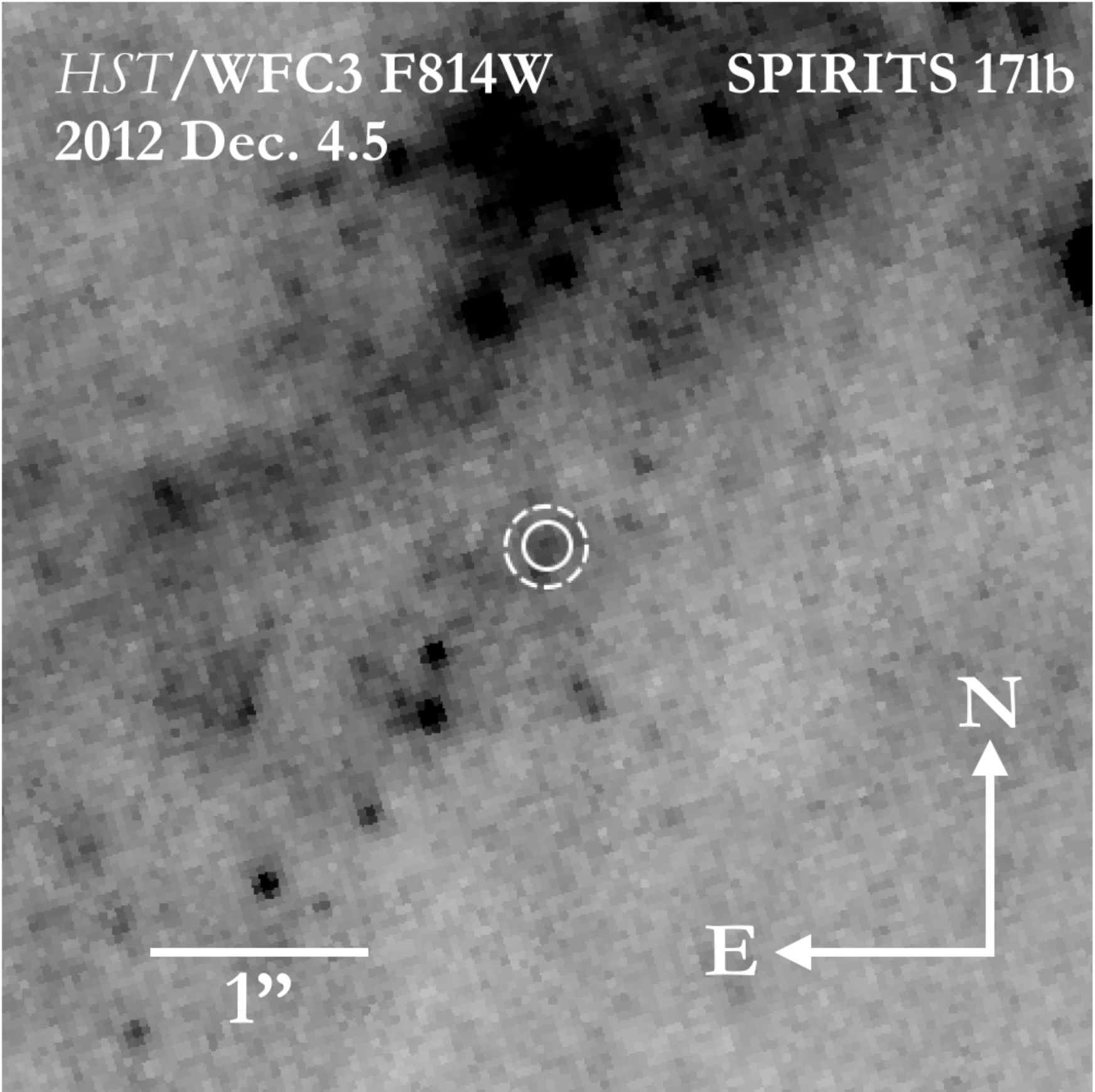}{0.33\textwidth}{}}
 \gridline{\fig{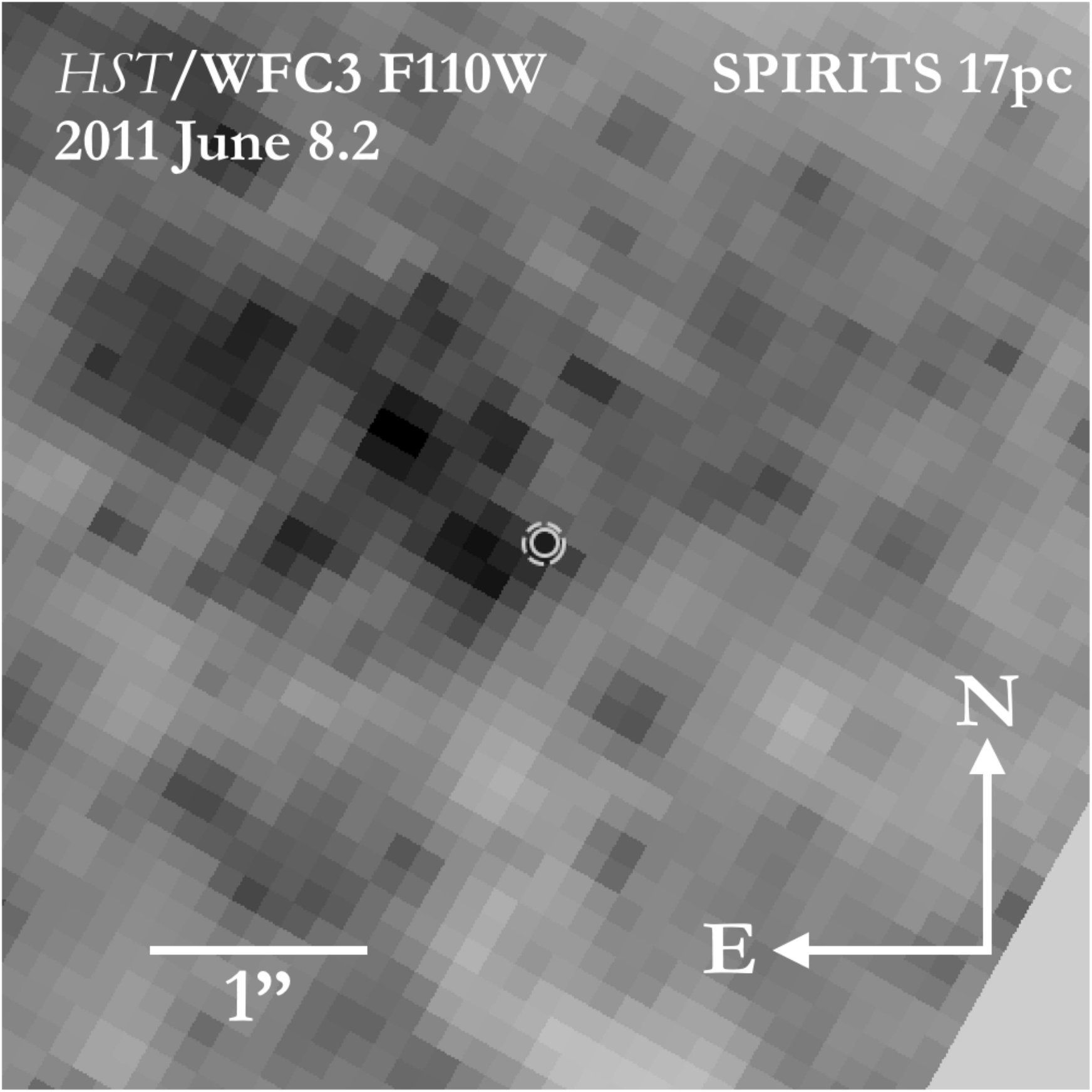}{0.33\textwidth}{}
           \fig{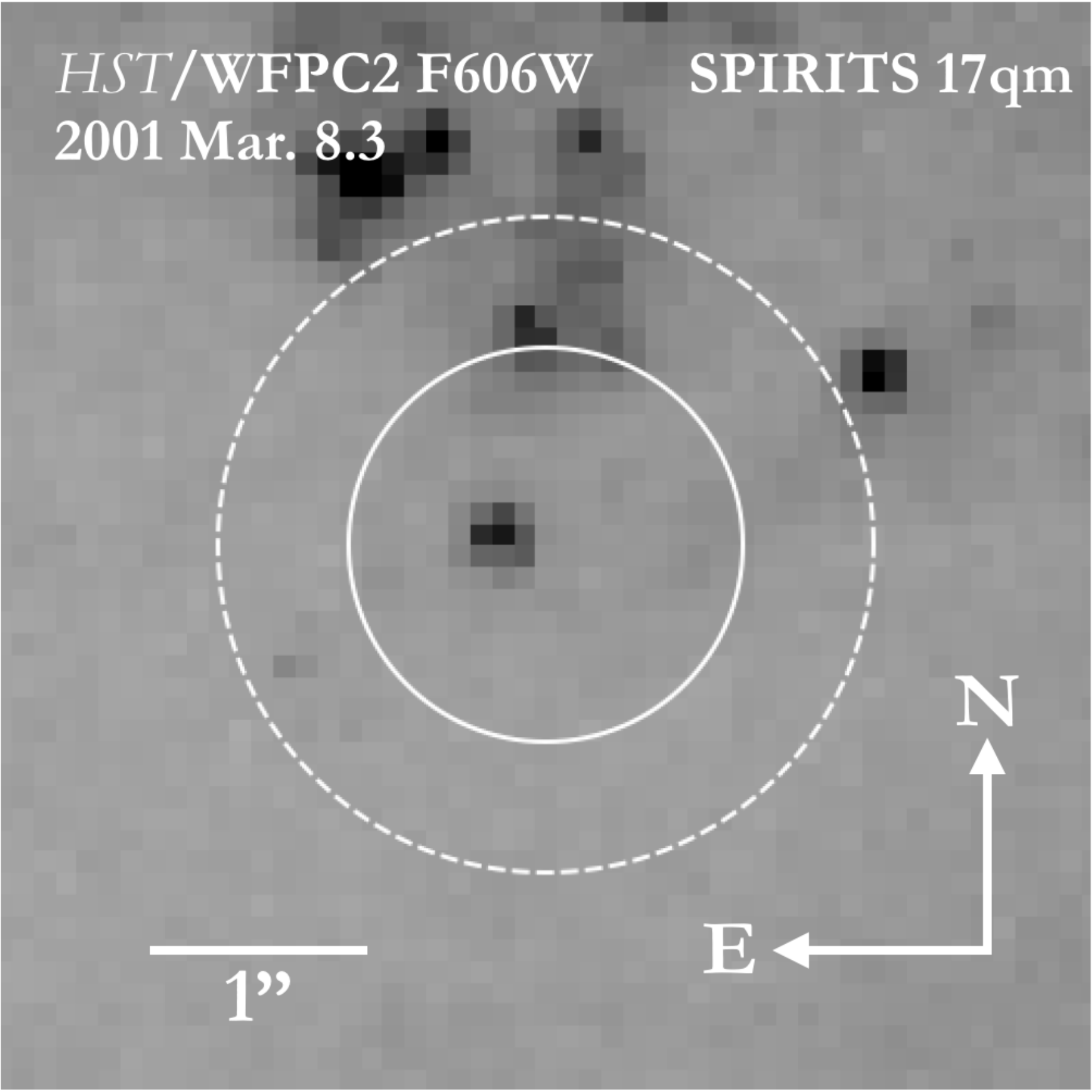}{0.33\textwidth}{}
           \fig{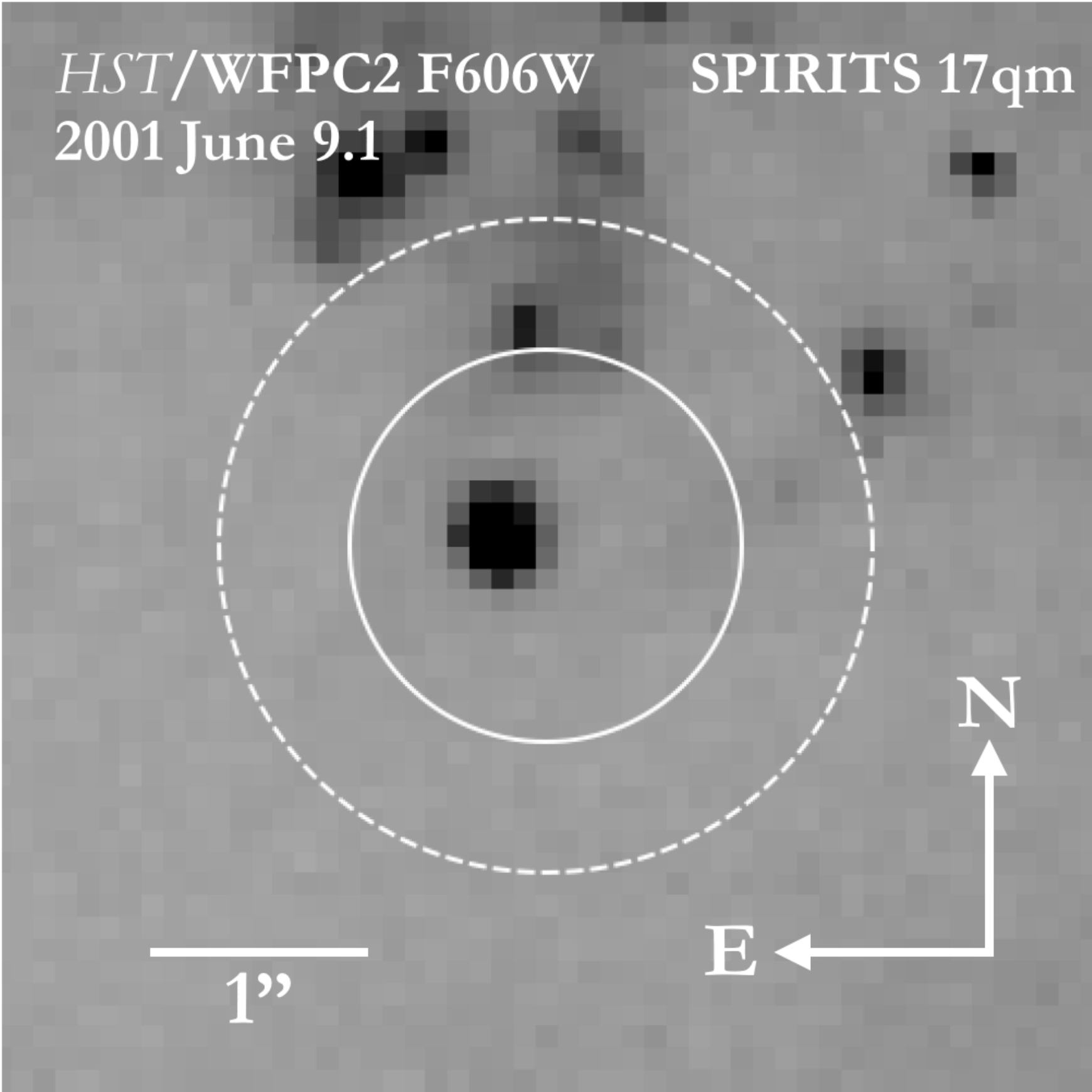}{0.33\textwidth}{}}
\caption{\label{fig:HST_images}
Archival \textit{HST} images of the locations, from left to right, of SPIRITS\,14azy, SPIRITS\,15ud, and SPIRITS\,17lb (top row), and SPIRITS\,17pc and the two epochs covering SPIRITS\,17qm (bottom row). The portion of each image shown is a $5\arcsec \times 5\arcsec$ box, oriented with north up and east to the left. The $3\sigma$ ($5\sigma$) error circles on the locations of the transients in each image are shown as the solid (dashed) white circles. Each image is labeled with the instrument/filter combination used and the date of the observation. 
}
\end{figure*}

The location of SPIRITS\,14azy was imaged with WFC3/\allowbreak UVIS in the F336W filter on 2010 October 28.8 with program SNAP-12229 (PI L. Smith), nearly 4\,yr before the discovery of the transient, which we registered with the \textit{Spitzer}/IRAC [4.5] discovery image of the transient. The rms uncertainty in the registration is 0.9 WFC3 pixels ($0\farcs036$) in both the $x$- and $y$-directions. As shown in Figure~\ref{fig:HST_images}, the location is within an apparent dust lane, completely devoid of stars consistent with the transient position. The limiting magnitude in the image is $U \gtrsim 25.5$, which we adopt as a limit on the progenitor flux, corresponding to a limit on the absolute magnitude of the progenitor star of $M_U \gtrsim -5.4$ (Galactic extinction correction only). 

SPIRITS\,15ud is located near the nucleus of M100 and was covered in several bands at several epochs with \textit{HST}/WFPC2 and WFC3. We selected the WFC3/UVIS F775W image from 2001 November 12.1 (PID GO-11646; PI A.\ Crotts) for registration with the \textit{Spitzer}/IRAC [4.5] discovery image, for which we obtained an rms uncertainty on the position of the transient of 0.84 WFC3 pixels ($0\farcs034$). The transient is located in a prominent, nearly opaque dust lane (see Figure~\ref{fig:HST_images}), and we detect no source consistent with the transient position to a limiting depth of $I \gtrsim 24.5$ ($M_I \gtrsim -6.3$; Galactic extinction only). We also examined this position in the ACS/HRC F814W frame from 2005 May 31.0 (PID GO-9776; PI D.\ Richstone), deriving a limit on the flux from the progenitor of $I \gtrsim 24.4$ ($M_I \gtrsim -6.4$). 

SPIRITS\,17lb is located near the edge of the southern spiral arm of IC\,2163, with possible extinction by the foreground spiral arm of the companion galaxy NGC~2207. This location was covered in WFC3/UVIS F625W and F814W images taken on 2012 December 4.5 (PID SNAP-13029; PI A. Filippenko). We selected the F814W image for registration with the \textit{Spitzer}/IRAC [4.5] discovery image, for which we obtained an rms uncertainty of 0.94 WFC3 pixels ($0\farcs038$). The transient location is coincident with a patch of unresolved, diffuse starlight, characterized by variable extinction in the vicinity. The limits on the progenitor flux from these images are $I \gtrsim 26.2$ and $R \gtrsim 26.8$, corresponding to limits in absolute magnitude of $M_I \gtrsim -8.9$ and $M_R \gtrsim -9.4$. 

We reported on the analysis of the WFC3/UVIS and IR frames covering the site of SPIRITS\,17pc and SPIRITS\,17qm in \citet{jencson18a} and \citet{jencson18b}, respectively. The site of SPIRITS\,17pc was covered in WFC3/UVIS and IR F336W, F438W, F814W, F110W, and F160W images taken on 2011 June 8 (PID GO-12185; PI J.\ Greene). We registered the F110W frame with a high-resolution, $J$-band image taken with the Keck II/NIRC2 adaptive optics image of the active transient from 2017 December 8.6, obtaining a registration rms uncertainty of 0.15 WFC3 pixels ($0\farcs02$). There are several blended sources near the location of SPIRITS\,17pc in the WFC3 frames. We analyzed the sources near the location using PSF-fitting photometry with \textsc{dolphot} \citep{dolphin00,dolphin16}. We identify a source consistent with the precise transient position and rated as a ``good star'' by \textsc{dolphot} in the F814W, F110W, and F160W images. There is coincident emission in the F336W and F438W frames, but it is not point-like and is blended with nearby objects. Our PSF photometry from \textsc{dolphot} on the candidate progenitor gives $\mathrm{F336W} >26.06$\,mag, $\mathrm{F438W} >26.18$\,mag, $\mathrm{F814W} = 24.21 \pm 0.05$\,mag, $\mathrm{F110W} =  21.76 \pm 0.04$\,mag, and $\mathrm{F160W} = 20.60\pm0.02$\,mag. At the distance to NGC~4388, the photometry can be well fit by a single blackbody component of $T = 1900 \pm 100$~K and $L = 2.1^{+0.4}_{-0.3}\times 10^{5}~L_{\odot}$. 

There are two epochs of WFPC2 F606W images taken on 2001 March 8.3 and 2001 June 9.1 (PID SNAP-8597; PI M.\ Regan) covering the site of SPIRITS\,17qm. We registered the first WFPC2 image with the \textit{Spitzer}/IRAC [3.6] discovery image and obtained an rms uncertainty of 3.0 WFPC2 pixels ($0\farcs3$). As shown in Figure~\ref{fig:HST_images}, there is a single, isolated point source consistent with the location of SPIRITS\,17qm in both WFPC2 frames. We obtained magnitudes from the Hubble Legacy Archive source catalogs of $V = 22.45$ (2001 March 8) and $V = 20.77$ (2001 June 9), indicating that the source is highly variable and brightened by ${\approx}1.7$~mag during the 3\,months between the two epochs. The absolute magnitudes then are $V \approx -9.3$ and $-11.0$ for each epoch, respectively, indicating that the star is highly luminous. While the localization of SPIRITS\,17qm in the WFPC2 frames is coarse, the marked variability of the coincident star strongly suggests a physical association. 

There is no archival \textit{HST} imaging available for the sites of SPIRITS\,15ade in NGC~5921 or SPIRITS\,16ix in NGC~4461. 

\subsection{\textnormal{Spitzer} light curves}\label{sec:IR_lcs}
The [4.5] light curves and $[3.6]{-}[4.5]$ color curves of the IR selected obscured SN candidate sample are shown in Figure~\ref{fig:IR_lcs}. In the left panel for each object, we compare to those of the optically discovered control sample. We define zero phase in a uniform way for each object across both samples as the time of peak brightness in the \textit{Spitzer}/IRAC bands. For SN~2017eaw, we adopt 2017 September 13.6 as the observed time of peak brightness; however, the SN was saturated in the images taken on this date, and we do not have a reliable flux measurement of the peak. 

The [4.5] light curves of the IR-selected sample appear broadly similar to those of the control sample, usually showing a luminous initial peak and subsequent fade over a timescale of ${\approx}200$--$600$~days. SPIRITS\,17pc and SPIRITS\,17qm represent notable exceptions, as both sources underwent previous luminous IR outbursts, up to nearly $1500$~days before the observed peak in the case of SPIRITS\,17qm. While both sources have begun to decline in IR brightness since their discovery, we will continue to monitor their IR with \textit{Spitzer} throughout Cycle 14 ending in 2020 January. 

The $[3.6]{-}[4.5]$ color curves are also qualitatively similar between the two samples. Most objects are found with red colors between $0 \lesssim [3.6]{-}[4.5] \lesssim 1$ throughout their evolution, with a few events achieving even redder colors up to ${\approx}3$\,mag. The reddest object we observe is SPIRITS\,15c with $[3.6]{-}[4.5] = 3.0 \pm 0.2$ at the time of the observed IR peak, rivaled by SN~2014bi with $[3.6]{-}[4.5] = 2.9 \pm 0.2$ at $225.5$~days post-peak. 

\begin{figure*}
\begin{minipage}[hpb]{180mm}
\centering
\gridline{\fig{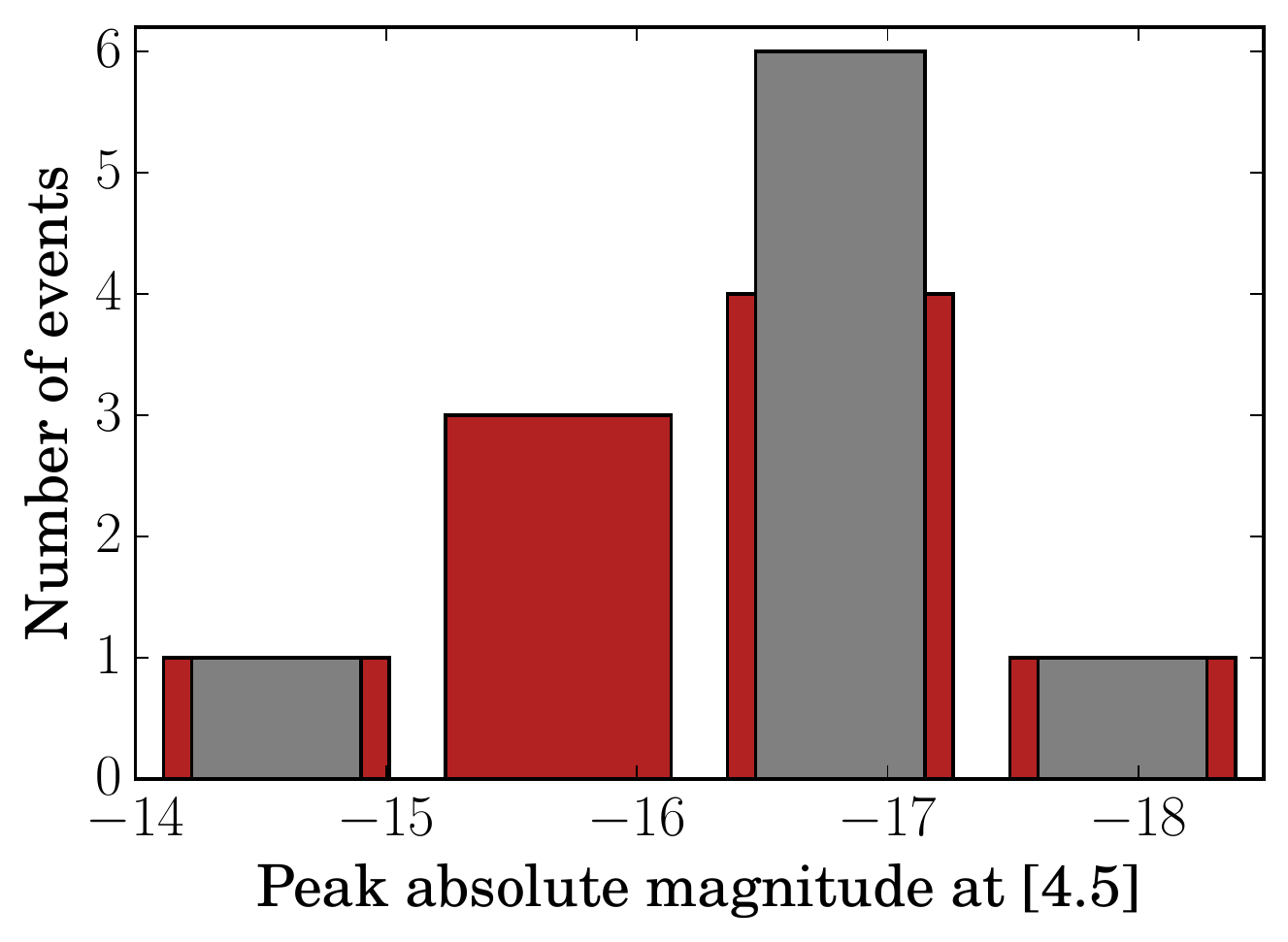}{0.33\textwidth}{} 
	\fig{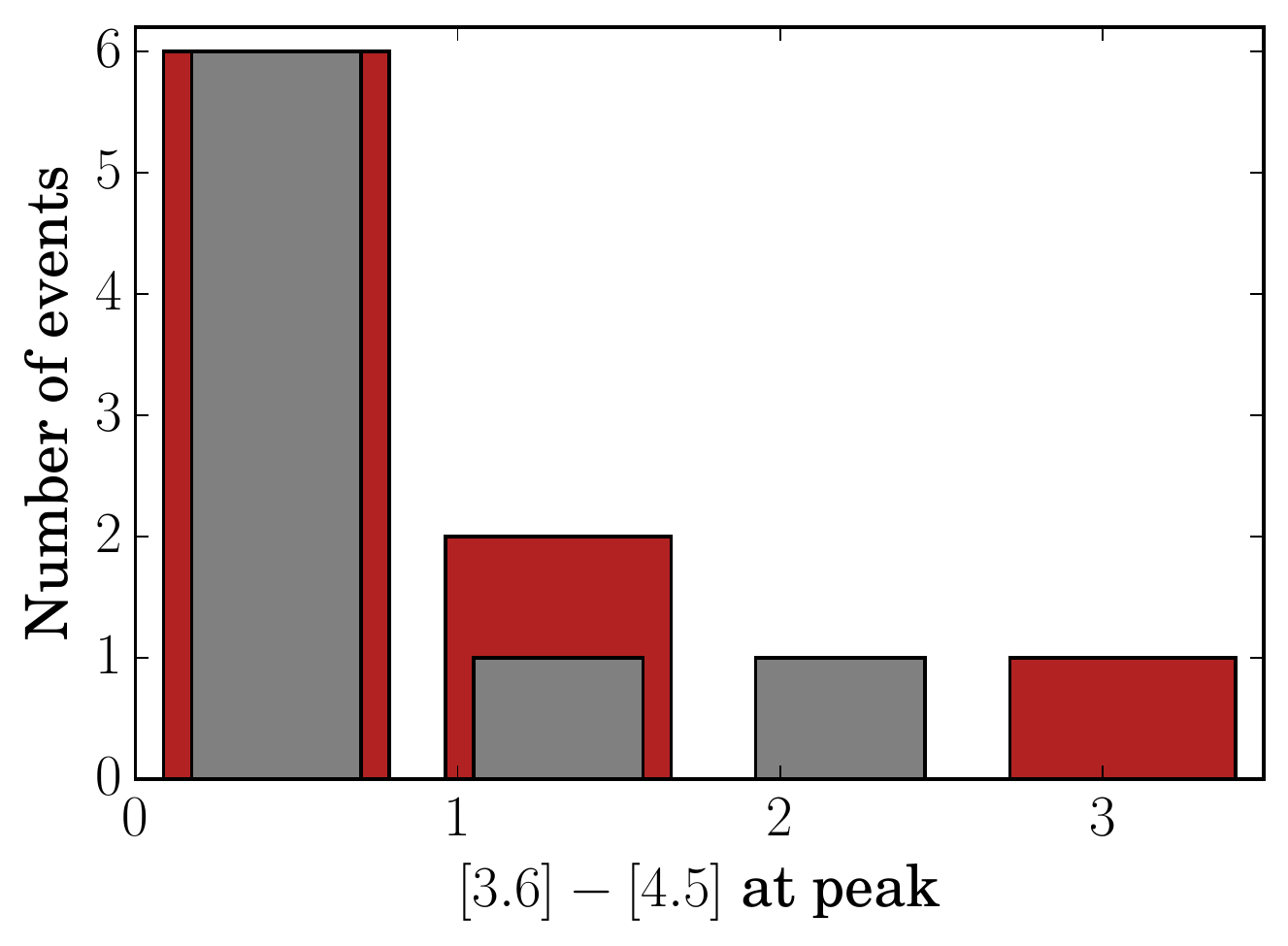}{0.33\textwidth}{}
    \fig{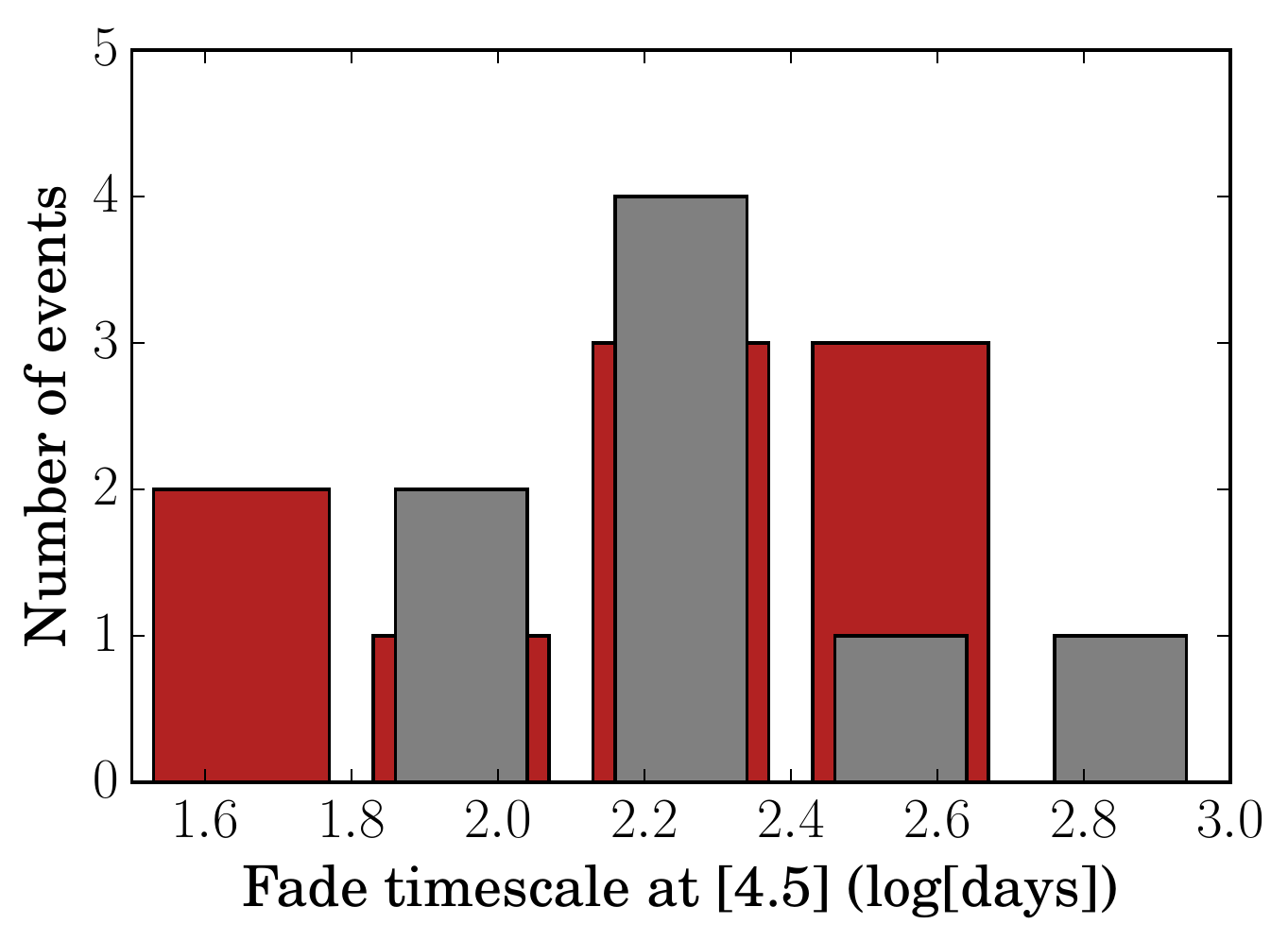}{0.33\textwidth}{}}
\caption{\label{fig:IR_hists}
Observed distributions of peak absolute magnitude at [4.5] (left), $[3.6]{-}[4.5]$ color at time of peak (middle), and fade timescale at [4.5] (right) for the IR-discovered sample (red) and the optically discovered control sample (gray).
}
\end{minipage}
\end{figure*}

We can make a more quantitative comparison between the IR-selected and optically selected control samples based on properties derived from the [3.6] and [4.5] light curves. In Figure~\ref{fig:IR_hists}, we show histograms for both samples of their peak absolute magnitudes at [4.5], $M_{[4.5],\mathrm{peak}}$, their $[3.6]{-}[4.5]$ colors at peak, and the characteristic fade timescales of their [4.5] light curve, $t_{\mathrm{fade},[4.5]}$, defined as the time in days for the light curve to decline by 1\,mag from a linear (in magnitudes) fit to the post-peak light curve. These properties for IR-selected events are also summarized in Table~\ref{table:lc_prop}.

\begin{deluxetable*}{lcccccccc}
\tablecaption{Properties Derived from Light Curves and Suggested Classifications\label{table:lc_prop}}
\tablehead{\colhead{Name} & \colhead{$t_0$} & \colhead{Max Age} & \colhead{$t_{\mathrm{peak}}$} & \colhead{$M_{[4.5],\mathrm{peak}}$} & \colhead{$[3.6]{-}[4.5]$\tablenotemark{a}} & \colhead{$t_{\mathrm{fade,[4.5]}}$} & \colhead{$A_{V}$} & \colhead{Classification} \\
\colhead{} & \colhead{(MJD)} & \colhead{days} & \colhead{(MJD)} & \colhead{(mag)} & \colhead{(mag)} & \colhead{(days)} & \colhead{(mag)} & \colhead{} }
\startdata
SPIRITS\,14azy & 56,772.0 & 27.9  & 57,245.1 & -14.4 & $0.55\pm0.04$ & 270     & 3.2             & LRN\tablenotemark{b}           \\
SPIRITS\,15c   & 56,890.4 & 27.0  & 57,057.4 & -17.1 & $3.0\pm0.2$ &  85     & 2.2             & SN~Ib/IIb      \\
SPIRITS\,15ud  & 57,271.7 & 381.4 & 57,271.7 & -16.4 & $0.5\pm0.1$ & 170     & $\gtrsim 3.7$   & SN~II\tablenotemark{b}         \\
SPIRITS\,15ade & 57,276.5 & 34.0  & 57,337.9 & -15.7 & $0.67\pm0.03$ & 220     & 2.7             & ILRT          \\
SPIRITS\,16ix  & 57,477.9 & 195.2 & 57,477.9 & -15.8 & $1.61\pm0.05$ &  55     & $\gtrsim 5.5$   & SN~II\tablenotemark{b}         \\
SPIRITS\,16tn  & 57,615.0 & 82.0  & 57,615.1 & -16.7 & $0.68\pm0.03$ &  55     & 7.8             & SN~II\tablenotemark{b}         \\
SPIRITS\,17lb  & 57,901.7 & 149.8 & 57,901.7 & -18.2 & $1.09\pm0.03$ & 160     & $\gtrsim 2.5$   & SN~II\tablenotemark{b,c}         \\
SPIRITS\,17pc  & 57,878.8 & 249.9 & 58,250.4 & -16.3 & $0.39\pm0.04$ & \nodata & $12.5$          & MSE          \\
SPIRITS\,17qm  & 58,062.2 & 234.0 & 58,062.2 & -15.9 & $0.21\pm0.04$ &  480    & $12.1$          & MSE/LBV \\
\enddata
\tablenotetext{a}{Measured at time $t_{\mathrm{peak}}$.}
\tablenotetext{b}{Suggested classification for these sources is not spectroscopically confirmed.}
\tablenotetext{c}{SPIRITS\,17lb was confirmed as a CCSN via the detection of its radio counterpart.}
\end{deluxetable*}

Both samples span a range in peak [4.5] luminosity, $M_{[4.5],\mathrm{peak}}$, between $-14$ and $-18.5$. The distributions in $M_{[4.5],\mathrm{peak}}$ appear similar, with both peaking between ${\approx}-16.3$ and $-17.4$. We note that there is a larger fraction of low-luminosity events in the IR-selected sample, i.e., three of nine objects (${\approx}33$\%) have $M_{[4.5],\mathrm{peak}}$ fainter than $-16$, while there is only one such object of eight (${\approx}13$\%) in the control sample. In $[3.6]{-}[4.5]$ color at peak, the two samples again show similar distributions that peak in the range $0.0 \lesssim [3.6]{-}[4.5] \lesssim 0.9$ with one-sided tails extending to redder colors. Finally, in the distributions of $t_{\mathrm{fade},[4.5]}$, we once again note the broad similarity between the two samples. The distributions extend between ${\approx}50$ and $500$~days, with peaks between ${\approx}130$ and $250$~days. Notably, though, the two most rapidly fading events in either sample are SPIRITS\,16ix and SPIRITS\,16tn, both from the IR-discovered sample, with $t_{\mathrm{fade},[4.5]} = 55$~days for both, while the next-fastest events are SPIRITS\,15c and SN~2016bau (Type~Ib) with $t_{\mathrm{fade},[4.5]} \approx 80$~days.

We performed Kolmogorov--Smirnov tests between the two samples in each of the light-curve-derived parameters discussed above. For the distributions in $M_{[4.5],\mathrm{peak}}$, $[3.6]{-}[4.5]$ color at peak, and $t_{\mathrm{fade},[4.5]}$, the tests return $p$-values of 0.31, 0.91, and 0.97, respectively. Thus, we are unable to reject the null hypotheses that the IR-selected and optically discovered samples are drawn from the same parent distribution in all three parameters. The overall similarity of the two samples in their IR properties supports the suggestion that the IR-selected objects presented in this work may represent a population of SNe that were systematically missed by optical transient searches. 


\subsubsection{Comparison to \textnormal{Spitzer} SNe}
In the right panel for each object in Figure~\ref{fig:IR_lcs}, we compare the SPIRITS transients to the entire sample of SNe so far detected by \textit{Spitzer}/IRAC. A large compilation including every available \textit{Spitzer}/IRAC detection of known SNe through 2014 was recently presented by \citet{szalai19}, including the previously published compilations of \citet{szalai13}, \citet{tinyanont16}, and \citet{johansson17}; detections of individual SNe originally reported by several authors; and new, previously unpublished detections. To this, we add new detections since 2014 of the optically discovered SNe in our control sample. 

We divide this large comparison sample by subtype into hydrogen-rich SNe~II (and interacting SNe~IIn), stripped-envelope SNe~IIb and SNe~Ib/c (and interacting SNe~Ibn), and thermonuclear SNe~Ia (and interacting SNe~Ia-CSM) to demonstrate the diagnostic utility of IR light curves. 

In the IR, as found by \citet{johansson17}, SNe~Ia are clearly separated from CCSNe by their rapidly declining [4.5] light curves and evolution to blue $[3.6]{-}[4.5]$ colors for the first 200~days. The IR emission of SNe~Ia is powered by the tail of the hot thermal component of the SN peaking in the optical, and \citet{johansson17} placed stringent limits on the presence of dust within $\lesssim 10^{17}$~cm of $M_{\mathrm{dust}} \lesssim 10^{-5}~M_{\odot}$ for the SNe~Ia SN~2014J, SN~2006X, and SN~2007le. Characterized by strong interaction with a dense circumstellar medium (CSM), SNe~Ia-CSM may display redder colors from $0.0 \lesssim [3.6]{-}[4.5] \lesssim 1.0$, and show a clear IR excess over a normal SN~Ia in their [4.5] light curves. Similarly, the unusual, dusty Type Iax SN~2014dt showed redder colors and developed a clear dust excess at [4.5] peaking at ${\approx}500$~days. With the exception of the ongoing outburst SPIRITS\,17pc that shows an initial rise, all of the SPIRITS transients are characterized by declining [4.5] light curves and red IR colors throughout their evolution, inconsistent with the characteristic evolution of SNe~Ia and largely dissimilar to the thermonuclear SN subtypes of SNe~Ia-CSM and SNe~Iax that have been now been characterized by \textit{Spitzer}. 

CCSNe, on the other hand, are characterized by more slowly declining IR light curves and redder IR colors with $0 \lesssim [3.6]{-}[4.5] \lesssim 3$. The IR emission of CCSNe may be powered by thermal emission from the SN itself, or warm circumstellar dust that may be newly formed in the ejecta or preexisting dust heated by the light from the explosion. Additionally, if CO has formed in the ejecta as in the case of \citet{banerjee18}, the emission at [4.5] may be additionally powered by the fundamental CO vibrational transition in this band, producing a significant [4.5] excess over the other IR bands. This likely contributes to the extreme observed $[3.6]{-}[4.5]$ colors in several cases. Among the sample of CCSNe, stripped-envelope events (including the interacting subtype Ibn) are relatively more homogeneous in their IR properties, characterized by monotonically declining [4.5] light curves with $M_{[4.5]}$ peaking between $-17.5$ and $-19$, and fading at a typical rate of ${\approx}0.01$~mag~day$^{-1}$. SNe~II show a larger spread in both peak luminosity, spanning $M_{[4.5]}$ between $-16$ and $-19$, and decay rates, with some objects having nearly constant [4.5] flux for the first ${\approx}500$~days. Strongly interacting SNe~IIn may exhibit extreme [4.5] luminosities of $M_{[4.5]}$ brighter than $-22$, though some SNe~IIn have IR light curves similar to more typical SNe~II. 

Now, for each IR-selected SPIRITS transient, we discuss inferences on possible classifications by comparing their IR light curves directly to the \textit{Spitzer}-observed SNe. Where appropriate, we also provide context for our final, suggested classifications based on all available observational data provided in the rightmost column of Table~\ref{table:lc_prop} and discussed in more detail in Section~\ref{sec:class}.

SPIRITS\,14azy, peaking at only $M_{[4.5]} = -14.4 \pm 0.2$, is fainter than any previously observed CCSN. While consistent with the luminosity of an SN Ia at a phase of ${\approx}120$~days, the [4.5] fade rate is much slower, and the source is overluminous compared to an SN~Ia by 400~days post-discovery. Furthermore, its color evolution is inconsistent with an SN~Ia. Based only on the IR light curves and considering only SN subtypes, SPIRITS\,14azy would be most consistent with a faint SN~II; however, as discussed below in Sections~\ref{sec:mb_lcs} and \ref{sec:class}, optical light curves bear strong similarity to the LRN M101 OT2015-1, arguing against SPIRITS\,14azy as a true CCSN. 

SPIRITS\,15c, spectroscopically confirmed as an SN~Ib/IIb in \citet{jencson17}, is fully consistent with the sample of stripped-envelope SNe in its [4.5] light curve. At $[3.6]{-}[4.5] = 3.0$ at peak light, it is the reddest stripped-envelope SN yet observed by \textit{Spitzer}. The extreme IR color is likely attributable to CO emission at [4.5], corroborated by the detection of emission from the CO $\Delta v = 2$ vibrational overtone transitions in the $K$ band also reported in \citet{jencson17}. 

The [4.5] light curve of SPIRITS\,15ud appears most similar to the sample of SNe~II. With an observed peak at $M_{[4.5]} = -16.4$, however, it may also be consistent with the class of SN~2008S-like transients (also including NGC~300 OT2008-1; \citealp{adams16a}). Given the large uncertainty in the phase of SPIRITS\,15ud of $\gtrsim 400$~days, however, it is likely that the transient was significantly more luminous than SN~2008S, but the IR peak was missed by our observations.

SPIRITS\,15ade again appears consistent with either a low-luminosity SN~II or an SN~2008S-like event in both its [4.5] light curve and $[3.6]{-}[4.5]$ color evolution. We examine the classification of this object as an SN~2008S-like event also considering its near-IR spectrum and limits on the presence of an IR progenitor star in Section~\ref{sec:class}. 

SPIRITS\,16ix and SPIRITS\,16tn, with largely similar IR properties, are unique among the SPIRITS IR transients presented here, and among all SNe previously observed by \textit{Spitzer}. Their light curves at [4.5] decline more rapidly than any CCSN yet observed and appear consistent with the decline of an SN~Ia. However, their red colors at $[3.6]{-}[4.5] \gtrsim 0.7$ rule out an SN~Ia scenario. In \citet{jencson18c}, we argued that the properties of SPIRITS\,16tn were most consistent with a weak SN~II, where the early bright IR emission was powered by a luminous dust echo, and the redder $[3.6]{-}[4.5]$~color at later times was likely attributable to a [4.5] excess from CO emission. This interpretation may also apply to SPIRITS\,16ix given the similarity of these objects in the \textit{Spitzer} bands. 

SPIRITS\,17lb is our most luminous transient at $M_{[4.5]} = -18.2 \pm 0.4$, and its light curve and color evolution are consistent with either a luminous SN~II or stripped-envelope SN~IIb or Ib/c. While we are unable to distinguish between CCSN subtypes based on the \textit{Spitzer} data alone, we discuss the likely classification of SPIRITS\,17lb as an SN~II based on the rest of our follow-up data at optical, near-IR and radio wavelengths in Section~\ref{sec:class}. 

SPIRITS\,17pc and SPIRITS\,17qm are remarkable among the SPIRITS sample owing to the presence of multiple IR outbursts in the \textit{Spitzer} light curves over the last ${\approx}1000$--$1500$~days. As reported in \citet{jencson18a}, SPIRITS\,17pc shows three distinct IR peaks at [3.6] and [4.5], growing progressively more luminous and longer in duration. During the current, ongoing outburst, SPIRITS\,17pc brightened to $M_{[4.5]} = -16.3 \pm 0.4$ over a period of at least $400$~days, with a fairly constant color between $[3.6]{-}[4.5] = 0.2$ and $0.4$. The increasing IR emission may indicate ongoing interaction of an SN blast wave with the surrounding CSM, or alternatively, active dust formation during a less extreme, nonterminal outburst.

SPIRITS\,17qm, as reported in \citet{jencson18b}, underwent a previous IR outburst ${\approx}1300$~days at $M_{[4.5]} = -15.2 \pm 0.1$ before the discovery by SPIRITS. The observed peak at discovery of $M_{[4.5]} = -15.9 \pm 0.1$ and subsequent slow decline are consistent with an SN~II, or even some previously observed interacting SN~IIn. Given its eruptive history, however, the discovery outburst of SPIRITS\,17qm may be due to a more intense nonterminal outburst, rather than a true SN explosion.

\subsection{Optical/Near-IR Spectroscopic Properties}\label{sec:spec}
\begin{figure*}
\plotone{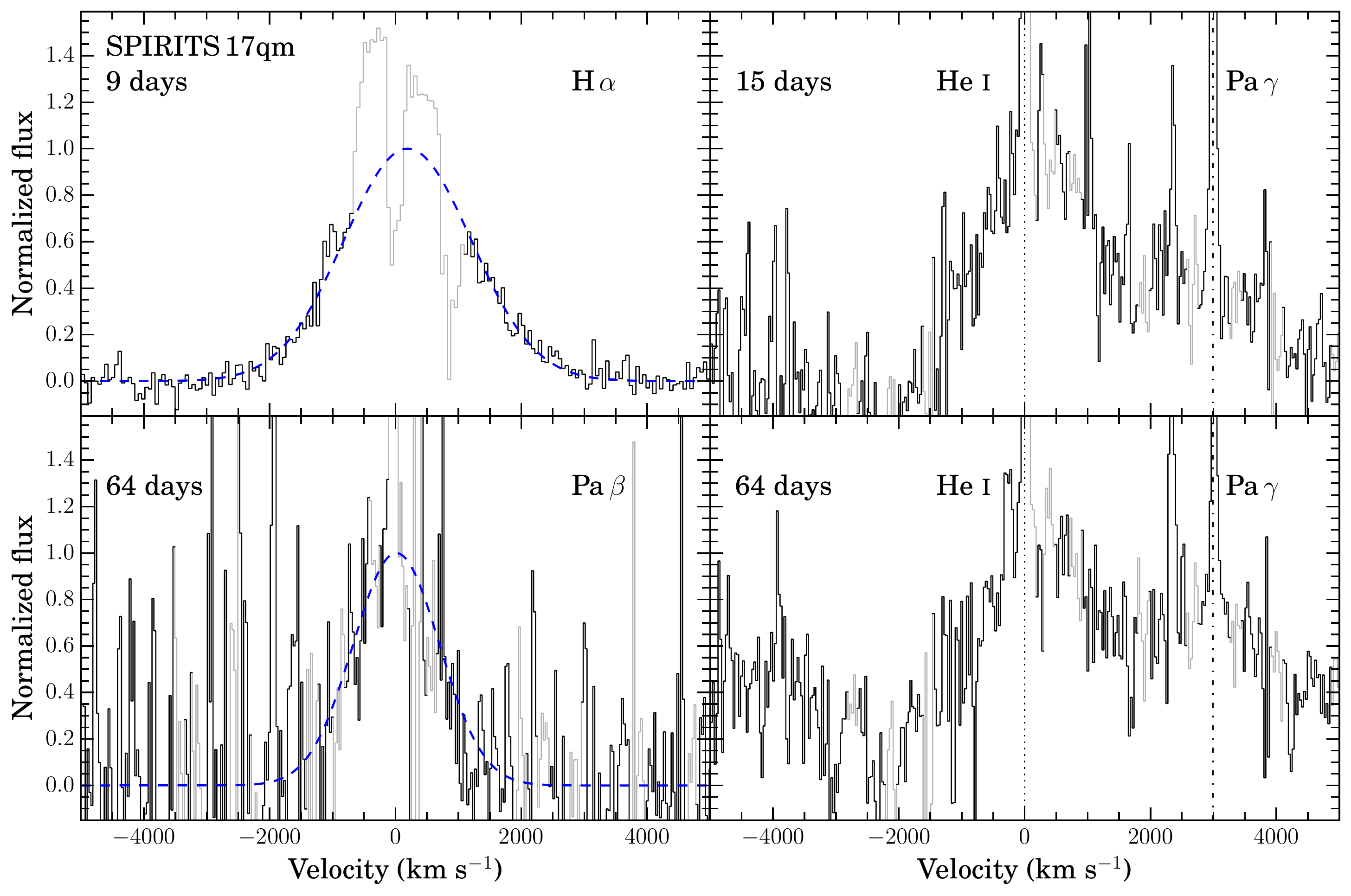}
\caption{\label{fig:H_He_17qm}
In the left panels we show the H$\alpha$ (top; 9 days) and Pa$\beta$ (bottom; 64 days) velocity profiles of SPIRITS\,17qm in black. Gaussian fits to the data with FWHM velocities of 2400 and 1600\,km~s$^{-1}$, respectively, are plotted as blue dashed curves, where spectral bins contaminated by emission features of the underlying star-forming region or bright OH airglow emission lines are shown in light gray and excluded from the fits. In the right 2 panels, we show the velocity profiles of He~\textsc{i} ($\lambda 10830$; dotted vertical lines) at 15 days (top) and 64 days (bottom), with FWHM velocities of ${\approx}2000$\,km~s$^{-1}$. The red wing of the line is blended with emission from Pa$\gamma$ (dashed-dotted vertical lines)}. 
\end{figure*}

Our optical spectroscopy of SPIRITS\,16tn (79.5~days), SPIRITS\,17pc (192.8~days), and SPIRITS\,17qm (9.3~days) is shown in Figure~\ref{fig:opt_spec}. The spectrum of SPIRITS\,16tn, previously presented in \citet{jencson18c}, shows only a featureless red continuum beyond ${\approx}8000$~\AA. The spectrum of SPIRITS\,17qm is dominated by strong H$\alpha$ emission with an FWHM velocity of 2400~km~s$^{-1}$ (top left panel of Figure~\ref{fig:H_He_17qm}), consistent with ejecta velocities of giant LBV eruptions. We do not detect any broader components that would indicate higher, explosive velocities of an interacting SN~IIn. Additionally, in SPIRITS\,17qm we detect weaker emission features of O~\textsc{i} ($\lambda \lambda 8446, 9266$), but with no clear detection of the O~\textsc{i} $\lambda7771$ line, as well as detections of the Ca~\textsc{ii} IR triplet ($\lambda\lambda8498, 8542, 8662$). In contrast, for SPIRITS\,17pc, we detect narrow, unresolved emission features of the underlying star-forming region (including H$\alpha$), along with the Ca~\textsc{ii} triplet in emission, but no significant O~\textsc{i} emission or broader components of H$\alpha$ associated with the transient. We confirmed the veracity of the weaker features reported here by close inspection of the reduced 2D spectra. 

\begin{figure}
\plotone{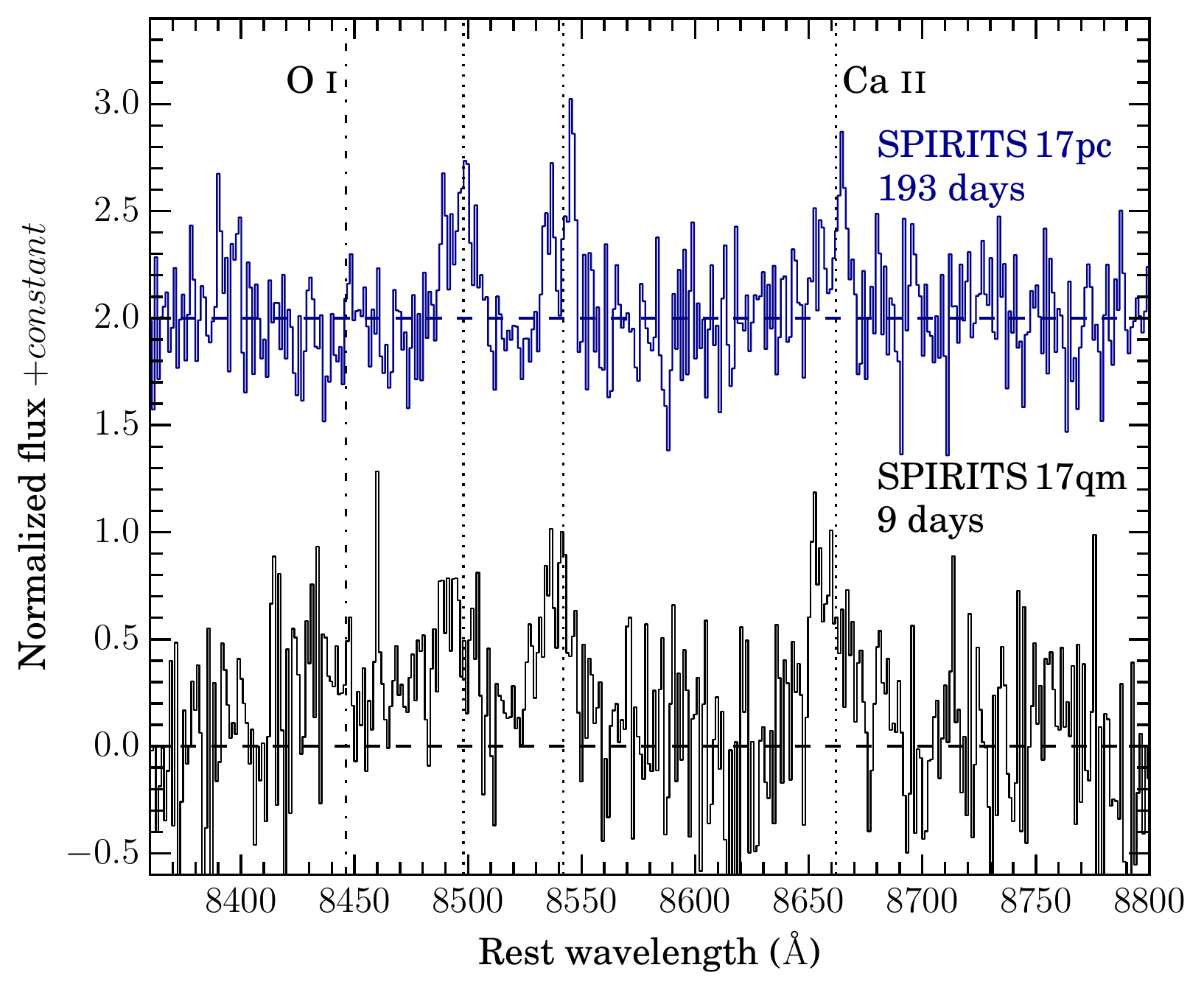}
\caption{\label{fig:CaII}
Comparison of the spectral region around O~\textsc{i} ($\lambda 8446$; dashed-dotted vertical line) and the Ca~\textsc{ii} IR triplet (dotted vertical lines) in the optical spectra of SPIRITS\,17pc (193 days) and SPIRITS\,17qm (9 days). A linear approximation to the continuum emission has been subtracted from the spectra. The spectra were then normalized in flux, and the SPIRITS\,17pc spectrum has been shifted vertically for clarity. The dashed blue and black horizontal lines show the zero-levels for SPIRITS\,17pc and SPIRITS\,17qm, respectively. 
}
\end{figure}

A zoom-in to the region around O~\textsc{i} ($\lambda 8446$) and the Ca~\textsc{ii} triplet in the spectra of SPIRITS\,17pc and SPIRITS\,17qm is shown in Figure~\ref{fig:CaII}. The Ca~\textsc{ii} lines in SPIRITS\,17pc are each double peaked, with a blueshifted component at ${\approx}-240$\,km~s$^{-1}$ and a redshifted component at ${\approx}100$\,km~s$^{-1}$. For SPIRITS\,17qm, the lines show a single component centered at a velocity consistent with the host and with an FWHM velocity of ${\approx}230$\,km~s$^{-1}$. We also note that the O~\textsc{i} line at 8446~\AA\ present in SPIRITS\,17qm, is absent or much weaker in SPIRITS\,17pc. 

\begin{figure}
\plotone{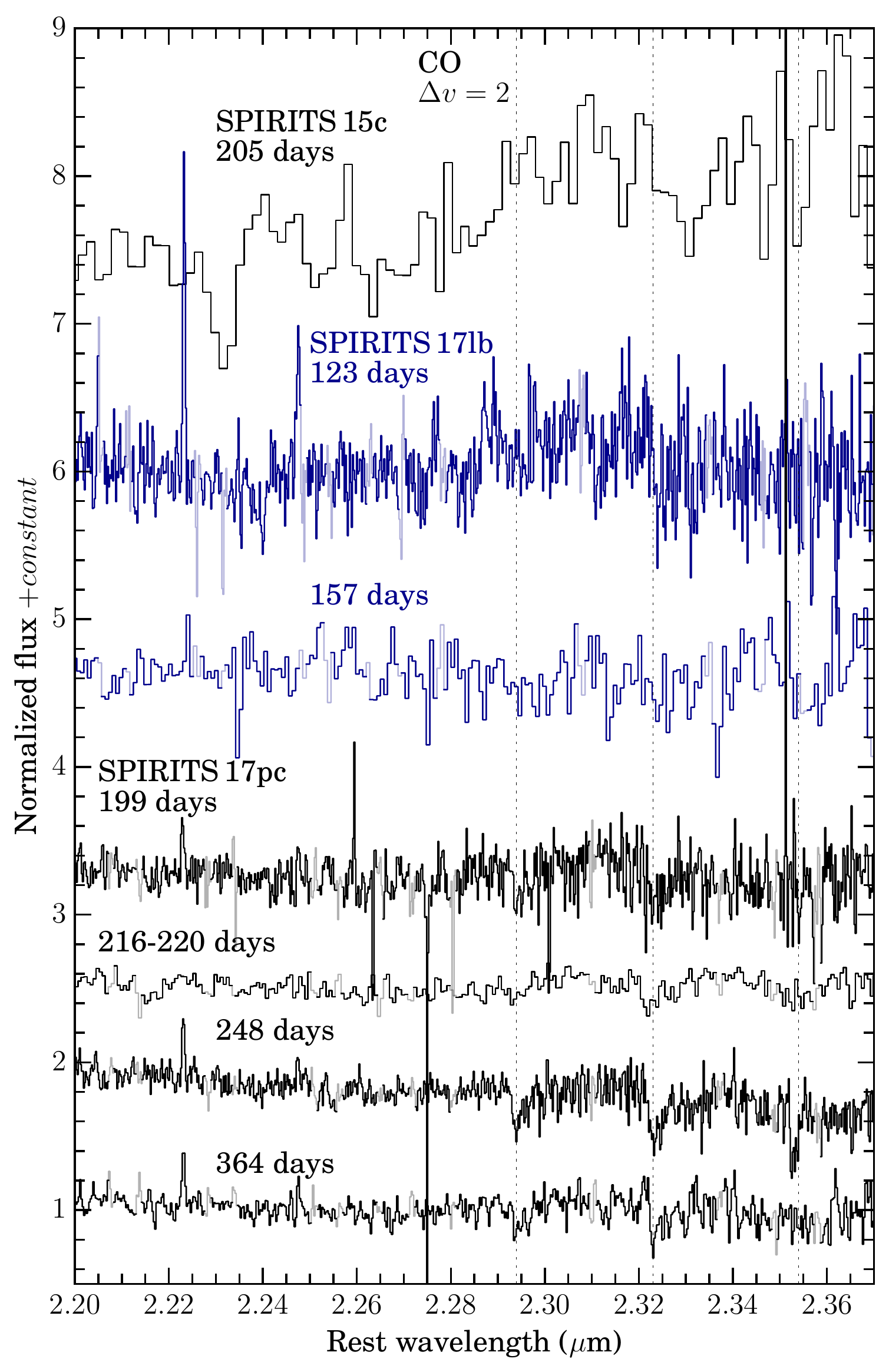}
\caption{\label{fig:CO}
Comparison of the region of the $K$ band between 2.2 and 2.37~\micron\ in the near-IR spectra of SPIRITS\,15c, SPIRITS\,17lb, and SPIRITS\,17pc. The spectra, shown in alternating colors for each object, are labeled on the figure along with their respective phases. Spectral bins of low S/N due to coincidence with bright OH airglow emission lines are plotted in lighter colors. The band heads of the $\Delta v=2$ vibrational transitions of CO are indicated by the dotted vertical lines.
}
\end{figure}

The full set of our near-IR spectroscopy is shown in Figure~\ref{fig:nir_spec}, including spectra of six of the nine IR-discovered transients. In the near-IR, these objects are spectroscopically diverse, showing a range of properties and features. Our clearest example of a spectroscopically confirmed CCSN is SPIRITS\,15c. In \citet{jencson17}, we compared the near-IR spectra of SPIRITS\,15c to those of the well-studied Type IIb SN~2011dh. We found strong similarities and identified several features in SPIRITS\,15c based on the comparison, including He~\textsc{i} (10830~\AA), emission features of neutral or singly ionized intermediate-mass elements and Fe, and emission from the $\Delta v=2$ vibrational overtone transitions of CO (see Figure~\ref{fig:CO} in comparison to SPIRITS\,17lb and SPIRITS\,17pc). We do not detect any hydrogen in the spectrum but cannot rule out the presence of hydrogen at earlier times. Thus, we find the spectrum to be consistent with a Type Ib or IIb classification. 

Our near-IR spectroscopy of SPIRITS\,16tn was first presented in \citet{jencson18c}, where the spectrum was characterized as a largely featureless red continuum. The spectrum, however, displays broad bumps in each of the $H$ and $K$ bands that we originally believed may have been due to difficulty in properly flux-calibrating low-S/N data, especially in the regions of strong telluric H$_2$O absorption where little to no flux from the transient is received through the atmosphere. Here, after carefully checking our calibrations, we suggest a real, astrophysical origin for these broad features as absorption by water vapor at higher temperatures than the narrower telluric features. Such features have been observed in the atmospheres of cool giants with spectral types no earlier than M6 \citep{rayner09}, where models of pulsating Mira variables show the formation of water in the dense, cool ($< 1000$~K) layers formed beyond periodic, outward-propagating shocks in their extended atmospheres \citep{bessell89,bessell96}. To our knowledge, this would be the first detection of water vapor absorption associated with a luminous transient or SN, and we discuss possible interpretations in Section~\ref{sec:class}. As we detected no unambiguous SN features, however, we cannot definitively classify SPIRITS\,16tn.

For SPIRITS\,17lb, we detect a red continuum in the $H$ and $K$ regions, plausibly attributable to emission from warm dust. We note only possible excess emission beyond 23000~\AA\ from CO in the 123-day spectrum, but which appears to fade by 157 days. As with SPIRITS\,16tn, there are no clear, broad features indicative of an SN. We find, however, that a lack of such features does not rule out an SN altogether for these objects. As discussed below in Section~\ref{sec:analysis_radio}, nonthermal synchrotron emission from the interaction of high-velocity ejecta with CSM was detected in SPIRITS\,17lb, confirming the core-collapse nature of this event. The strength of the radio emission suggests an SN~II classification, possibly indicating that SNe~II may be characterized by only weak or absent near-IR features at late phases. It is relevant to note here that we have not attempted to remove contamination from the host galaxy background from our spectra, and that the absence of strong, clear features may also be attributable to the inclusion of galaxy light as sources become faint at late times. 

\begin{figure}
\plotone{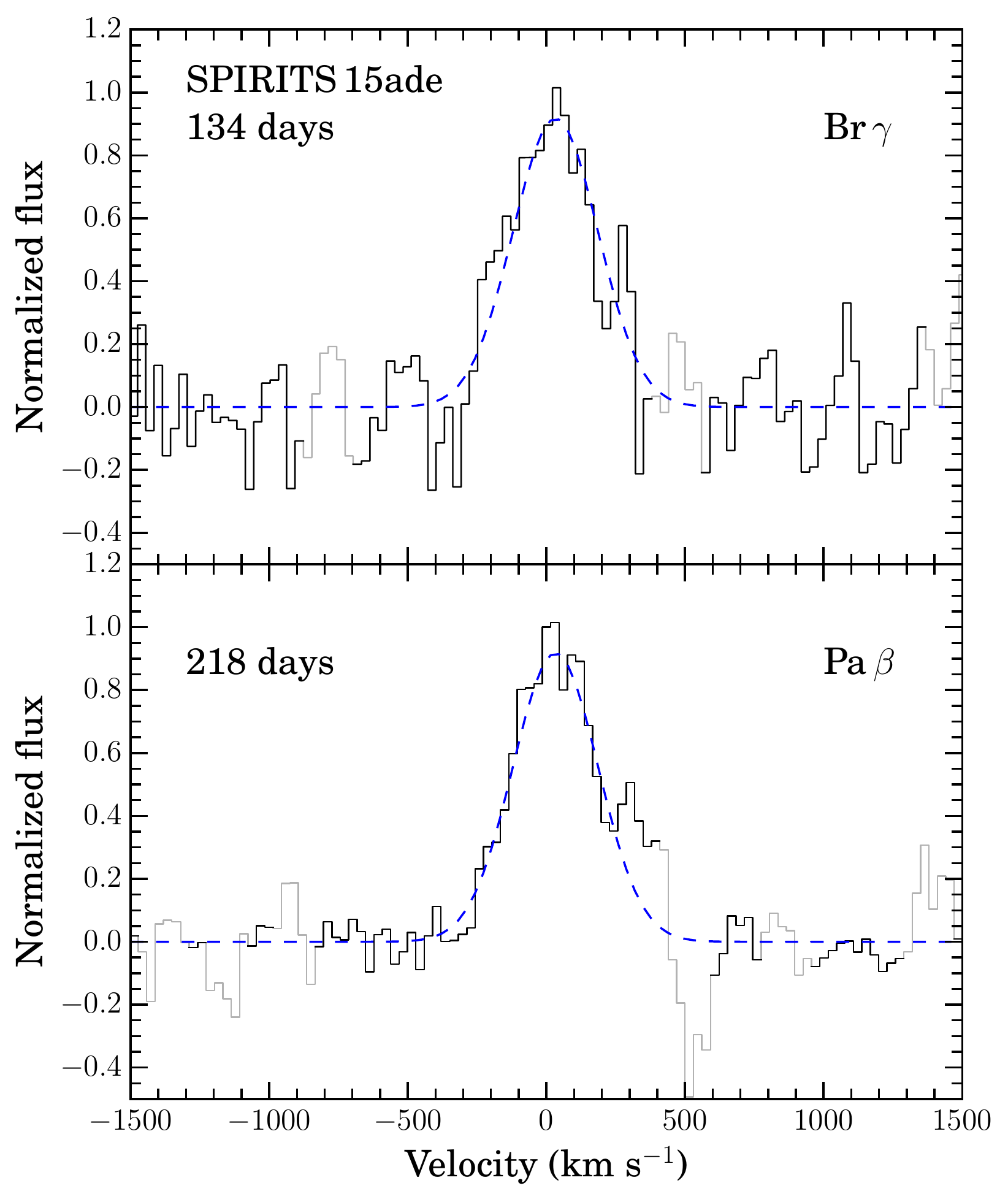}
\caption{\label{fig:H_15ade}
Velocity profiles of Br$\gamma$ (134 days) and Pa$\beta$ (218 days) in the spectra of SPIRITS\,15ade are shown as the black solid lines in the top and bottom panels, respectively, in the rest frame of the host galaxy, NGC~5921. Gaussian fits to the profiles, excluding velocity bins of low S/N due to coincidence with bright OH airglow emission lines (plotted in light gray), are shown as the blue dashed curves.  
}
\end{figure}

The spectra of SPIRITS\,15ade show only the H~\textsc{i} recombination lines Br$\gamma$ in the $K$ band and Pa$\beta$ in the $J$ band. As shown in Figure~\ref{fig:H_15ade}, the peak velocities are consistent with the recession velocity of the host, and the velocity profiles can be approximated by simple Gaussians with FWHM velocities of ${\approx}360$ and $390$\,km~s$^{-1}$, respectively. These are similar to the low expansion velocities seen in the H~\textsc{i} lines in optical spectra of the prototypical members of the class of ILRTs SN~2008S \citep[e.g.,][]{botticella09} and NGC~300 OT2008-1 \citep{bond09,humphreys11}. The similarity of the [4.5] light curve of SPIRITS\,15ade to that of SN~2008S (Section~\ref{sec:IR_lcs}) strengthens its association with this class. There is also an apparent secondary peak in the Br$\gamma$ and Pa$\beta$ velocity profiles in SPIRITS\,15ade at ${\approx}300$\,km~s$^{-1}$, possibly indicative of a bipolar or toroidal outflow geometry. Double-peaked profiles were also seen in the H~\textsc{i} and Ca~\textsc{ii} emission features in NGC~300 OT2008-1 \citep{bond09}, but at lower velocities of ${\approx}70$--$80$\,km~s$^{-1}$. 

The near-IR spectra of SPIRITS\,17pc, taken between 199 and 364~days, show a relatively smooth continuum, with a few notable features. We identify an absorption band of molecular CN at 1.1~\micron\ along with the CO $\Delta v=2$ band heads beyond 2.3~\micron\ (shown in more detail in Figure~\ref{fig:CO}). These features are characteristic of mid-G to early K-type stellar spectra, and the spectra bear particular resemblance to G6 Ia--Ib supergiants (see the NASA Infrared Telescope Facility; IRTF spectral library of cool stars; \citealp{rayner09}). Despite good coverage of several near-IR H~\textsc{i} recombination lines, we detect only narrow H lines from the underlying star-forming region and no broader features associated with the transient (also for He~\textsc{i} $\lambda\lambda$10830, 20581). As described above, the Ca~\textsc{ii} IR triplet emission (seen in absorption in mid-G to early K-type stellar spectra) suggests an outflow with velocities of ${\approx}100$--$240$\,km~s$^{-1}$. While we do not detect H recombination features from the transient, this is consistent with a temperature of ${\approx}5400$--$5600$~K inferred by the presence of CN/CO absorption and the presence of Ca~\textsc{ii} in the spectrum, as such temperatures are too low for a significant fraction of the hydrogen to be ionized. The relatively low outflow velocities observed for SPIRITS\,17pc and cooler, stellar-like spectrum suggest that this event is likely associated with nonterminal outbursts or eruptions of its progenitor, rather than a true SN. 

In the near-IR, SPIRITS\,17qm shows emission features of H~\textsc{i}, as well as the He~\textsc{i} $\lambda10830$ line. As shown in Figure~\ref{fig:H_He_17qm}, these lines are relatively broad. The Pa$\gamma$ line at 64~days, well approximated by a gaussian with an FWHM velocity of $1600$\,km~s$^{-1}$, is somewhat narrower than the H$\alpha$ emission line observed at an earlier phase of 9~days. The spectra are consistent with a giant eruption from an LBV, and while a low-energy SN~IIn may also be possible, the lack of higher-velocity features argues against the terminal explosion of the progenitor. 

\subsection{Multi-band Light Curves and Extinction Estimates}\label{sec:mb_lcs}
\begin{figure*}
\begin{minipage}[hpb]{180mm}
\centering
\gridline{\fig{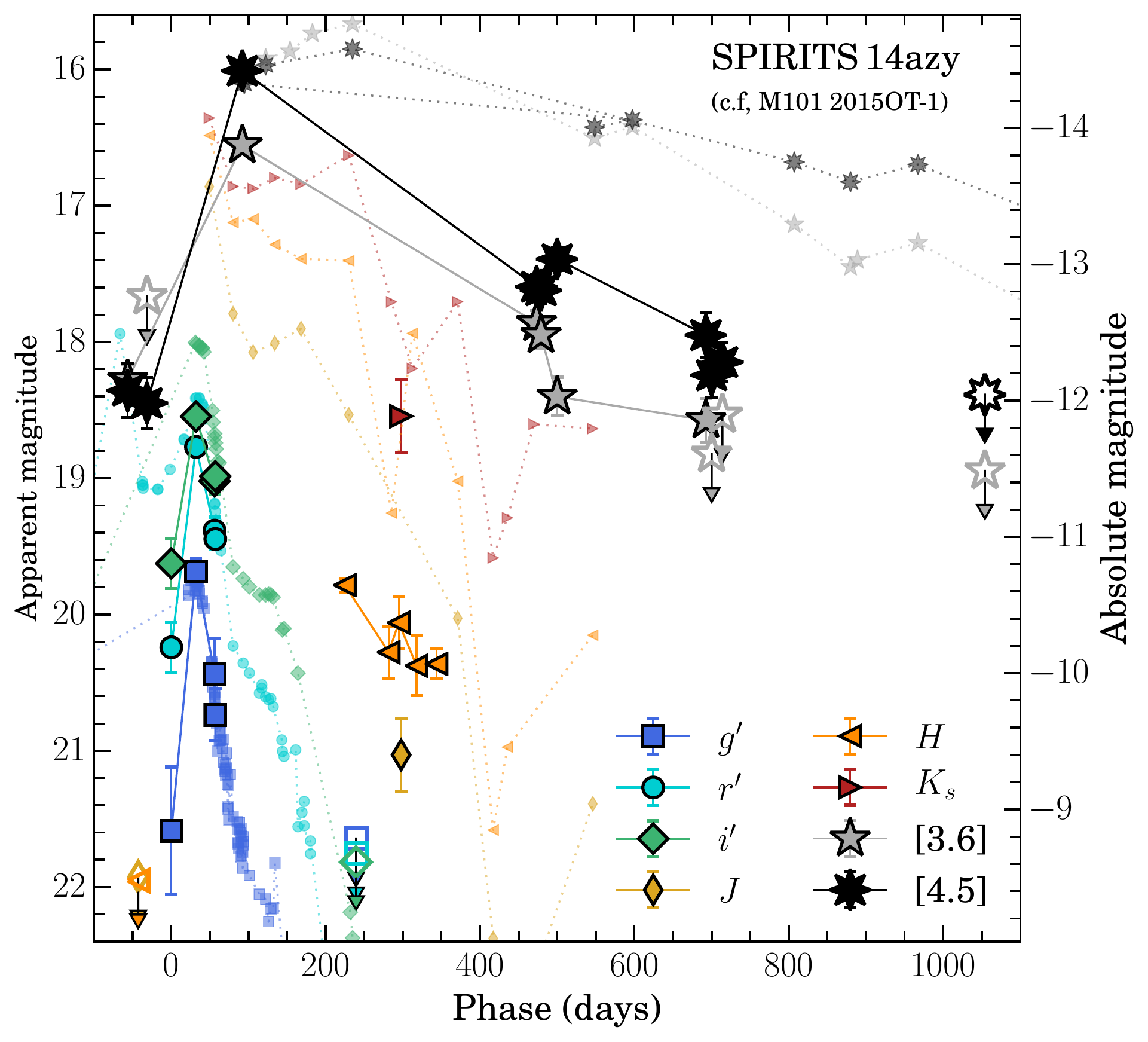}{0.45\textwidth}{}
          \fig{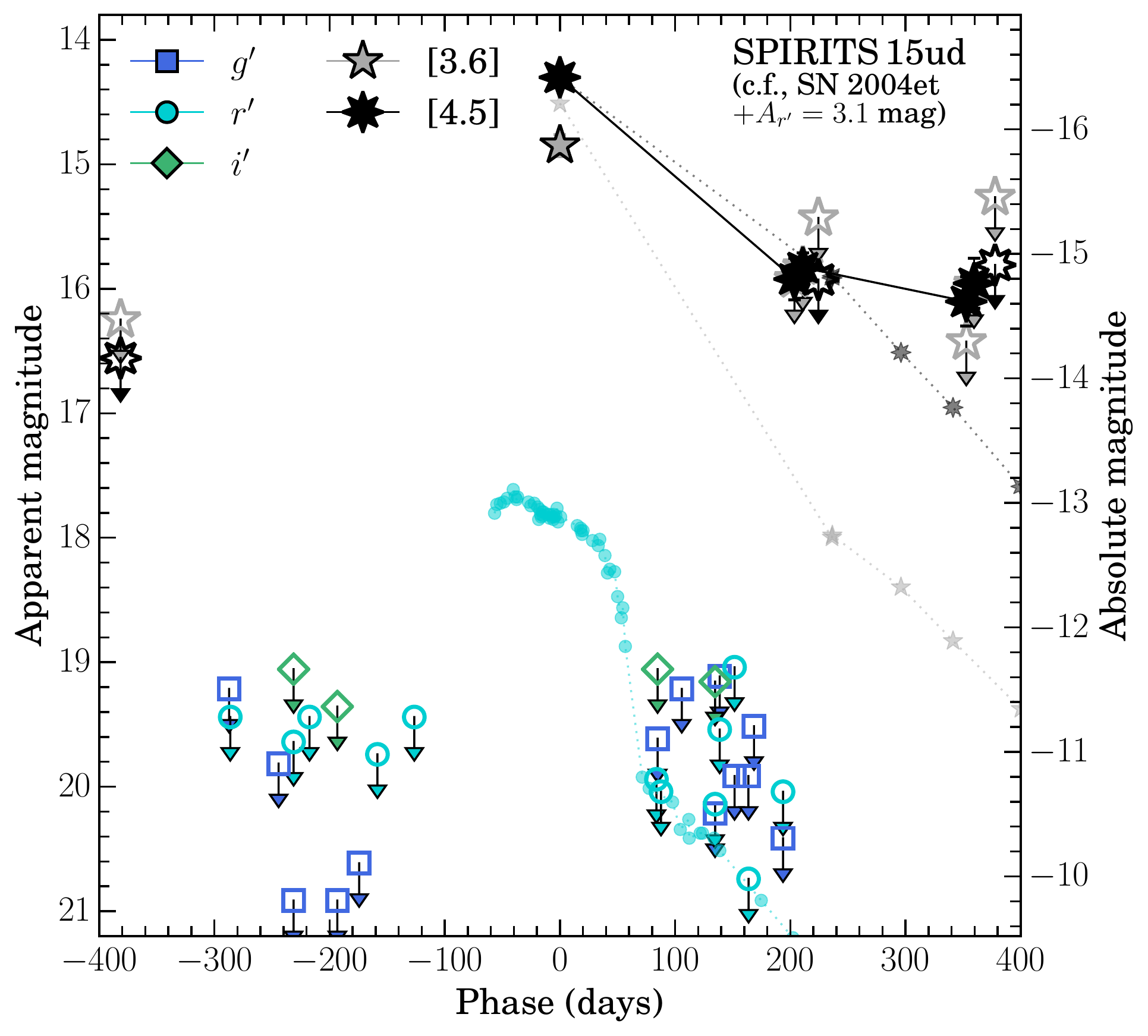}{0.45\textwidth}{}}
\gridline{\fig{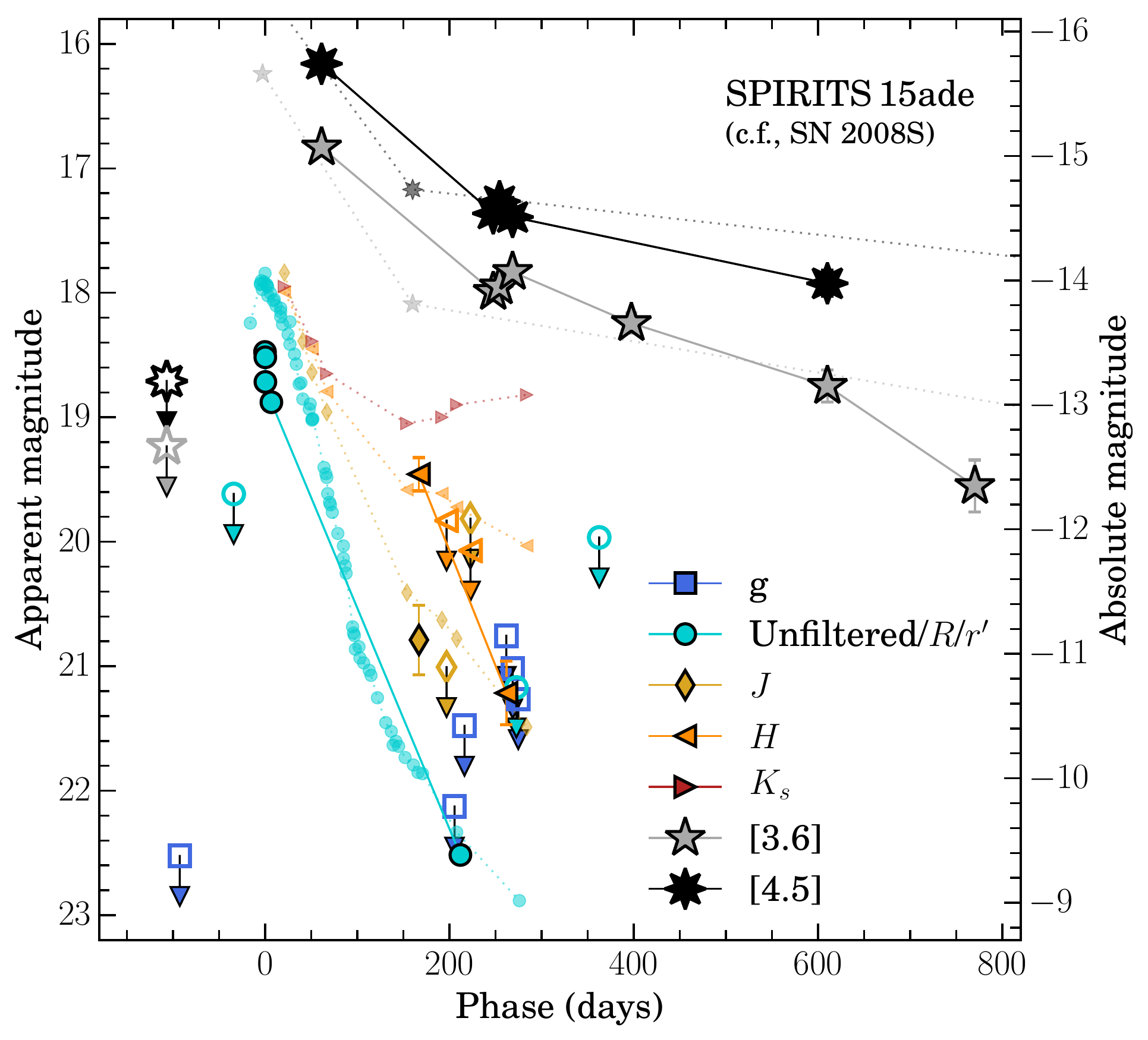}{0.45\textwidth}{}
          \fig{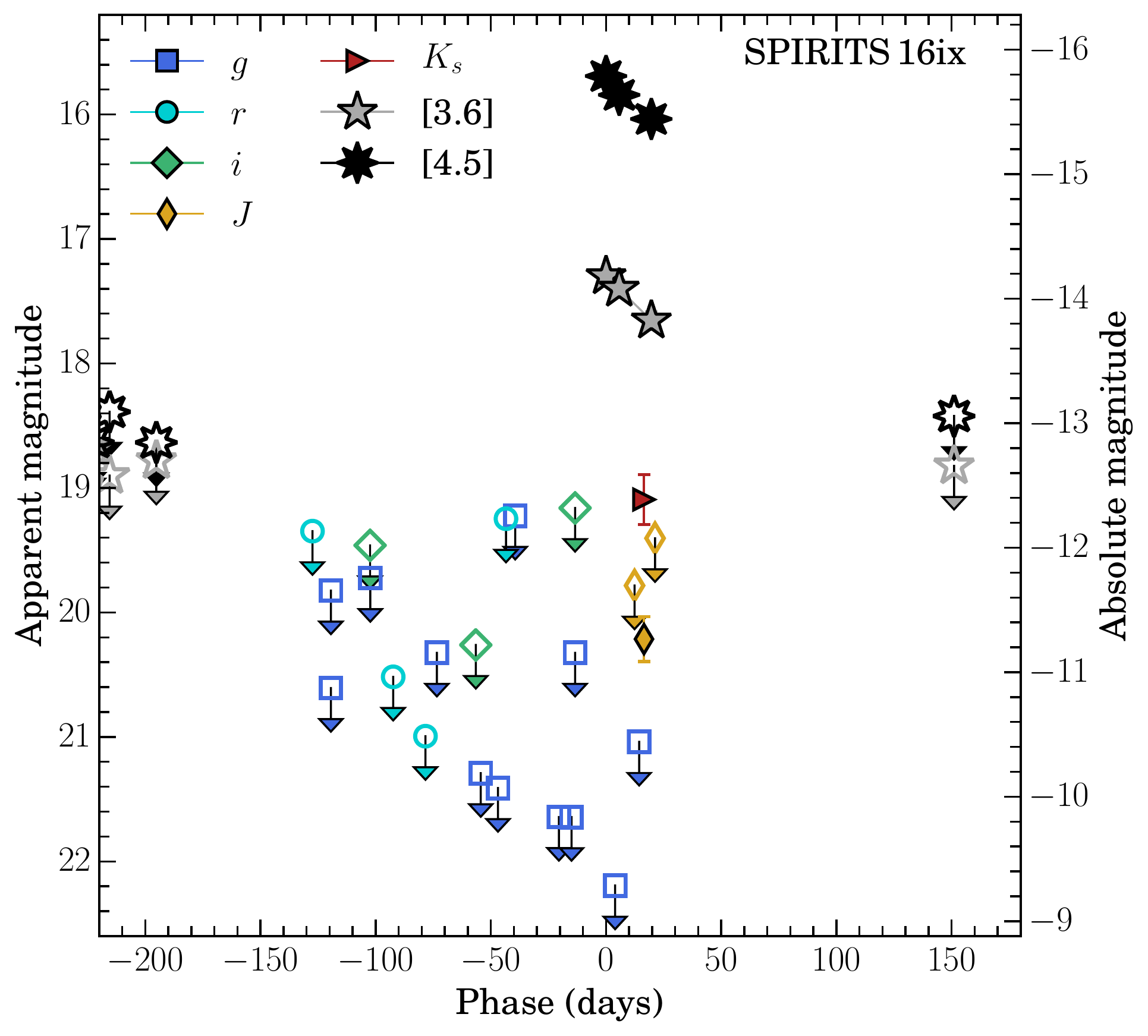}{0.45\textwidth}{}}
\caption{\label{fig:multiband_lcs}
Multiband light curves of the SPIRITS sample of luminous, IR-discovered transients. The large, black-outlined, filled symbols are detections of the SPIRITS transient indicated in the upper right corner of each panel. Upper limits from nondetections are shown as the large, open symbols with downward-pointing arrows. The smaller, faint symbols show corresponding light curves of a well-studied comparison object, also listed in parentheses in the upper right corner of each panel. All light curves have been corrected for Galactic extinction, and comparison light curves are also corrected for additional host/intrinsic reddening as described in the text. Comparison light curves are then shifted in apparent magnitude to the distance of their corresponding SPIRITS event. The SN\,2004et light curves, in comparison to SPIRITS\,15ud, have additionally been shifted in apparent magnitude to match in [4.5] peak luminosity and then reddened by $A_V = 3.7$ to be consistent with our $r^{\prime}$-band limits. Similarly, in comparison to SPIRITS\,17lb, the SN\,2004et light curves have been reddened by $A_V = 2.5$ to be consistent with our $V$-band limit.
}
\end{minipage}
\end{figure*}

Using available data, we compiled multiband light curves from the optical to near-IR for each luminous SPIRITS transient in our sample. Theses light curves are shown in Figure~\ref{fig:multiband_lcs}. We then estimate the visible extinction, $A_{V}$ (after correcting for the Galactic contribution), which may come from the foreground interstellar medium (ISM) of the host galaxy, from the local circumstellar environment of the progenitor, or by dust formed in the event itself. We provide our estimates, based on the analysis below, in Table~\ref{table:lc_prop}. In most cases, we are unable to make strong statements about the origin of the extinction, and throughout this section we assume a standard Milky Way ISM extinction law with $R_V = 3.1$ \citep{fitzpatrick99}. 
To obtain $A_V$ estimates where we have a reasonably secure classification of the transient, we attempt to make a direct comparison with template light curves from a well-studied object. Where the classification of a transient is less secure, or where good template light curves were not available, we adopt $m_{X} - m_{[4.5]}$ near the observed peak of the transient as an estimate of $A_{X}$, the extinction in broadband filter $X$ (preferably optical), and convert to $A_V$ with our assumed extinction law.

\begin{figure*}
\figurenum{\ref{fig:multiband_lcs}, continued}
 \gridline{\fig{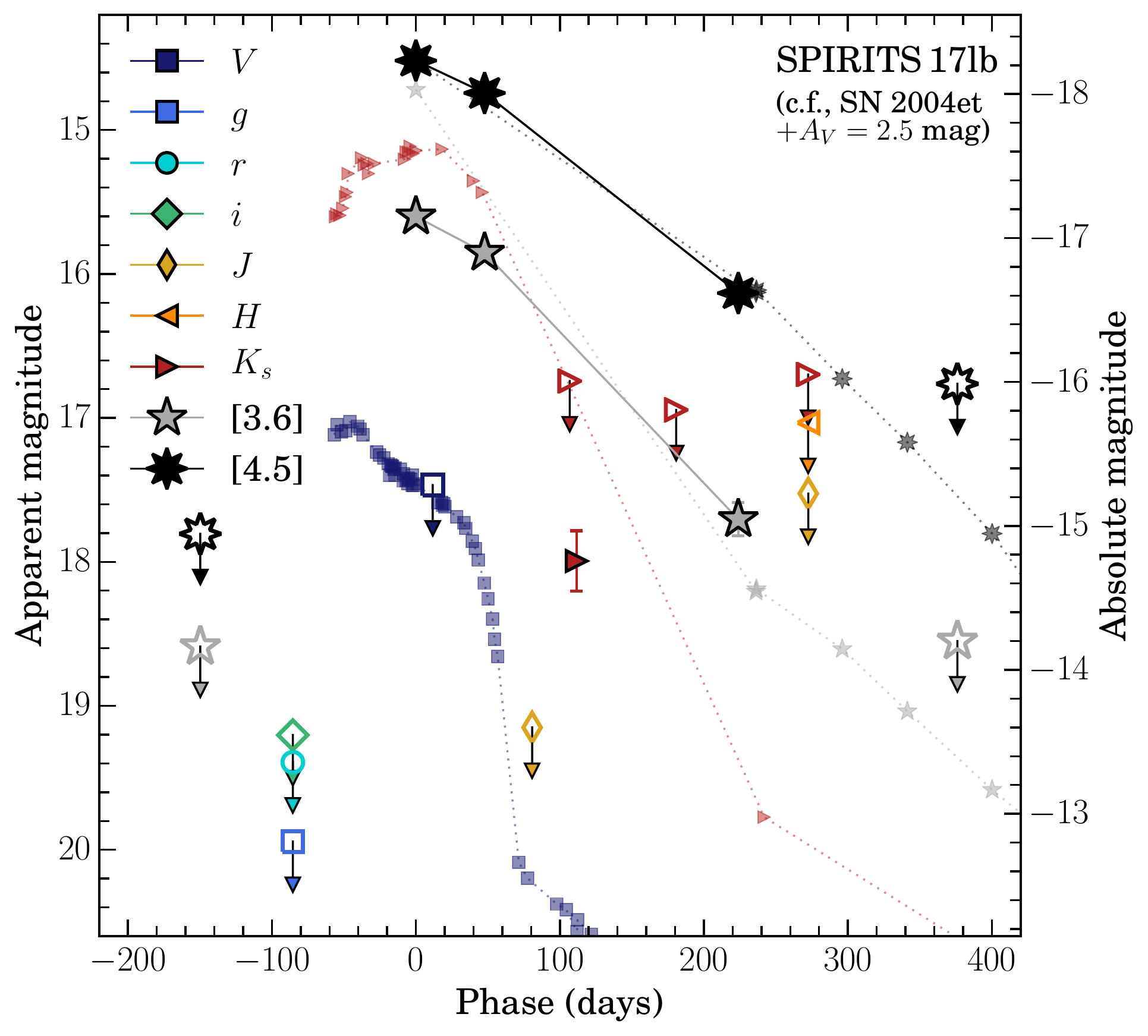}{0.45\textwidth}{}
           \fig{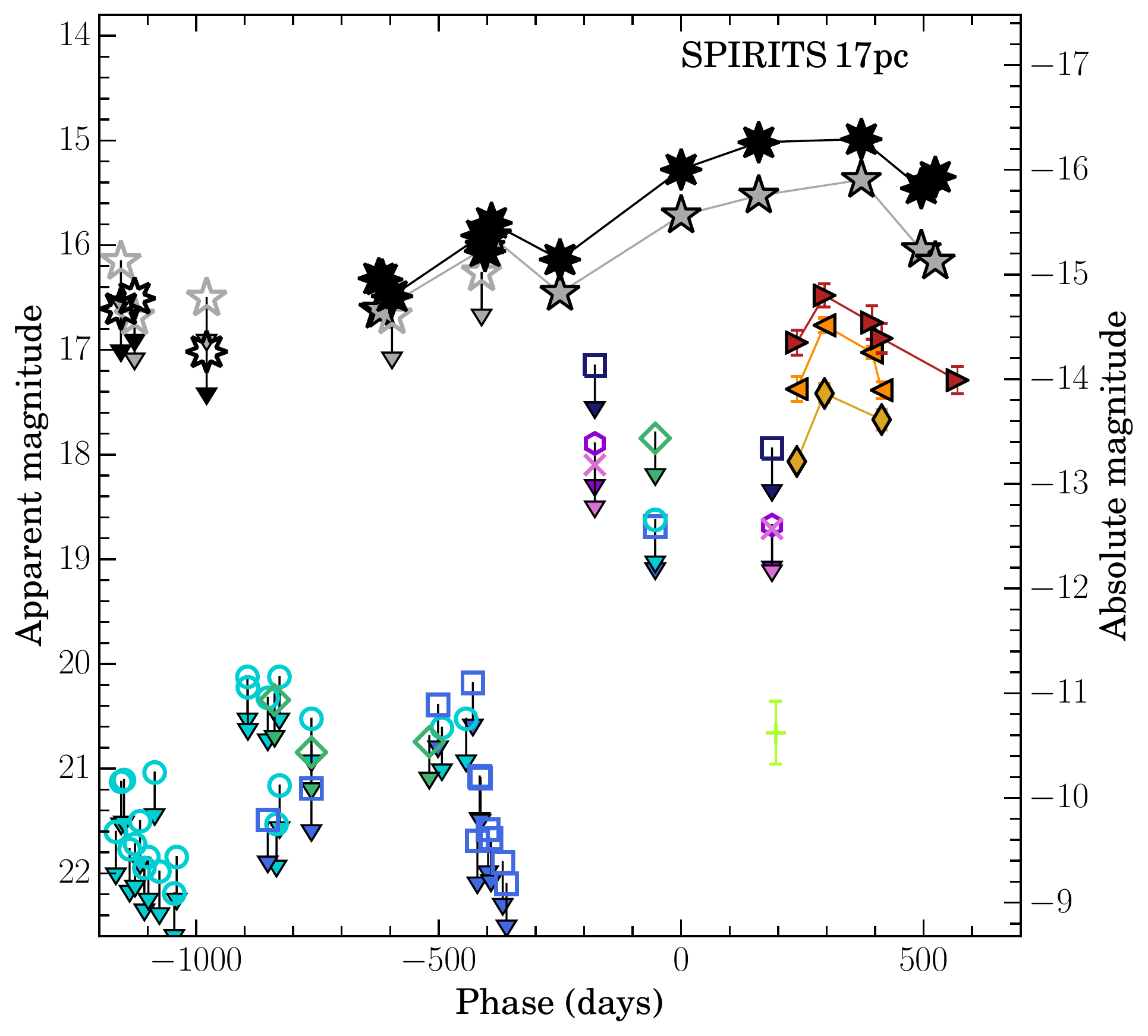}{0.45\textwidth}{}}
 \gridline{\fig{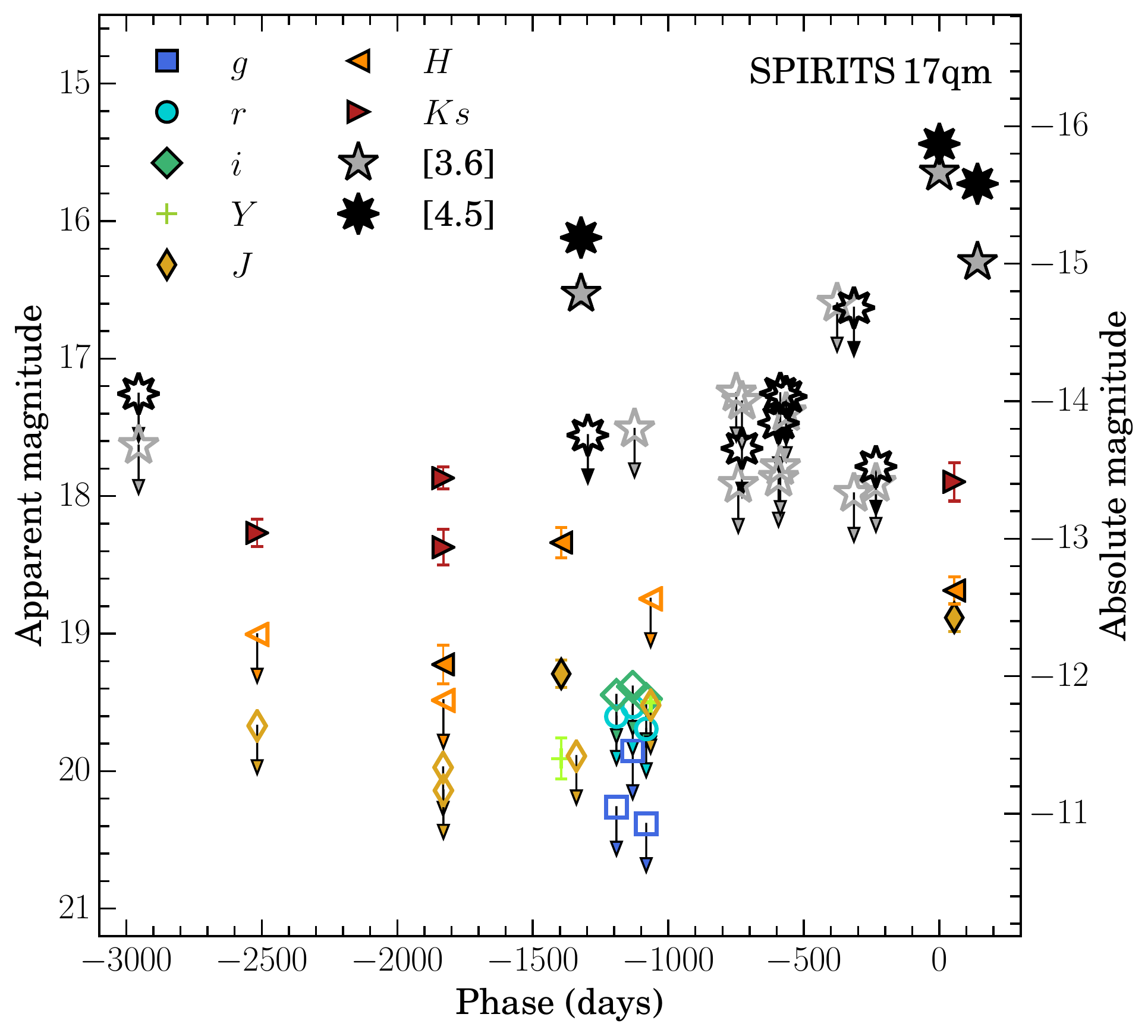}{0.45\textwidth}{}}
\caption{}
\end{figure*}

The light curves of SPIRITS\,15c were discussed in detail in \citet{jencson17}, where we estimated $A_V \approx 2.2$ based on comparison to the well-studied Type IIb SN~2011dh. Similarly, for SPIRITS\,16tn we estimated $A_V \approx 7.8$ based on a comparison to the low-luminosity Type IIP SN~2005cs \citep{jencson18c}. 

The light curves of SPIRITS\,14azy are shown in Figure~\ref{fig:multiband_lcs} in comparison to the 2015 LRN in M101 (M101 OT2015-1), a well-studied example of a common-envelope ejection in a merging stellar binary system \citep[][, and 2019, in preparation]{blagorodnova17}. SPIRITS\,14azy, shows a short-lived, optical counterpart peaking at $g^{\prime} = 20.0$ ($M_{g^{\prime}} = -10.7)$, which rises to peak and fades within $\lesssim 60$~days. The optical, broadband colors near peak are red, with $g^{\prime}-r^{\prime} = 0.9$ and $g^{\prime}-i^{\prime} = 1.1$. The transient displays a longer-lived, IR excess detected out to $\gtrsim 300$~days in the near-IR and to $\gtrsim 700$~days in the \textit{Spitzer}/IRAC bands. Shifted in apparent magnitude to the distance of SPIRITS\,14azy in NGC~2997, we see that the light curves of M101 OT2015-1 are similar to SPIRITS\,14azy both in the absolute brightness of the transient in $g^{\prime}$ and the \textit{Spitzer}/IRAC IR bands and in the timescale of the optical transient. While M101 OT2015-1 is characterized by somewhat redder optical colors and an even longer-lived IR excess, the overall similarity of the two events is readily apparent. We characterize the reddening for SPIRITS\,14azy as $A_{g^{\prime}} \approx g^{\prime}_{\mathrm{peak}} - [4.5]_{\mathrm{peak}} = 3.7$, corresponding to $A_V = 3.2$. We caution that this estimate should not be directly interpreted as a measurement of the extinction from the host galaxy or local environment, and it may also be indicative of ``intrinsic'' reddening of the transient due to dust formation or the relatively cool effective temperatures associated with LRN-like events.

SPIRITS\,15ud was inaccessible for ground-based observing owing to its proximity to the sun from Earth at the time of the \textit{Spitzer} discovery, and thus our constraints on any optical emission associated with the transient are only from phases before/after ${\approx}\pm100$~days. As discussed below in Section~\ref{sec:class}, we find the most likely interpretation of the IR light curves of this event to be an SN~II. Our deepest limits on the post-peak optical flux of SPIRITS\,15ud are in the $r^{\prime}$ band, and we compare to the light curve in this band of the Type IIP SN~2004et from \citet{maguire10} to obtain a constraint on the extinction. We shift the light curves in phase to match the time of the observed [4.5] peak of SPIRITS\,15ud and in apparent magnitude to match the observed [4.5] brightness, and we find that our optical limits then require $A_{r^\prime} \gtrsim 3.1$ to be consistent with the declining light curve of SN~2004et. This corresponds to $A_V \gtrsim 3.7$, which we adopt as a lower limit on the extinction to SPIRITS\,15ud. 

The \textit{Spitzer} light curves of SPIRITS\,15ade, as discussed above in Section~\ref{sec:IR_lcs}, show a distinct similarity those of SN~2008S, a prototype of the class of ILRTs, in both IR luminosity at $M_{[4.5],\mathrm{peak}} \approx -16$ and the subsequent IR light curve decline. The first detection of the source was in in an unfiltered optical CCD image taken on 2015 September 11.5, where the transient was identified as PSN\,J15220552+\allowbreak 0503160 by M.\ Aoki. Calibrated to the SDSS $r$ band, our photometry yields an observed optical peak at $r = 18.6 \pm 0.2$ ($M_r = -13.4$) 61.3 days before the first detection by SPIRITS. The optical luminosity is similar to the $R$-band peak of SN~2008S (from \citealp{botticella09}, assuming $A_V \approx 1$ in excess of the Galactic contribution as in \citealp{botticella09} and \citealp{szczygiel12}), and our late-time optical measurements are consistent with the optical fading of SN~2008S within ${\approx}200$~days. We detect a near-IR excess in the $J$, $H$, and $K_s$ bands at a phase of ${\approx}170$~days, again similar to that observed in SN~2008S. We characterize the reddening to SPIRITS\,15ade, though it may be intrinsic to the source, as $A_{r} = R_{\mathrm{peak}} - [4.5]_{\mathrm{peak}} = 2.3$, corresponding to $A_V = 2.7$.

SPIRITS\,16ix is a near twin to SPIRITS\,16tn in both their observed peak IR luminosities at $M_{[4.5],\mathrm{peak}} \approx -16$ and their rapid IR decline rates (Section~\ref{sec:IR_lcs}). Furthermore, both objects are remarkably red, as shown for SPIRITS\,16ix in Figure~\ref{fig:multiband_lcs}. Our deepest optical limit during the IR observed peak of SPIRITS\,16ix reached $g \gtrsim 22.2$ ($M_g \gtrsim -9.3$), indicating $A_{g} \gtrsim 6.3$. Corresponding to $A_V \gtrsim 5.5$, we adopt this as a lower limit on the extinction to SPIRITS\,16ix. 

Similar to our analysis for SPIRITS\,15ud, we compare our photometric data for SPIRITS\,17lb to those of SN~2004et. Shifting SN~2004et in phase to match the time of the observed [4.5] peak and in apparent magnitude to the distance of SPIRITS\,17lb, we find that their [4.5] light curves track each other remarkably well, though SPIRITS\,17lb is significantly redder than SN\,2004et in $[3.6]{-}[4.5]$ until $\gtrsim 150$~days. Applying reddening with $A_V \gtrsim 2.5$ is then required for the SN~2004et light curves to be consistent with our $V$-band limit at $11.7$~days, which we adopt as a lower limit on the extinction to SPIRITS\,17lb. 

As discussed in Section~\ref{sec:IR_lcs}, the \textit{Spitzer} light curves of SPIRITS\,17pc are characterized by multiple outbursts over the past ${\approx}4$\,yr that appear to be progressively increasing in both brightness and duration. Deep optical limits constrain the optical variability of the source between ${\approx}-1200$ and $-300$~days, and we note that the IR outburst peaking at $[4.5] = 15.8$ near $-400$~days was extremely red with $g^{\prime} - [4.5] > 5.9$. During the brightest, longest-duration IR outburst seen in the \textit{Spitzer}/IRAC bands lasting $\gtrsim 500$~days, we detected underlying near-IR variability peaking at $K_s = 16.5$ at $295.6$~days with a red near-IR color of $J-K_s = 0.9$ with a comparatively shorter duration of $\lesssim 200$~days. We also obtained our only optical detection of SPIRITS\,17pc at $Z = 20.7$ at $195.1$~days, indicating $A_Z = Z - [4.5] = 5.7$. To estimate the extinction in Table~\ref{table:lc_prop}, we convert this to $A_V = 12.5$ but again caution that the extreme $\mathrm{optical}{-}\mathrm{IR}$ color may be intrinsic to the source and is not necessarily indicative of extinction by the host ISM or local environment of the progenitor. 

The light curves of SPIRITS\,17qm also indicate multiple epochs of significant variability across the near-IR and \textit{Spitzer}/IRAC bands extending back to at least a phase of $-2500$~days before the discovery detection and observed IR peak at $[4.5] = 15.4$ ($M_{[4.5]} = -15.9$). During the pre-discovery outburst detected by \textit{Sptizer}, our photometry indicates a notably red SED with $Y - [4.5] = 3.8$. Similarly, our near-IR follow-up observations during the post-peak decline at $t=55.1$~days indicate $A_J = J - [4.5] = 3.3$, corresponding to $A_V = 12.1$. As with SPIRITS\,17pc, rather than indicating host or environmental extinction, the extreme colors we observe may be intrinsic to the SED of the source or caused by dust formation. 

\subsection{Radio constraints}\label{sec:analysis_radio}
Radio observations of CCSNe probe the nonthermal emission generated when the fastest ejecta interact with the CSM produced by the pre-explosion stellar wind of the progenitor. The spectrum is dominated by synchrotron emission as the blast wave propagates through the CSM, where turbulent instabilities may accelerate electrons to relativistic energies and amplify magnetic fields \citep{chevalier82}. At early times, the synchrotron spectrum is self-absorbed at high frequencies, with possible additional contributions from internal free-free absorption, and free-free absorption by the external, pre-shocked, ionized CSM \citep[e.g.,][]{chevalier82,chevalier98}. As the shock propagates out into the CSM, the peak in the spectrum, below which the source is opaque, shifts to lower frequencies. As no SN~Ia has been detected in the radio to deep limits in radio luminosity of $L_{\nu} \lesssim 10^{24}$~erg~s$^{-1}$~Hz$^{-1}$, detection of this characteristic signature could provide strong confirmation our sources as CCSNe. 

\begin{figure*}
\centering
\gridline{\fig{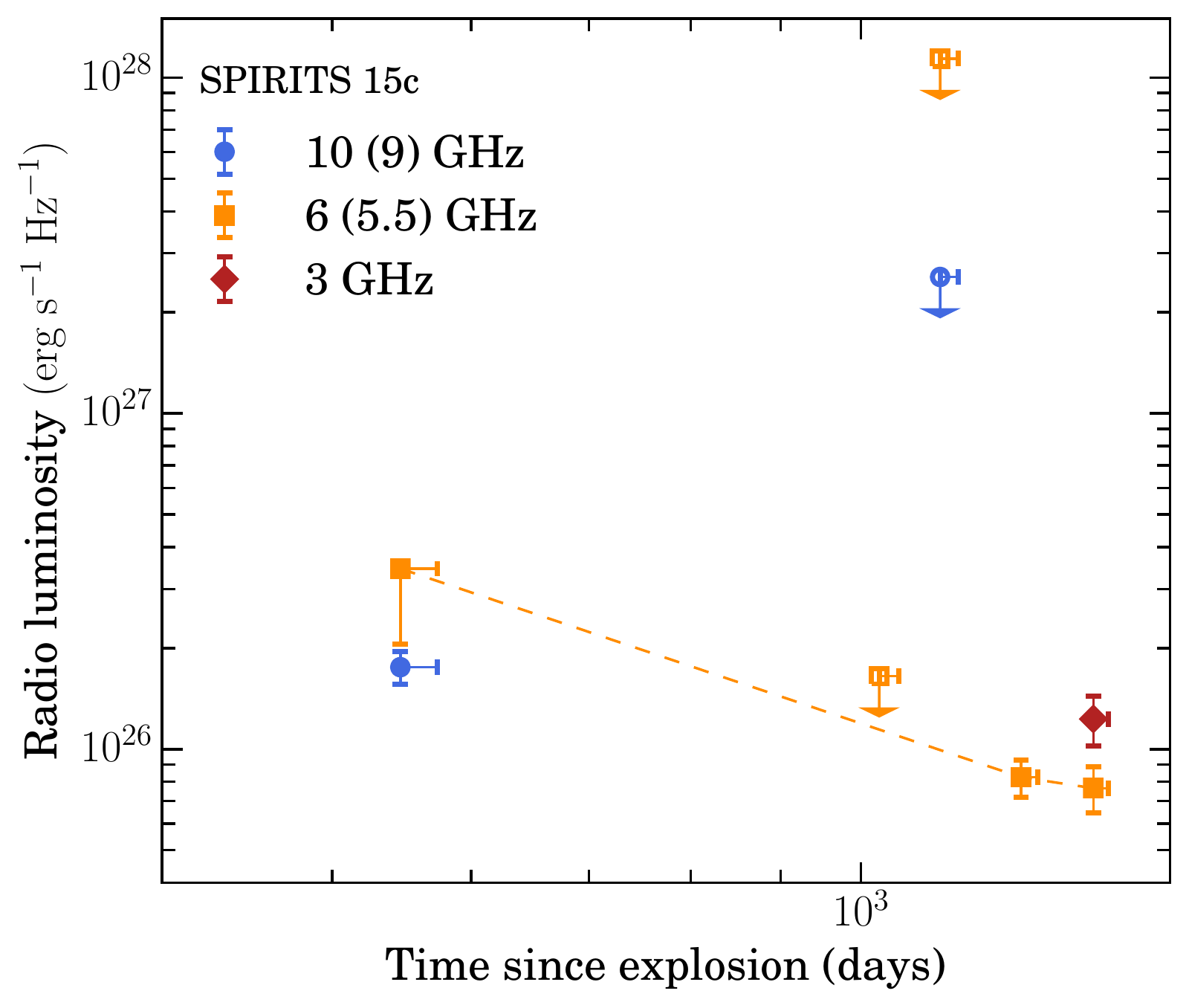}{0.45\textwidth}{}
          \fig{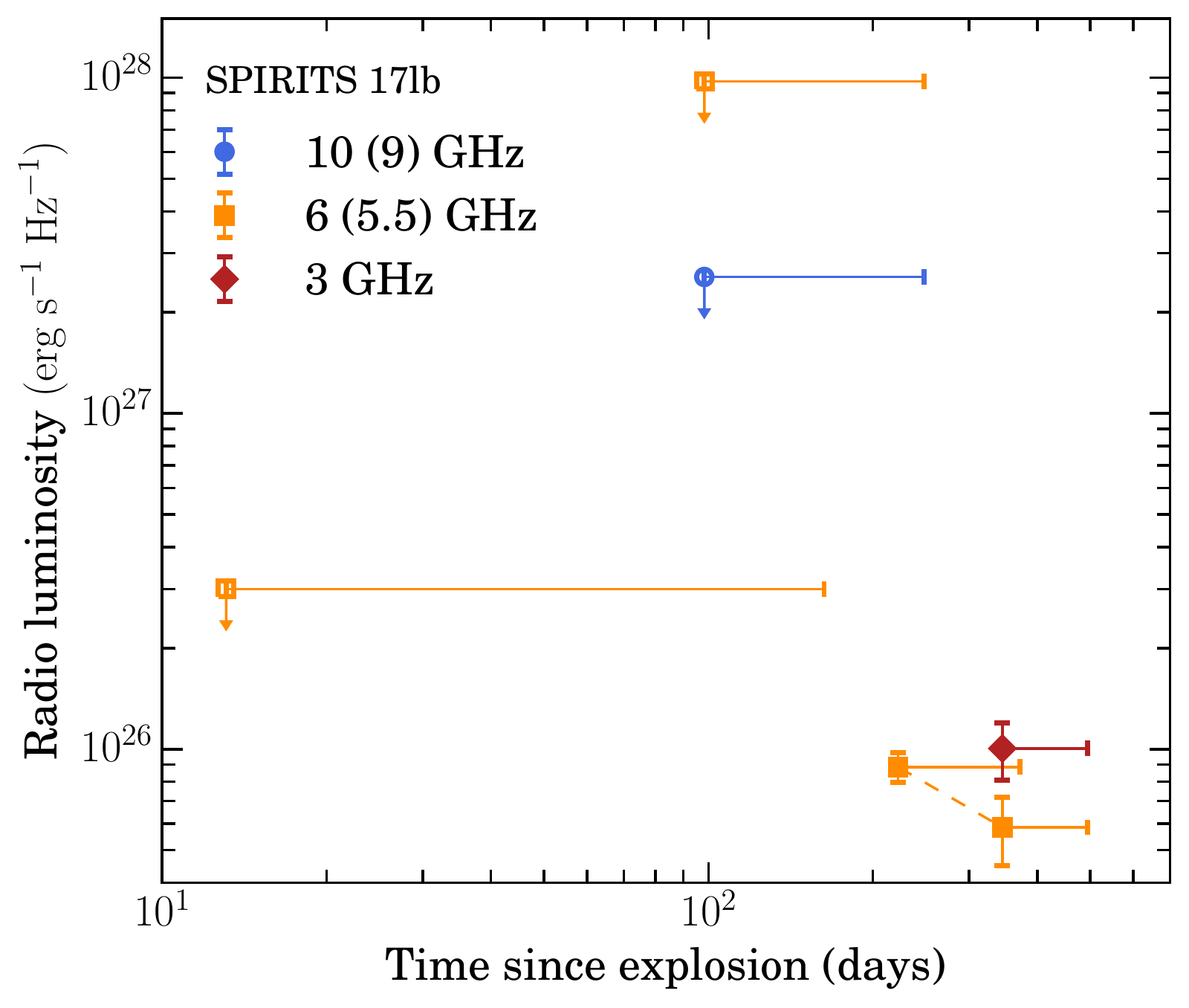}{0.45\textwidth}{}}
\caption{\label{fig:radio_lcs}
Radio light curves of SPIRITS\,15c (left) and SPIRITS\,17lb (right) in multiple frequency bands. Detections are indicated by filled symbols, while upper limits from non-detection are indicated by unfilled symbols with downward-pointing arrows. Horizontal error bars indicate the uncertainty in the absolute phase since explosion for these events. 
}
\end{figure*}

In Figure~\ref{fig:radio_lcs}, we show the radio light curves of SPIRITS\,15c and SPIRITS\,17lb. For both sources, we detect a declining radio source consistent with optically thin synchrotron emission. At $t=546.7$~days for SPIRITS\,15c, the transient is detected at both 10 and 6\,GHz, with a spectral index $\alpha = 1.32^{+0.05}_{-1.01}$, where the flux density is given by $S_{\nu} \propto \nu^{-\alpha}$. The source is observed to fade at 6\,GHz, with a spectral index of $\alpha = 0.7\pm0.1$ between 3 and 6\,GHz at $t=1356.6$~days.  We infer from these observations that SPIRITS\,15c peaked at a 6\,GHz radio luminosity of $L_{\nu} \gtrsim 3.5\times10^{26}$~erg~s$^{-1}$~Hz$^{-1}$ at time $t \lesssim 546.7$~days (maximum age $\lesssim 573.7$~days). For SPIRITS\,17lb, the source fades at 6\,GHz between $t = 222.5$ and $345.3$~days, with a spectral index of $\alpha = 0.78 \pm 0.13$ between 3 and 6\,GHz at the later epoch. SPIRITS\,17lb thus peaked at 6\,GHz at $L_{\nu} \gtrsim 8.8\times10^{25}$~erg~s$^{-1}$~Hz$^{-1}$ at time $t \lesssim 222.5$~days (maximum age $\lesssim 372.3$~days). The remaining SPIRITS events were undetected in our radio follow-up observations.

\begin{figure*}
\centering
\gridline{\fig{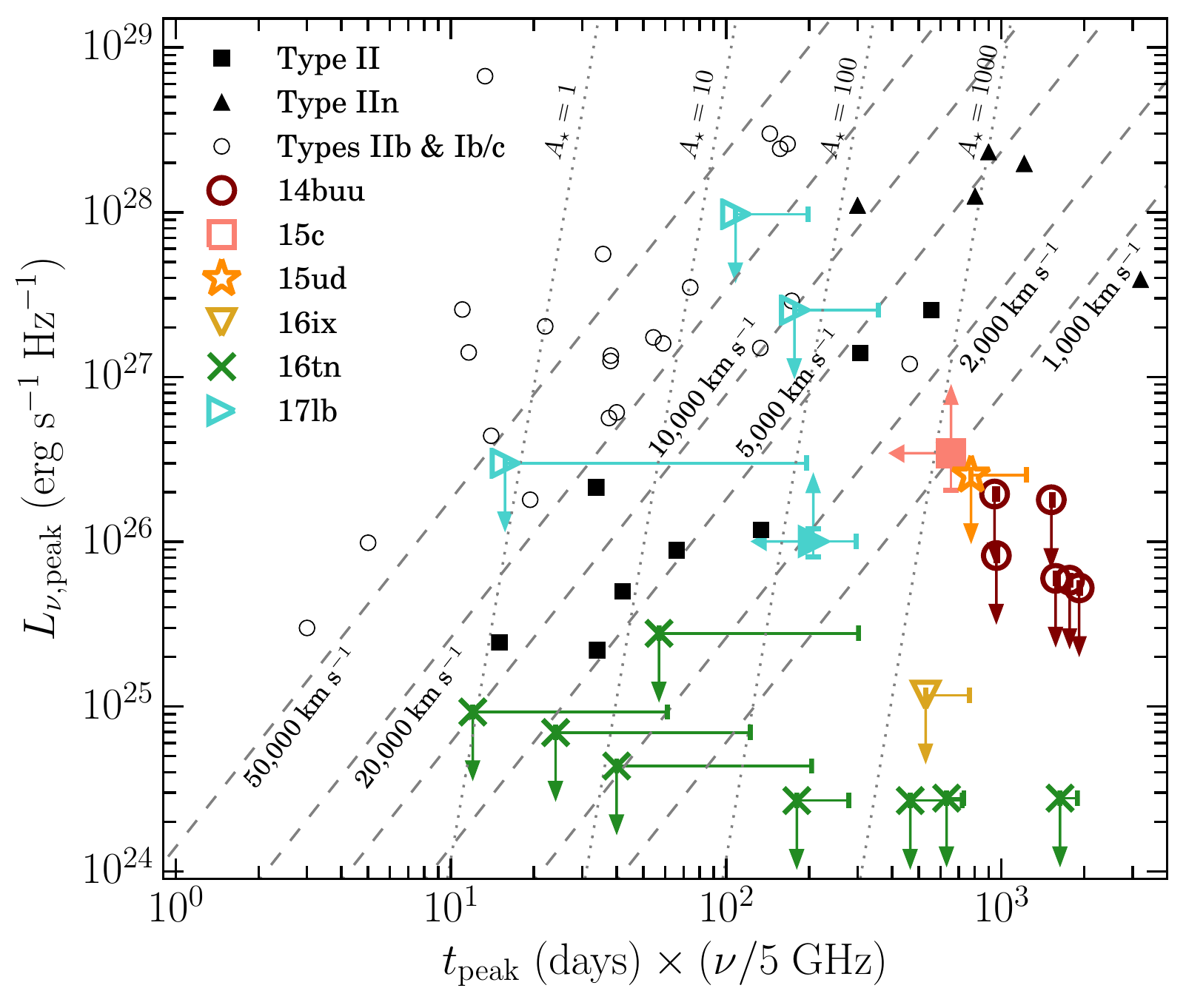}{0.45\textwidth}{}
          \fig{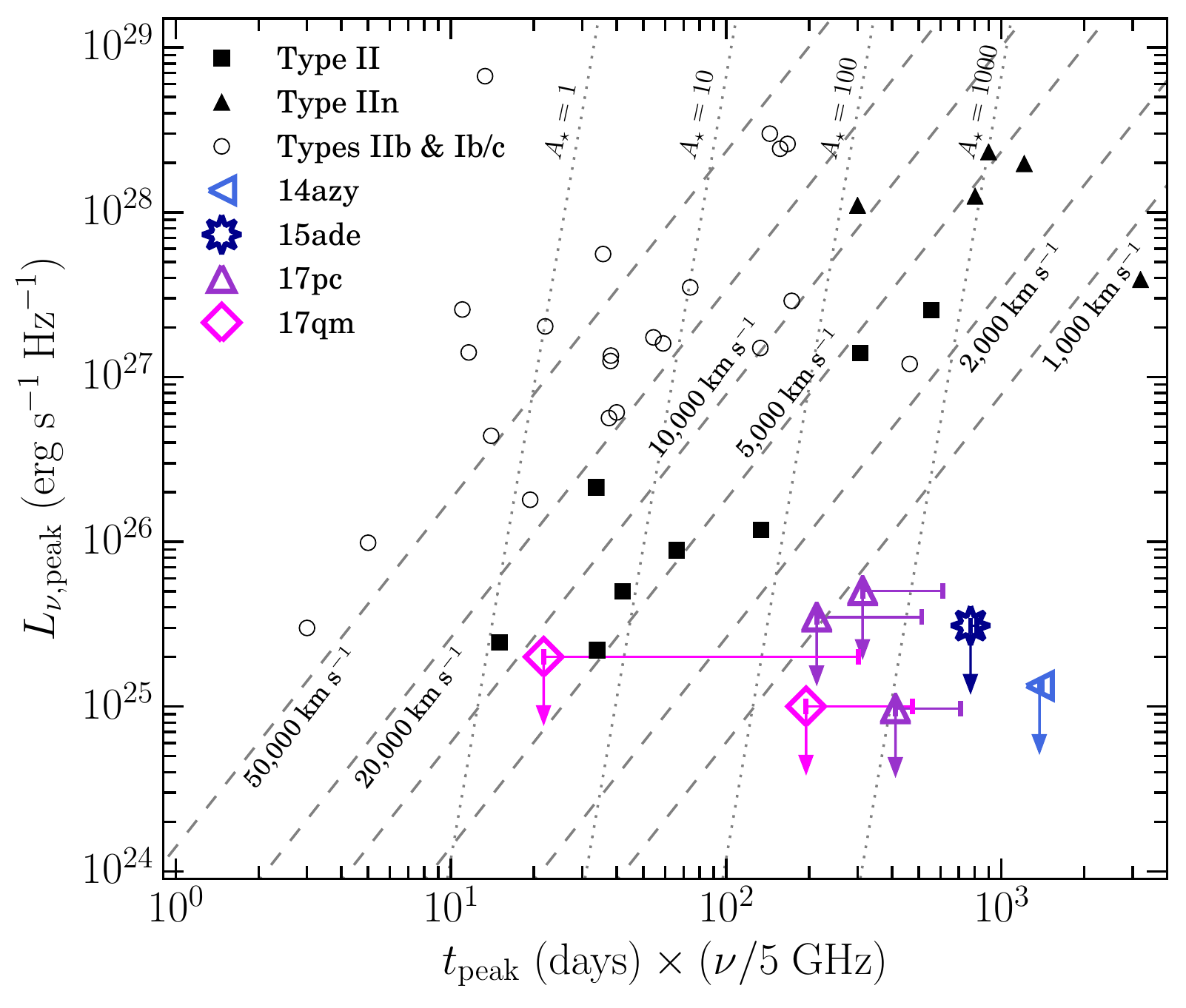}{0.45\textwidth}{}}
\caption{\label{fig:radio_lims}
In each panel, we show the peak radio luminosity vs. time of peak times the frequency of observation for radio CCSNe adapted from, e.g., \citet{chevalier06a} and \citet{romero-canizales14}. This is an updated version of Figure~10 from \citet{jencson18c}. SNe~II are shown as black squares, and strongly interacting SNe~IIn are shown as black triangles. Open circles represent stripped-envelope SNe~IIb and Ib/c. Upper limits on the radio luminosity of SPIRITS transients at a given phase are shown as multicolor, open symbols with downward-pointing arrows, where the horizontal error bars represent our uncertainty in the absolute phase since explosion. The possible CCSNe are shown in the left panel, while the non-SN events are shown in the right panel. Filled symbols for SPIRITS\,15c and SPIRITS\,17lb represent constraints on $t_\mathrm{peak}$ and $L_{\nu,\mathrm{peak}}$ from detections of the transients. Symbols corresponding to each object are labeled in the legend on the left side of the figure. Assuming an SSA model with an electron distribution with p = 3 for the shock wave propagating through the CSM, one can infer the shock velocity (dashed lines) and CSM density parameter ($A_{*}$; dotted lines) from the position on this diagram.}
\end{figure*}

In Figure~\ref{fig:radio_lims}, we show the peak luminosities of radio CCSNe of various subtypes and the time to peak times the frequency of observation compared to our constraints for SPIRITS\,15c and SPIRITS\,17lb, as well as our limits on the radio luminosity of the rest of the sample. Following our discussion in Section~4.2.1 of \citet{jencson18c}, we use the self-similar solution for the propagation of the SN blast wave into the CSM described in \citet{chevalier98} and \citet{chevalier06b}. If synchrotron self-absorption (SSA) is the dominant absorption mechanism, and assuming that the emitting electron population is described by a power law with an energy spectral index $p = 3$, then, as calculated by \citet{chevalier98}, the size of the radio-emitting region is given by

\begin{equation}\label{eq:Rs}
\begin{split}
	R_{\mathrm{s}} = 4.0 \times 10^{14} q^{-1/19} \left(\frac{f}{0.5} \right)^{-1/19} \left(\frac{F_{\mathrm{p}}}{\mathrm{mJy}} \right)^{9/19} \\
	     \times \left(\frac{D}{\mathrm{Mpc}}\right)^{18/19} \left(\frac{\nu}{5~\mathrm{GHz}}\right)^{-1} \mathrm{cm},
\end{split}
\end{equation}

\noindent where $q \equiv \epsilon_e/\epsilon_B$ is the ratio of the energy density in relativistic electrons to that in the magnetic field, $f$ is the filling factor of the radio-emitting region, $F_p$ is the peak flux at frequency $\nu$, and $D$ is the distance to the source. We show the inferred shock velocities assuming energy equipartition ($q = 1$) and $f = 0.5$ (as estimated by \citealp{chevalier06b}) as the dashed lines in Figure~\ref{fig:radio_lims}.

Assuming that the CSM was produced by a steady pre-SN stellar wind, its density profile as a function of radius, $r$, will be given by $\rho_{\mathrm{w}} = A/r^2 \equiv \dot M / (4 \pi r^2 v_{\mathrm{w}})$, where $\dot M$ is the mass-loss rate and $v_{\mathrm{w}}$ is the wind velocity. We define $A \equiv \dot M / (4 \pi v_{\mathrm{w}})$ as the normalization of the CSM density profile, and $A_{\star} \equiv A/(5 \times 10^{11}~\mathrm{g}~\mathrm{cm}^{-1})$ is a dimensionless proxy for $A$ as in \citet{chevalier82}. The radio emission at time $t$ since explosion is then sensitive to the density profile of the CSM as

\begin{equation}\label{eq:Astar}
\begin{split}
	A_{\star} \epsilon_{B-1} q^{8/19} = 1.0 \left(\frac{f}{0.5} \right)^{-8/19} \left(\frac{F_{\mathrm{p}}}{\mathrm{mJy}} \right)^{-4/19} \\
	     \times \left(\frac{D}{\mathrm{Mpc}}\right)^{-8/19} \left(\frac{\nu}{5~\mathrm{GHz}}\right)^{2} \left(\frac{t}{10~\mathrm{days}}\right)^{2},
\end{split}
\end{equation}

\noindent where $\epsilon_{B-1} \equiv \epsilon_B/0.1$ \citep{chevalier06b}. The inferred value of $A_*$ depends very sensitively on $t_{peak}$, shown in Figure~\ref{fig:radio_lims} as the nearly vertical dotted lines. 

Our constraint on the peak radio luminosity for SPIRITS\,15c is consistent with its spectroscopic classification as an SN~IIb or SN~Ib, which tend to be more luminous radio sources ($10^{26} \lesssim L_{\nu,\mathrm{peak}} \lesssim 10^{28}$~erg~s$^{-1}$~Hz$^{-1}$) peaking on timescales between 10 and 100\,days. Because our radio observations were post-peak, we are only able to place limits on the relevant parameters. Our observations require $v_s \gtrsim 1000$\,km~s$^{-1}$ and $A_* \lesssim 1400$. This translates to a limit on the pre-SN mass-loss rate of $\dot{M} \lesssim 1.4\times10^{-3} \left(\frac{\epsilon_{B}}{0.1}\right) \left(\frac{v_w}{100~\mathrm{km~s}^{-1}}\right) M_{\odot}$~yr$^{-1}$.

SPIRITS\,17lb, despite its younger phase at the time of first observation with the VLA, is notably less luminous, more consistent with the population of radio SNe~II. We find $v_s \gtrsim 800$\,km~s$^{-1}$ and $A_* \lesssim 800$, corresponding to a limit on the pre-SN mass-loss rate of $\dot{M} \lesssim 8.0\times10^{-4} \left(\frac{\epsilon_{B}}{0.1}\right) \left(\frac{v_w}{100~\mathrm{km~s}^{-1}}\right) M_{\odot}$~yr$^{-1}$.

As discussed in \citet{jencson18c}, our deep radio nondetections for SPIRITS\,16tn rule out a stripped-envelope classification, except possibly the most rapidly evolving, high-velocity events that may have fast-peaking radio light curves. The new late-time limits beyond $t \gtrsim 500$~days presented here also argue against a strongly interacting SN~IIn, unless the radio emission is still strongly self-absorbed at this phase. A weak SN~II is the most consistent with our observations for SPIRITS\,16tn, and specifically, our inferred constraint on the pre-SN mass-loss rate of $\dot{M} \lesssim 2.4\times10^{-5} \left(\frac{\epsilon_{B}}{0.1}\right) \left(\frac{v_w}{100~\mathrm{km~s}^{-1}}\right) M_{\odot}$~yr$^{-1}$ may suggest a RSG progenitor of lower initial mass (10--15~$M_{\odot}$). Similarly, the late-time limit for SPIRITS\,16ix is significantly deeper than observed radio-luminous SN~IIn and stripped-envelope SNe, and is most consistent with an SN~II. Our weaker constraint for SPIRITS\,15ud may be consistent with either an SN~II or stripped-envelope classification. 

As discussed below in Section~\ref{sec:class}, based on the sum of all available observational data, we do not believe SPIRITS\,14azy, SPIRITS\,15ade, SPIRITS\,17pc, and SPIRITS\,17qm were terminal CCSN explosions. Thus, we do not expect strong radio counterparts for these events, but we show our radio limits in the right panel of Figure~\ref{fig:radio_lims} for completeness. Specifically for SPIRITS\,17qm, while the most likely interpretation is a giant LBV eruption, an SN~IIn is consistent with the optical/IR data but may be disfavored by the lack of radio emission detected thus far.

\section{Putting it all together: suggested transient classifications}\label{sec:class}
Here we discuss suggested classifications for each SPIRITS event in our sample based on the combined observational constraints across radio, IR, and optical wavelengths. We find that five of our nine IR events are confirmed or plausible CCSNe, while the remaining four are more likely non-SN massive-star outbursts of various origins. Our suggested classification for each event is given in Table~\ref{table:lc_prop}.

\subsection{The confirmed and plausible CCSNe}
As discussed in detail in \citet{jencson17}, SPIRITS\,15c is a clear example of a confirmed, moderately obscured CCSN. Our assessment is based primarily on near-IR spectroscopy showing distinct similarity to the Type IIb SN\,2011dh, and in particular the presence of a broad (${\approx}8400$\,km~s$^{−1}$) double-peaked emission line of He~\textsc{i} at 1.0830~$\mu$m. The optical/near-IR light curves were well matched to those of SN~2011dh assuming $A_V = 2.2$ and a standard Milky Way ISM extinction law with $R_V = 3.1$. In this work we present new radio observations of SPIRITS\,15c, which are characterized by the detection of a declining, optically thin synchrotron source consistent with the radio counterparts of other, previously observed stripped-envelope CCSNe. 

SPIRITS\,17lb, our most luminous IR transient at $M_{[4.5]} = -18.2$, clearly falls in the IR luminosity range of CCSNe (see Figure~\ref{fig:IR_lcs}), and is more luminous than any other class of known IR transient. The IR light curves and color evolution are consistent with either a hydrogen-rich Type II or stripped-envelope Type Ib/c or IIb. The near-IR spectra, taken at phases of $123$ and $157$~days, show no strong features indicative of a CCSN or specific SN subtype. There is a possible weak detection of the CO $\Delta v = 2$ features in the $123$~day $K$-band spectrum, which is not present at $157$~days. We note that some SNe~II, including the recent SN~2017eaw, have shown strong CO $K$-band features that may fade on timescales of months \citep[e.g.,][]{tinyanont19}, but this does not provide a definitive classification for SPIRITS\,17lb. As discussed in Section~\ref{sec:analysis_radio}, we detected a declining, optically thin synchrotron source at the location of SPIRITS\,17lb, confirming this source as a CCSN. Our constraints on the peak radio luminosity and time of the synchrotron peak are most consistent with previously observed radio SNe of Type IIP. This rules out a strongly interacting SN~IIn, as the observed emission is optically thin and thus not strongly self-absorbed by a dense wind, and disfavors a typically more luminous stripped-envelope event. Assuming a steady pre-SN wind, we derive constraints on the mass-loss rate of $\dot{M} \lesssim 8.0\times10^{-4} \left(\frac{\epsilon_{B}}{0.1}\right) \left(\frac{v_w}{100~\mathrm{km~s}^{-1}}\right) M_{\odot}$~yr$^{-1}$, consistent with an RSG progenitor. We infer a lower limit on the extinction to SPIRITS\,17lb of $A_V \gtrsim 2.5$ in Section~\ref{fig:multiband_lcs} comparing to the Type~IIP SN\,2004et. 

The evidence for SPIRITS\,15ud, SPIRITS\,16ix, and SPIRITS\,16tn is more circumstantial, but we find a reddened CCSN to be a plausible interpretation for each of these events. For SPIRITS\,15ud, at $M_{[4.5],\mathrm{peak}} = -16.4$ the IR light curves are broadly consistent with an SN~II, stripped-envelope Type Ib/c or IIb, or alternatively, an SN~2008S-like ILRT. Given the large uncertainty in the age of SPIRITS\,15ud at discovery ($< 381.4$~days), it is likely that our \textit{Spitzer} observations missed the peak of this event and that its true peak luminosity was substantially higher. This would place SPIRITS\,15ud firmly in the IR luminosity range characteristic of CCSNe and disfavors the ILRT interpretation. Despite the large uncertainty in its age in Section~\ref{fig:multiband_lcs} and under the assumption that SPIRITS\,15ud was an SN~II, we placed a lower limit on the extinction of $A_V \gtrsim 3.7$. This interpretation is further supported by the location of SPIRITS\,15ud in a prominent, nearly opaque dust lane in archival \textit{HST} imaging. 

We examined the observational constraints for SPIRITS\,16tn in \citet{jencson18c} and again found an SN~II, possibly a low-luminosity event similar to SN~2005cs, heavily obscured by $A_V = 7-9$ to be the most likely interpretation of this event. A new piece of the puzzle is the possible identification of water vapor absorption in the near-IR spectrum of SPIRITS\,16tn (Section~\ref{sec:spec}). In the CCSN scenario, a possibility is that the progenitor was encased in a dense molecular cloud, and water in the vicinity was heated by the explosion to produce the observed, broad absorption. The high extinction for SPIRITS\,16tn is consistent with this interpretation. This would constitute the first direct identification of a CCSN associated with a molecular cloud, though Galactic candidates for SN remnants interacting with molecular gas have been previously noted \citep[e.g., W44 and IC~443,][]{chevalier99}. Interestingly, \citet{chevalier99} argued that the progenitors of such events must be relatively low mass, i.e., early B-type stars on the main sequence (8--12~$M_{\odot}$), as the ionizing flux from a more massive O-type progenitor will clear a region ${\approx}15$~pc in radius of molecular material. Our inferred limit on the pre-SN mass-loss rate from radio observations of SPIRITS\,16tn was also consistent with a lower-mass RSG progenitor, which may add further support to this hypothesis. 

Alternatively, given the apparent similarity of the water absorption features to those of late-type Mira variables, SPIRITS\,16tn may represent a previously unknown class of transient associated with such stars. It is unclear, however, that water in the atmosphere would survive a luminous outburst or mass-loss event. Finally, an additional, though unlikely, possibility is chance positional coincidence between SPIRITS\,16tn and an unrelated Mira variable. Continued monitoring for ongoing Mira-like variability in the near-IR is required to confirm or rule out this possibility. 

SPIRITS\,16ix is a near twin of SPIRITS\,16tn in their IR properties, and thus we also consider a CCSN, possibly another weak or low-luminosity SN~II, as a plausible interpretation. SPIRITS\,16tn was clearly associated with active star-formation, indicating a likely massive-star origin further suggestive of a CCSN. We note that for SPIRITS\,16ix, however, there is no clear evidence of star-formation at the site, and given the classification of the host as a lenticular S0 galaxy, the association is more ambiguous. We noted in \citet{jencson18c} that given the low explosion energy inferred for SPIRITS\,16tn and lack of spectroscopic features distinctive of a CCSN, a nonterminal ``impostor''  explosion or eruption remains a viable scenario. Neither SPIRITS\,16ix nor SPIRITS\,16tn has shown any evidence of prior or subsequent IR variability other than the singular events described in this work. This further supports our interpretation that they were isolated, and possibly terminal, events in the lives of their progenitors. It is particularly notable that SPIRITS\,16ix and SPIRITS\,16tn, while distinctly similar to each other in their IR light curves, are also unique compared to the rest of the IR- and optically discovered transients. Furthermore, they are the two most severely reddened possible CCSNe ($A_V \gtrsim 5.5$) presented in this work. We thus speculate that they may represent a new class of IR-dominated transients, possibly associated with low-energy CCSNe arising preferentially in particularly extinguished environments and/or from progenitors that are directly associated with or encased in dense molecular clouds.

\subsection{ILRT: SPIRITS\,15ade}
As a distinct class of IR-dominated events, the nature of ILRTs remains unclear. The prototypical objects are the ``impostor'' SN~2008S and NGC~300 2008OT-1 \citep{bond09}. ILRTs are observed to have dust-obscured, IR-luminous ($M_{[4.5]} < -10$; $\nu L_{\nu} > 3.7\times10^{3}$) pre-explosion counterparts, suggested to be extreme AGB stars of intermediate mass (${\approx}10$--$15~M_{\odot}$) self-obscured by a dusty wind \citep{prieto08,bond09,thompson09}. The recently discovered transient in M51 (M51~OT2019-1) was also proposed as a new member of this class. Its likely progenitor was a similar dust-obscured star, with significant IR variability detected in archival \textit{Spitzer}/IRAC imaging over the past ${\approx}12$\,yr \citep{jencson19}. Less luminous than typical CCSNe at peak, emission lines in their spectra (including H recombination and strong [Ca~\textsc{ii}] and Ca~\textsc{ii} features) also indicate lower velocities up to few ${\times}100$\,km~s$^{-1}$ \citep[e.g.,][]{bond09,botticella09,humphreys11}. A suggested physical scenario involves weak explosion, possibly an electron-capture SN (disfavored for the slow rise and extended phase of early dust destruction observed in M51~OT2019-1; \citealp{jencson19}), or massive stellar eruption that initially destroys the obscuring, circumstellar dust and produces a short-lived optical transient. In the aftermath, the development of a significant IR excess indicates the recondensation of dust and consequent reobscuration of the transient \citep{thompson09,kochanek11,szczygiel12}. There is currently no evidence that the progenitor stars survive such events, as both prototypes have now faded below their pre-explosion luminosities in the IR \citep{adams16a}.

SPIRITS\,15ade shares several properties with this class. While we do not directly detect a progenitor star, our limit from archival \textit{Spitzer}/IRAC imaging at $M_{[4.5]}$ fainter than $-12.8$ is consistent with the IR progenitors of SN\,2008S and NGC~300 OT2008-1. As we found in Section~\ref{fig:multiband_lcs}, the IR light curves (peaking at $M_{[4.5]} = -15.7$) and the observed peak and duration of the associated optical transient are also very similar to known ILRTs. Finally, we detect H recombination emission in our near-IR spectra, suggestive of an outflow at 300--400\,km~s$^{-1}$, similar to the low expansion velocities of known ILRTs \citep[e.g.,][]{bond09,botticella09,humphreys11}.

\subsection{Possible LRN: SPIRITS\,14azy}
The class of extragalactic transients referred to as luminous red novae (LRNe) are believed to be more massive analogs of the population of stellar mergers observed in the Galaxy, including the striking example of the ${\approx}1$--$3~M_{\odot}$ contact binary merger V1309~Sco \citep{tylenda11,mccollum14} and the B-type stellar merger V838~Mon \citep{bond03,sparks08}. While sharing several properties with ILRTs, LRNe are distinctly characterized by multipeaked, irregular light curves. At early times, their spectra show H emission features, but they develop red optical colors, atomic and molecular absorption features, and significant IR excesses as they evolve. Recent examples include the 2011 transient in NGC~4490 (NGC~4490 OT2011-1; \citealp{smith16}), and the 2015 event in M101 (M101 OT2015-1; \citealp{blagorodnova17}). Specifically in the case of NGC~4490 OT2011-1, the late-time IR excess was too luminous to be explained as an IR echo, indicating the presence of a surviving, merged remnant. With unobscured, directly detected progenitor systems in archival imaging estimated at $20$--$30~M_{\odot}$ for NGC~4490 OT2011-1 and ${\approx}18~M_{\odot}$ for M101 OT2015-1, these events extend a correlation noted by \citet{kochanek14} for LRNe between progenitor masses and the transient peak luminosity. Specifically, the progenitor of M101 OT2015-1 was identified in archival imaging as having properties of an F-type supergiant with $L \approx 8.7 \times 10^4~L_{\odot}$ and $T_{\mathrm{eff}} \approx 7000$~K, likely having recently evolved off the main sequence and crossing the Hertzsprung gap as it expands. 

Based primarily on our analysis of its multiband light curves, we find that SPIRITS\,14azy is most consistent with the LRN class of IR transients, and in particular bears strong similarity to M101 OT2015-1 in its peak luminosity at both [4.5] and $g^{\prime}$. Thus, we suggest that it may have a similar mass progenitor system. In Section~\ref{sec:archival_imaging}, we placed a limit on the flux of the progenitor in the HST/WFC3 UVIS F336W filter corresponding to $M_U$ fainter than $-5.4$. As the transient is located in a prominent dust lane of the host, the foreground extinction in $U$ band may be significant. Even in the absence of host extinction, and assuming bolometric corrections from \citet{flower96}, with corrections by \citet{torres10} using a consistent choice of $M_{\mathrm{bol},\odot} = 4.73$ and \citet{crowther97} filter transformations, this limit is approximately consistent with a zero-age main-sequence O9 star ($T_{\mathrm{eff}} \approx 30,000$~K, $L\approx40,000~L_{\odot}$) of similar mass to the progenitor of M101 OT2015-1. 

\subsection{MSEs: SPIRITS\,17pc and SPIRITS\,17qm}
SPIRITS\,17pc has undergone multiple, IR-dominated outbursts over a time period spanning at least $1100$~days. These outbursts are extremely red (we infer $A_V \approx 12.5$), possibly indicative of copious dust formation. Near-IR spectra taken during the most recent, longest-duration, and most luminous outburst show features similar to a mid- to late G-type supergiant spectrum, including CN and CO absorption. An optical spectrum shows double-peaked \ion{Ca}{2} emission, but no H or He features, indicative of a cool, few ${\times}100$\,km~s$^{-1}$ outflow. We identified a luminous ($L \approx 2\times10^5~L_{\odot}$) star detected in 2011 archival, multiband \textit{HST} imaging at the precise location of SPIRITS\,17qm as a likely progenitor. The luminosity is lower than that of a classical LBV but still likely indicates a very massive progenitor. The progenitor photometry is consistent with a blackbody temperature of $T = 1900 \pm 100$~K, likely indicating that active dust formation was occurring around the underlying star. Given the previous history of strong IR variability, cool spectral features, low outflow velocities, and lack of strong radio emission, we argue that the most recent outburst of SPIRITS\,17pc represents a more intense mass-loss event or eruption, rather than a terminal SN explosion. 

Similar to SPIRITS\,17pc, SPIRITS\,17qm has also undergone multiple IR outbursts in SPIRITS, and furthermore was also extremely red (estimated $A_V = 12.1$). However, we highlight several notable differences. The spectra of SPIRITS\,17qm were dominated by broader ($v \approx 1500$-$2000$\,km~$^{-1}$) features of H and He and also showed prominent \ion{Ca}{2} emission, consistent with a giant LBV eruption. Again, the lack of strong radio emission argues against the interpretation of SPIRITS\,17qm as a terminal, Type~IIn CCSN explosion. The presumed progenitor of SPIRITS\,17qm detected in multiepoch $HST$ imaging was highly luminous and variable ($L \approx 2\times10^{6}~L_{\odot}$, $\Delta V = 1.7$), also consistent with the properties of an LBV. Assuming that the progenitor was undergoing S Doradus-like variability, the foreground extinction (host or circumstellar) was likely no larger than $A_V \approx 0.5$ at the time of the $\textit{HST}$ observations in 2001. Thus, we suggest that the extreme optical/IR color observed for SPIRITS\,17qm may be due to the formation of copious dust in the intervening years and/or during the observed giant eruption. Incorporating ongoing monitoring, the physical details of SPIRITS\,17qm and SPIRITS\,17pc as self-obscuring MSEs will be explored more fully in a future publication. 

\section{The ``Redness'' Distribution of Luminous IR transients}\label{sec:redness}
\begin{figure}
\plotone{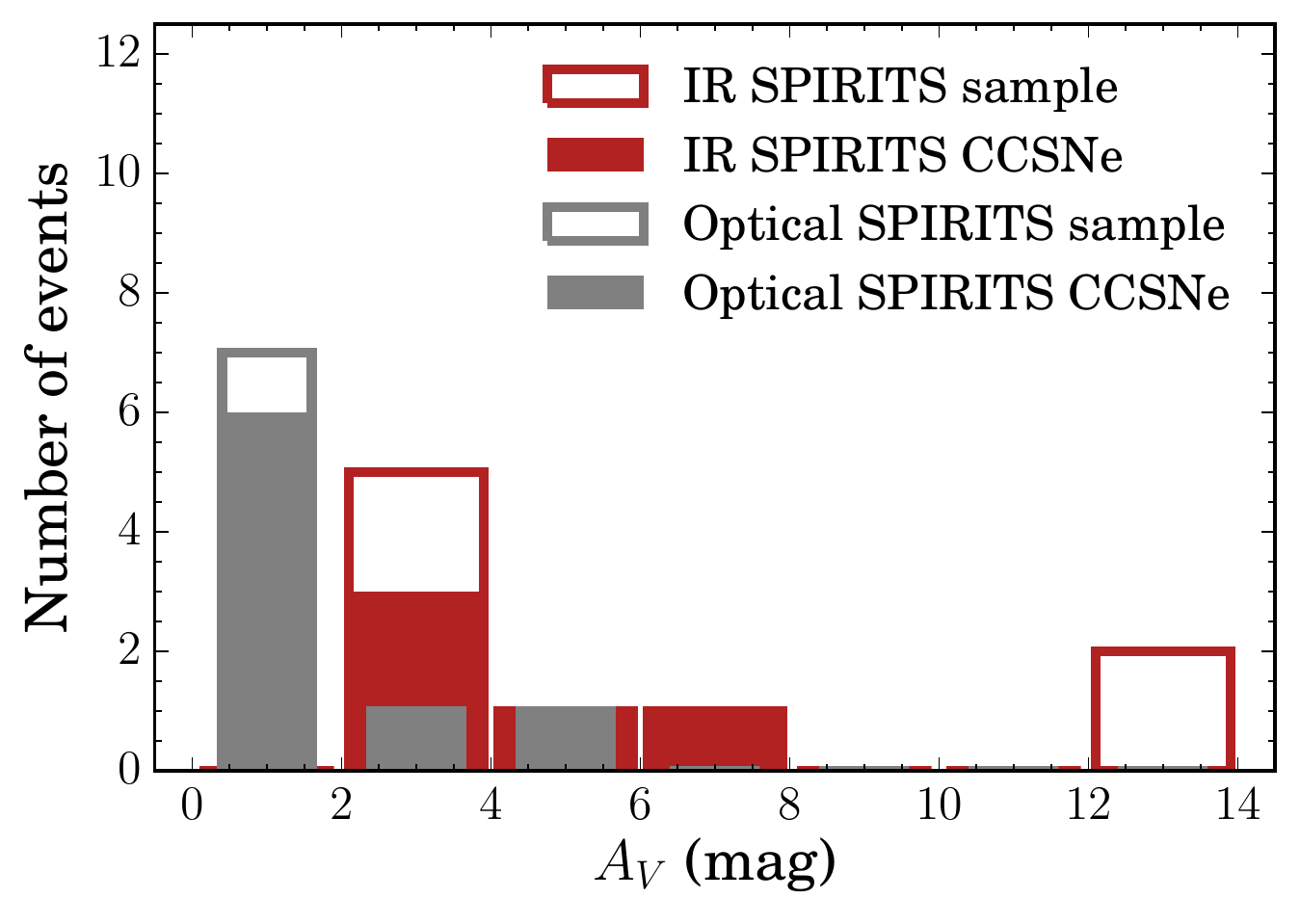}
\caption{\label{fig:AV_control_hist}
Distribution of $A_V$ inferred for the IR-selected sample of luminous SPIRITS transients (red) and the optically selected control sample (gray). The filled regions of the histograms represent confirmed or possible CCSNe. The unfilled regions include the Type~Iax SN~2014dt from the control sample and the non-SN stellar outbursts from the IR sample. 
}
\end{figure}

In Figure~\ref{fig:AV_control_hist}, we show the distribution of $A_V$ (including lower limits) inferred for our sample of luminous SPIRITS transients. This includes confirmed and plausible CCSNe as the filled portions of the histograms, as well the Type Iax SN~2014dt from the optical control sample and the non-SN IR transients as the unfilled portions. For the non-SN transients, we emphasize that our $A_V$ estimates are only a proxy for the observed $\mathrm{optical}{-}\mathrm{IR}$ color and may be indicative of internal reddening (due to dust formation) or the intrinsically cool SEDs, rather than external extinction. We note that our IR-discovered events are overall much redder, all nine objects having $A_V \gtrsim 2$, compared to the optical control sample with seven of nine objects having $A_V \lesssim 2$. In particular, the MSEs SPIRITS\,17pc and SPIRITS\,17qm stand out as the reddest events in our sample with $A_V > 12$. For the 5 IR-discovered events interpreted as possible CCSNe, the estimates range from $2.2 \lesssim A_V \lesssim 7.8$. The optically discovered CCSNe in our control sample are again notably less reddened, ranging from $A_V \sim 0$ up to $4.3$, with six of the eight events having $A_V \lesssim 2$. This suggests that the IR-selected sample of luminous SPIRITS transients may represent populations of much redder events that are largely inaccessible to optical searches, namely, heavily obscured CCSNe and exceptionally red or self-obscuring transients associated with massive-star outbursts of various origins. 

\begin{figure*}
\begin{minipage}[hpb]{180mm}
\centering
\gridline{\fig{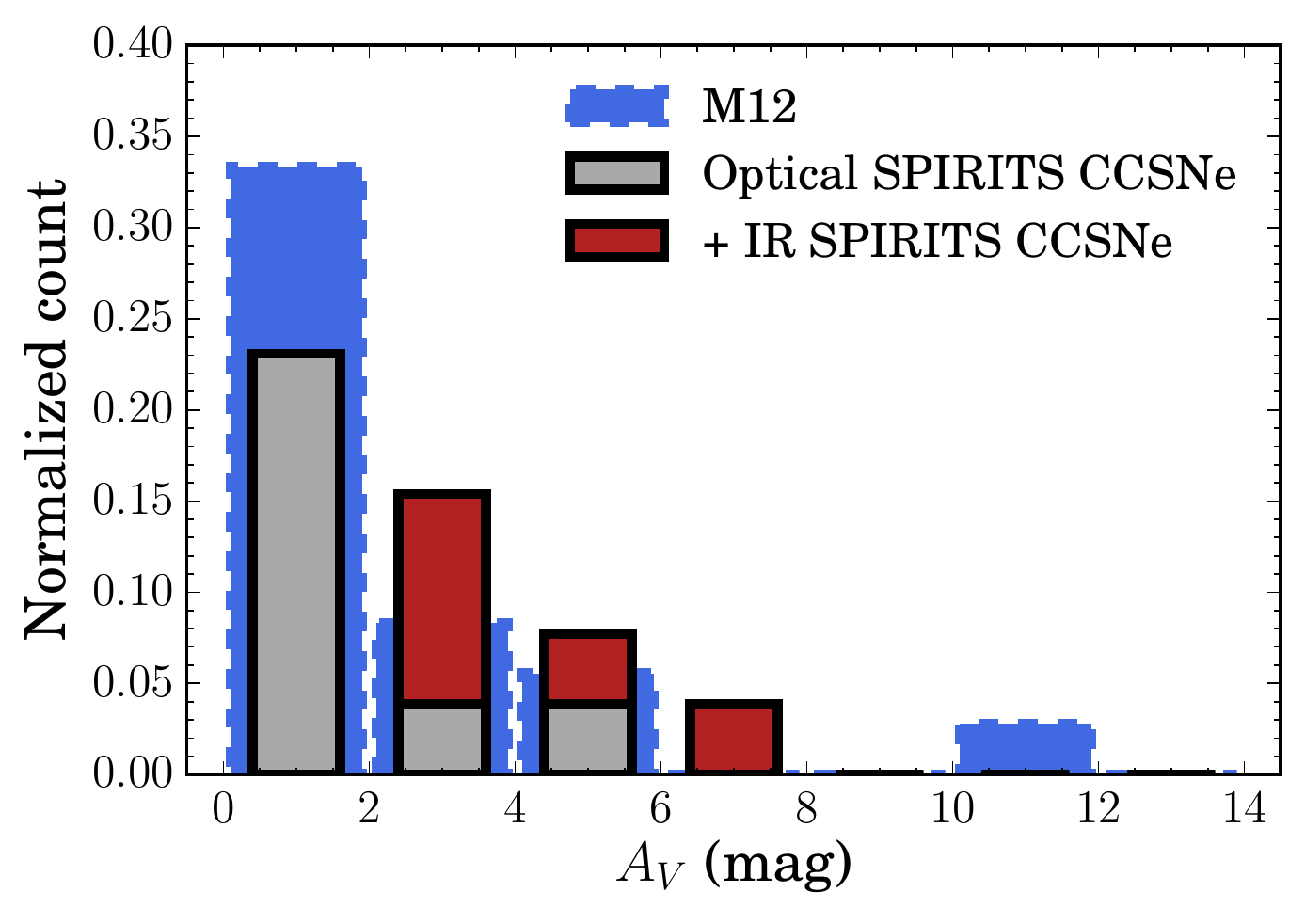}{0.5\textwidth}{}
          \fig{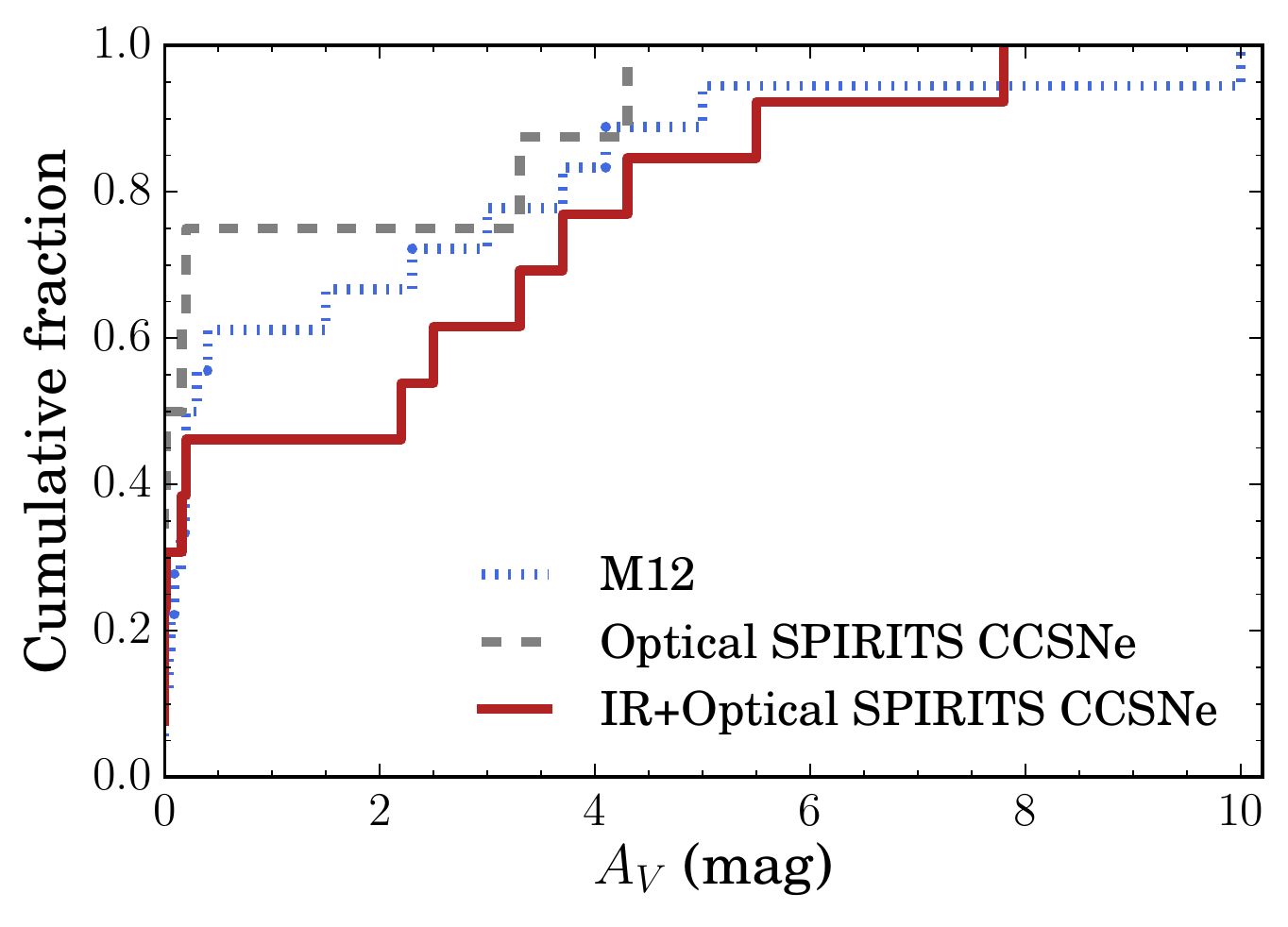}{0.5\textwidth}{}}
\caption{\label{fig:AV_hists_M12}
\textbf{Left}: normalized distribution of $A_V$ for CCSNe from M12 sample, and the sample of confirmed/possible CCSNe in SPIRITS (filled histogram), including events from the optically discovered control sample (gray) and IR-discovered events (red). \textbf{Right:} cumulative distribution functions in host $A_V$ estimates for M12 (blue short-dashed curve), optically known SPIRITS CCSNe from the control sample (gray dashed curve), and the full sample of likely SPIRITS CCSNe including both optically known and IR-discovered events (red solid curve).
}
\end{minipage}
\end{figure*}

The sample of all known CCSNe discovered between 2000 and the end of 2011 hosted by galaxies within 12 Mpc along with literature estimates of their host extinction compiled by \citet[][hereafter M12]{mattila12} provides another useful comparison for the SPIRITS sample. In Figure~\ref{fig:AV_hists_M12}, we directly compare the distribution of host $A_V$ for the SPIRITS sample of confirmed and plausible CCSNe (both optically and IR-selected events) to the M12 sample. We note that M12 excluded five SNe in galaxies with inclinations $>$60$^{\circ}$ from their primary analysis, as they believed that their volume-limited sample was incomplete for SNe in highly inclined hosts. We first include these events here for a more direct comparison to the SPIRITS sample, which includes some highly inclined galaxies. In the SPIRITS sample, 7 of the 13 CCSNe (58.3\%) have $A_V \gtrsim 2$, and the median value is $2.2$. For M12, only 6 of 18 events (33.3\%) have $A_V \gtrsim 2$, and the median is significantly lower at $A_V = 0.25$. Shown as cumulative distributions in the right panel of Figure~\ref{fig:AV_hists_M12}, we find that the M12 sample is bracketed between the optically known CCSNe in SPIRITS and the full SPIRITS sample including the IR-discovered events, where again a larger fraction of events were found with large host extinction. Here it is particularly important to reemphasize that for three of the five IR-discovered CCSNe in the SPIRITS sample, our $A_V$ estimates are only lower limits, meaning that the true distribution may be even more skewed to large host extinctions. 

If we now compare only those events in normal galaxies at inclinations $<$60$^{\circ}$, the M12 sample contained 2 of 13 SNe (15\%) that appeared as outliers above the $A_V$ distribution expected for dust smoothly distributed in disk galaxies with $A_V \gtrsim 3.5$\,mag. The fraction in the SPIRITS sample (excluding events in highly inclined hosts and IC\,2163, which is part of an interacting pair and a borderline LIRG; see our discussion of galaxy selection biases below in Section~\ref{sec:rates}) is twice as high, with two of six events (33\%) with $A_V > 3.5$\,mag (SPIRITS\,15ud and SN\,2016bau). It is difficult to draw strong conclusions from this comparison given the small number of events included. It may hint, however, that the fraction of CCSNe suffering high host extinction, above that expected solely from inclination effects, and thus missing from optical searches, may be higher than previously estimated.

\section{SPIRITS Constraints on the Optically Missed Fraction of Nearby CCSNe}\label{sec:rates}
We now consider constraints from our sample on the fraction of optically missed CCSNe in nearby galaxies, an important consideration for measurements of the local CCSN rate. $N_{\mathrm{CCSNe}} = 13$ confirmed/plausible CCSNe were recovered by SPIRITS and pass our selection criteria, with $N_{\mathrm{mCCSNe}} = 5$ events unreported by any optical search. If we assume that the completeness of our IR survey to optically discovered and optically missed SNe is the same (we examine this assumption in more detail below), then the fraction of missing CCSNe, $f$, is given by the binomial distribution

\begin{equation}
    P\left(f\right) \propto \left(1-f\right)^{N_{\mathrm{CCSNe}}} f^{N_{\mathrm{mCCSNe}}}
\end{equation}

\noindent and the requirement that $\int_{0}^{1} P\left(f\right) df = 1$ sets the normalization. The nominal, observed missed fraction is $0.385$ with a 90\% confidence interval of $0.166<f<0.645$.

Since 2 of 10 optically discovered and classified CCSNe, SN\,2014bc and SN\,2016adj, were not recovered in SPIRITS and selected as part of our sample with the criteria outlined in Sections~\ref{sec:obsc_sample} and \ref{sec:control_samp}, our selection efficiency for optically known CCSNe is $0.8$. SN\,2016adj in Cen A was itself flagged as a saturated source in our transient identification pipeline. An optically obscured CCSN in a galaxy as near as Cen~A at a similar IR brightness would have been similarly flagged, and it is natural to assume that our survey incompleteness to optically bright and optically obscured CCSNe in the nearest galaxies ($D \lesssim 5$\,Mpc) is the same. SN\,2014bc was located in the saturated core of NGC~4285, and again we may expect our incompleteness to optically missed events located near the core of their host galaxies to be the same. Still, we may obtain a more conservative estimate of $f$ if we assume that our sample was complete in optically missed events. With $N_{\mathrm{CCSNe}} = 15$ total CCSNe in SPIRITS galaxies, and again with $N_{\mathrm{mCCSNe}} = 5$, we obtain a missed fraction of 0.333 with a 90\% confidence interval of $0.142<f<0.577$. 

Next, we consider the effects of selection biases in our galaxy sample. For the first 3\,yr (2014--2016), the full sample of 190 SPIRITS galaxies was more representative across galaxy types, though primarily composed of luminous and massive galaxies beyond 5\,Mpc. IC\,2163, producing both SPIRITS\,15c and SPIRITS\,17lb, also stands out from the sample in that with recent distance estimates it lies much farther than the rest of the sample at $35.5$\,Mpc, is currently undergoing strong tidal interactions with its companion galaxy, NGC~2207, and is a borderline LIRG at $L_{\mathrm{IR}} \approx 10^{11}~L_{\odot}$ \citep{sanders03}. Including only events discovered during the original 3\,yr survey, and further excluding SPIRITS\,15c in IC\,2163, we have $N_{\mathrm{CCSNe}} = 8$ and $N_{\mathrm{mCCSNe}} = 3$, giving $f=0.375$ with a 90\% confidence interval of $0.111<f<0.711$. This is notably similar to our nominal estimate from the full sample, but with a larger 90\% confidence interval due to the smaller number of events. 

Finally, we consider the possibility that some events interpreted here as CCSNe have been misclassified. SPIRITS\,16tn and SPIRITS\,16ix have uniquely fast-fading IR light curves compared to the full sample of CCSNe observed by \textit{Spitzer}, and while SPIRITS\,16tn is clearly associated with a region of active star formation, there is not such a clear association for SPIRITS\,16ix. Excluding these two events, we have $N_{\mathrm{CCSNe}} = 11$ and $N_{\mathrm{mCCSNe}} = 3$, giving $f=0.273$ with a 90\% confidence interval of $0.079<f<0.564$. 

Our data suggest an optically missed fraction of about $1/3$ and cannot rule out a value twice as high. Such a large missing fraction could help explain the claimed factor of 2 discrepancy by \citet{horiuchi11} between the measured rate of CCSNe and rate of cosmic star formation at $z=0$, but the precision of our measurement is limited by our small sample size and remaining ambiguous classifications. We also note that more recent volumetric CCSN rate estimates from the Supernova Diversity and Rate Evolution (SUDARE; \citealp{botticella13}) have challenged the claim by \citet{horiuchi11} and suggest better agreement between CCSN and star-formation rates at medium redshift ($0.2 < z < 0.8$; \citealp{cappellaro15}). In the other direction, some local studies have even suggested that CCSNe may be overproduced compared to H$\alpha$- and UV-inferred star formation rates \citep{botticella12,horiuchi13,xiao15}, and interestingly, accounting for the population of heavily obscured nearby events uncovered by SPIRITS may further increase this tension.

We also now place our estimate in the context of other recent work to constrain the missing CCSN fraction and uncover events buried in densely obscured environments. M12 estimated that locally $18.9^{+19.2}_{-9.5}$\% of CCSNe are missed by optical surveys, substantially higher than previous estimates by \citet{mannucci07} at 5\%--10\%. Our nominal estimate of $38.5^{+26.0}_{-21.9}$\% of CCSNe missed in nearby galaxies is even higher, falling at the upper end of the allowed range by M12, but with substantial overlap with our 90\% confidence interval. The estimates of \citet{mannucci07} and M12 rise steeply to $>30$\% and $\sim 40$\% by $z = 1$, respectively, and substantial work has also been dedicated to uncovering missing supernovae at higher redshift, particularly in the densely obscured and highly star-forming nuclear regions of starburst galaxies and (U)LIRGs. With seeing-limited imaging \citep[e.g.][]{mannucci03,miluzio13} and high-resolution space- or ground-based adaptive optics imaging \citep{cresci07,mattila07,kankare08,kankare12,kool18}, a total of 16 individual CCSNe have been uncovered in the IR in these galaxies. Radio very long baseline interferometry (VLBI) imaging of multiple (U)LIRGs \citep[e.g.,][and references therein]{lonsdale06,perez-torres09, romero-canizales11,romero-canizales14,bondi12,varenius17} has also revealed scores of radio sources interpreted as SN and SN remnants. Even with the resolution of the upcoming \textit{James Webb Space Telescope}, probing the innermost nuclear regions of such systems will remain challenging, and recently \citet{yan18} undertook a successful pilot study examining the spatially integrated IR light curves of high-redshift (U)LIRGS for variability suggestive of ongoing SN explosions. Finally, deep, targeted optical searches of nearby galaxies may also achieve the sensitivity necessary to uncover heavily extinguished events, e.g., SN~2016ija uncovered by the $D < 40$\,Mpc
(DLT40) SN search in NGC 1532 with $A_V \approx 6$ \citep{tartaglia18}. In this context, we emphasize that our work represents the first attempt to uncover the obscured population and directly constrain the missing fraction of CCSNe in nearby galaxies in the IR. 

\section{Summary and Conclusions}\label{sec:summary}
We have presented a sample of nine luminous IR transients discovered in nearby ($D \lesssim 35$\,Mpc) galaxies by the SPIRITS survey between 2014 and 2018. These events were selected as having $M_{\mathrm{IR}}$ brighter than $-14$ ($\nu L_{\nu} > 1.5\times10^{5}~L_{\odot}$ at [4.5]) in either the [3.6] or [4.5] channels of \textit{Spitzer}/IRAC, and we required at least two SPIRITS detections and that these events were not present in the first epoch of SPIRITS imaging. Combining archival imaging constraints from \textit{Spitzer} and \textit{HST}, concomitant monitoring of SPIRITS galaxies with several ground-based telescopes, coverage in the iPTF survey from Palomar Observatory, and a dedicated ground- and space-based follow-up effort for our transients at optical, near-IR, and radio wavelengths, we construct detailed observational characterizations and attempt to determine their nature. Here we summarize our analysis of the sample and primary conclusions:

\begin{enumerate}
\item The IR-discovered sample of transients were predominantly found in the spiral arms of star-forming hosts, except for one event, SPIRITS\,16ix, in a lenticular S0 galaxy. This strongly suggests an association with young stellar populations and massive stars. They span IR luminosities with $M_{[4.5],\mathrm{peak}}$ between $-14$ and $-18.2$, show IR colors between $0.2 < [3.6]{-}[4.5] < 3.0$, and fade on timescales between $55$~days~$< t_{\mathrm{fade}} < 480$~days. 

\item We define a control sample of optically discovered and classified transients recovered in SPIRITS using the same selection criteria (as outlined in Section~\ref{sec:obsc_sample}). The control sample also consists of nine events, including known eight CCSNe and one SN~Iax. Overall, the control sample and the IR-selected sample are similar in their IR properties, with similar distributions in $M_{[4.5],\mathrm{peak}}$, their $[3.6]{-}[4.5]$ colors at peak, and $t_{\mathrm{fade}}$.

\item Of the nine IR-discovered events, we suggest that five may be significantly dust-obscured CCSNe. SPIRITS\,15c was confirmed via a near-IR spectrum similar to the Type~IIb SN~2011dh and detection of a radio counterpart consistent with an SN~Ib/IIb \citep{jencson17}. SPIRITS\,17lb, as the most luminous IR transient in our sample, and despite a lack of unambiguous SN features in its near-IR spectrum, was confirmed via its radio counterpart most consistent with an SN~II. Our radio data constrain the pre-SN mass-loss rates of the progenitors of SPIRITS\,15c and SPIRITS\,17lb to $\dot{M} \lesssim 1.4\times10^{-3} \left(\frac{\epsilon_{B}}{0.1}\right) \left(\frac{v_w}{100~\mathrm{km~s}^{-1}}\right) M_{\odot}$~yr$^{-1}$ and $\dot{M} \lesssim 8.0\times10^{-4} \left(\frac{\epsilon_{B}}{0.1}\right) \left(\frac{v_w}{100~\mathrm{km~s}^{-1}}\right) M_{\odot}$~yr$^{-1}$, respectively. The IR light curve of SPIRITS\,15ud is also most consistent with a CCSN classification. SPIRITS\,16ix and SPIRITS\,16tn may constitute a previously unknown class of rapidly fading, luminous IR transients occurring in the most densely obscured environments. We suggest that these two events are also CCSNe, possibly directly associated with or occurring within dense molecular clouds. 

\item The four remaining luminous IR SPIRITS transients represent a diverse array of IR-dominated events arising from massive-star progenitor systems. The multiband optical and IR light curves of SPIRITS\,14azy and SPIRITS\,15ade are most similar to an LRN (possible massive stellar merger similar to M101 OT2015-1) and ILRT (possible weak or electron-capture SN similar to SN~2008S), respectively. SPIRITS\,17pc underwent multiple, extremely red outbursts over several years, now showing G-type spectral features in the near-IR and double-peaked \ion{Ca}{2} emission indicative of a few ${\times}100$\,km~s$^{-1}$ outflow. We identified a luminous ($L \approx 2\times10^5~L_{\odot}$) star detected in multiband archival \textit{HST} imaging as a likely progenitor. SPIRITS\,17qm, also undergoing multiple IR-dominated outbursts, showed high-velocity ($\sim 1500$--$2000$\,km~s$^{-1}$) H and He features consistent with a giant LBV eruption. We identified a visually luminous ($L \approx 10^6~L_{\odot}$), highly variable progenitor in archival \textit{HST} imaging. Future observations and continued monitoring of SPIRITS\,17pc and SPIRITS\,17qm may determine the physical origin of their IR outbursts and constrain the ultimate fate of their progenitors. 

\item We estimated the visual extinction, $A_V$---either as a direct estimate of the host extinction (in the case of a CCSN) or as a proxy for the $\mathrm{optical}{-}\mathrm{IR}$ color for non-SN transients that may be self-obscured or intrinsically red---for each event in our sample based on their multiband optical and IR light curves. All nine IR-discovered events, including the five confirmed/plausible CCSNe, have $A_V \gtrsim 2$, while only two of the nine objects in the optically discovered control sample have extinction estimates so high. Compared to a volume- and time-limited sample of CCSNe within 12\,Mpc compiled by \citet{mattila12}, we find that the SPIRITS sample of optically known and newly IR-discovered CCSNe contains a larger fraction of significantly obscured events ($A_V \gtrsim 2$). Furthermore, 7 out of the 13 plausible CCSNe detected by SPIRITS were significantly obscured. This makes it likely that optical transient/SN searches at larger distances are very incomplete because they miss highly obscured events.
    
\item A total of 13 confirmed/plausible CCSNe recovered in SPIRITS passed our selection criteria, including 8 optically discovered events in the control sample and 5 IR-discovered events presented here. For our galaxy sample, this implies that a large fraction, as high as $38.5^{+26.0}_{-21.9}$\% (90\% confidence), of CCSNe are being missed by optical surveys in nearby galaxies. 
\end{enumerate}

\acknowledgments
We thank the anonymous referee for their thorough read of the paper and helpful comments. We thank C.\ Contreras and the staff of Las Campanas Observatory for help with conducting observations and data reduction. We thank E.\ Hsiao for helpful commentary and advice in the analysis and discussion of the spectra. We thank K.\ Mooley, D.\ Dong, M.\ Anderson, M.\ Eastwood, and A.\ Horesh for helpful advice and assistance with VLA data reduction. We thank S.\ Mattila and T.\ Reynolds for valuable discussions in revising this work.

This material is based on work supported by the National Science Foundation Graduate Research Fellowship under grant No. DGE-1144469. H.E.B.\ acknowledges that support for \textit{HST} program Nos. GO-13935, GO-14258, and AR-15005 was provided by NASA through grants from the Space Telescope Science Institute, which is operated by the Association of Universities for Research in Astronomy, Incorporated, under NASA contract NAS5-26555. R.D.G. was supported by NASA and the United States Air Force. This work is part of the research program VENI, with project No. 016.192.277, which is (partly) financed by the Netherlands Organisation for Scientific Research (NWO).

This work is based in part on observations made with the {\it Spitzer Space Telescope\/}, which is operated by the Jet Propulsion Laboratory, California Institute of Technology, under a contract with NASA. This work is based in part on observations with the NASA/ESA {\it Hubble Space Telescope\/} obtained at the Space Telescope Science Institute and from the Mikulski Archive for Space Telescopes at STScI, which are operated by the Association of Universities for Research in Astronomy, Inc., under NASA contract NAS5-26555. We acknowledge the use of public data from the \textit{Swift} data archive.

Some of the data presented herein were obtained at the W. M. Keck Observatory, which is operated as a scientific partnership among the California Institute of Technology, the University of California, and the National Aeronautics and Space Administration. The Observatory was made possible by the generous financial support of the W. M. Keck Foundation. The authors wish to recognize and acknowledge the very significant cultural role and reverence that the summit of Maunakea has always had within the indigenous Hawaiian community.  We are most fortunate to have the opportunity to conduct observations from this mountain. 

Part of the optical and near-infrared photometric data included in this paper were obtained by the Carnegie Supernova Project through the support of NSF grant AST-1008343.

Based on observations obtained at the Gemini Observatory acquired through the Gemini Observatory Archive and processed using the Gemini IRAF package, which is operated by the Association of Universities for Research in Astronomy, Inc., under a cooperative agreement with the NSF on behalf of the Gemini partnership: the National Science Foundation (United States), National Research Council (Canada), CONICYT (Chile), Ministerio de Ciencia, Tecnolog\'{i}a e Innovaci\'{o}n Productiva (Argentina), Minist\'{e}rio da Ci\^{e}ncia, Tecnologia e Inova\c{c}\~{a}o (Brazil), and Korea Astronomy and Space Science Institute (Republic of Korea).

UKIRT is owned by the University of Hawaii (UH) and operated by the UH Institute for Astronomy; operations are enabled through the cooperation of the East Asian Observatory. When the data reported here were acquired, UKIRT was supported by NASA and operated under an agreement among the University of Hawaii, the University of Arizona, and Lockheed Martin Advanced Technology Center; operations were enabled through the cooperation of the East Asian Observatory. 

Some of the data used in this paper were acquired with the RATIR instrument, funded by the University of California and NASA Goddard Space Flight Center, and the 1.5\,m Harold L.\ Johnson telescope at the Observatorio Astron\'omico Nacional on the Sierra de San Pedro M\'artir, operated and maintained by the Observatorio Astron\'omico Nacional and the Instituto de Astronom{\'\i}a of the Universidad Nacional Aut\'onoma de M\'exico. We acknowledge the contribution of Leonid Georgiev and Neil Gehrels to the development of RATIR.

The National Radio Astronomy Observatory is a facility of the National Science Foundation operated under cooperative agreement by Associated Universities, Inc. 

The Australia Telescope Compact Array is part of the Australia Telescope National Facility, which is funded by the Australian Government for operation as a National Facility managed by CSIRO. This paper includes archived data obtained through the Australia Telescope Online Archive (\url{http://atoa.atnf.csiro.au}).

The Digitized Sky Surveys were produced at the Space Telescope Science Institute under U.S. Government grant NAG W-2166. The images of these surveys are based on photographic data obtained using the Oschin Schmidt Telescope on Palomar Mountain and the UK Schmidt Telescope. The plates were processed into the present compressed digital form with the permission of these institutions. The Second Palomar Observatory Sky Survey (POSS-II) was made by the California Institute of Technology with funds from the National Science Foundation, the National Geographic Society, the Sloan Foundation, the Samuel Oschin Foundation, and the Eastman Kodak Corporation. The Oschin Schmidt Telescope is operated by the California Institute of Technology and Palomar Observatory. The UK Schmidt Telescope was operated by the Royal Observatory Edinburgh, with funding from the UK Science and Engineering Research Council (later the UK Particle Physics and Astronomy Research Council), until 1988 June, and thereafter by the Anglo-Australian Observatory. The blue plates of the southern Sky Atlas and its Equatorial Extension (together known as the SERC-J), as well as the Equatorial Red (ER) and the Second Epoch [red] Survey (SES), were all taken with the UK Schmidt. Supplemental funding for sky-survey work at the STScI is provided by the European Southern Observatory.

Funding for SDSS-III has been provided by the Alfred P. Sloan Foundation, the Participating Institutions, the National Science Foundation, and the U.S. Department of Energy Office of Science. The SDSS-III website is \url{http://www.sdss3.org/}. SDSS-III is managed by the Astrophysical Research Consortium for the Participating Institutions of the SDSS-III Collaboration, including the University of Arizona, the Brazilian Participation Group, Brookhaven National Laboratory, Carnegie Mellon University, University of Florida, the French Participation Group, the German Participation Group, Harvard University, the Instituto de Astrofisica de Canarias, the Michigan State/Notre Dame/JINA Participation Group, Johns Hopkins University, Lawrence Berkeley National Laboratory, Max Planck Institute for Astrophysics, Max Planck Institute for Extraterrestrial Physics, New Mexico State University, New York University, Ohio State University, Pennsylvania State University, University of Portsmouth, Princeton University, the Spanish Participation Group, University of Tokyo, University of Utah, Vanderbilt University, University of Virginia, University of Washington, and Yale University.

This publication makes use of data products from the Two Micron All Sky Survey, which is a joint project of the University of Massachusetts and the Infrared Processing and Analysis Center/California Institute of Technology, funded by the National Aeronautics and Space Administration and the National Science Foundation.

\facilities{\textit{Spitzer} (IRAC), \textit{HST} (WFPC2, WFC3), \textit{Swift} (UVOT), PO:1.5m (CCD), 	PO:1.2m, Swope (CCD), Du Pont (RetroCam), Magellan:Baade (FourStar, FIRE), Keck:I (LRIS, MOSFIRE), Keck:II (DEIMOS, NIRES), Hale (WIRC), Gemini:Gillett (GNIRS), Gemini:South (FLAMINGOS-2), UKIRT (WFCAM), OANSPM:HJT (RATIR), EVLA, ATCA, AMI}

\software{DOLPHOT \citep{dolphin00,dolphin16},\\ DAOPHOT/ALLSTAR package \citep{stetson87}, IRAF \citep{tody86,tody93}, Gemini IRAF package (\url{http://www.gemini.edu/sciops/data-and-results/processing-software}), LPipe \citep[][\url{http://www.astro.caltech.edu/~dperley/programs/lpipe.html}]{perley19}, MOSFIRE Data Reduction Pipeline (\url{https://keck-datareductionpipelines.github.io/MosfireDRP/}), Spextool \citep{cushing04}, CASA \citep{mcmullin07}.}

\bibliographystyle{yahapj}
\bibliography{SPIRITS_lum_transient_sample_arXivV2}


\end{document}